\documentclass[review=false, screen,acmsmall,nonacm]{acmart}
\AtBeginDocument{%
  \providecommand\BibTeX{{%
    \normalfont B\kern-0.5em{\scshape i\kern-0.25em b}\kern-0.8em\TeX}}}

\usepackage[colorinlistoftodos]{todonotes}
\usepackage{soul,xcolor}
\usepackage{booktabs}
\usepackage{multirow}
\usepackage{multicol}
\usepackage{bm}
\usepackage[capitalize,noabbrev]{cleveref}
\usepackage{tikz}
\usepackage{verbatim}
\usetikzlibrary{arrows.meta,shapes,positioning,shadows,trees}
\usepackage{url}
\usepackage{amsmath}
\usepackage{graphicx}
\usepackage{longtable}
\usepackage{enumitem}
\usepackage{titlesec}
\newcounter{mynode}
\tikzset{step node/.code={\stepcounter{mynode}
}}

\titleformat{\subsubsection}
{\bfseries}{\thesubsubsection}{1em}{}

\renewcommand{\thesubsubsection}{\thesubsection.\arabic{subsubsection}}

\DeclareMathOperator*{\Attn}{Attention}
\DeclareMathOperator*{\Softmax}{softmax}
\DeclareMathOperator*{\MultiHead}{MultiHeadAttn}
\DeclareMathOperator*{\Concat}{Concat}
\DeclareMathOperator*{\Head}{head}
\DeclareMathOperator*{\ReLU}{ReLU}
\DeclareMathOperator*{\SelfAttention}{SelfAttention}
\DeclareMathOperator*{\LayerNorm}{LayerNorm}
\DeclareMathOperator*{\FFN}{FFN}
\DeclareMathOperator*{\EncoderPreNet}{Encoder-PreNet}
\DeclareMathOperator*{\EncoderMain}{Encoder-Main}
\DeclareMathOperator*{\DecoderPreNet}{Decoder-PreNet}
\DeclareMathOperator*{\DecoderMain}{Decoder-Main}
\DeclareMathOperator*{\DecoderPostNet}{Decoder-PostNet}
\usepackage{xstring}
\titleformat{\paragraph}[runin]{\normalfont\normalsize\bfseries}{\theparagraph}{1em}{\hspace{1em}}[.]
\titlespacing{\paragraph}{0pt}{3.25ex plus 1ex minus .2ex}{1em}

\let\realcite\cite
\renewcommand{\cite}[1]{\ifx.#1.\hl{[?]}\else\realcite{#1}\fi}

\tikzset{
  basic/.style  = {draw, text width=2cm, drop shadow, font=\sffamily, rectangle},
  root/.style   = {basic, rounded corners=2pt, thin, align=center,
                   fill=green!30},
  level 2/.style = {basic, rounded corners=6pt, thin,align=center, fill=green!60,
                   text width=8em},
  level 3/.style = {basic, thin, align=left, fill=pink!60, text width=6.5em}
}
\setcopyright{none}
\renewcommand\footnotetextcopyrightpermission[1]{}
\begin{document}



\title{A Review of Deep Learning Techniques for Speech Processing}

\author{Ambuj Mehrish}
\email{ambuj\_mehrish@sutd.edu.sg}
\affiliation{%
  \institution{Singapore University of Technology and Design}
   \country{Singapore}
}

\author{Navonil Majumder}
\affiliation{%
  \institution{Singapore University of Technology and Design}
   \country{Singapore}
}

\author{Rishabh Bhardwaj}
\affiliation{%
  \institution{Singapore University of Technology and Design}
   \country{Singapore}
}

\author{Rada Mihalcea}
\affiliation{%
  \institution{University of Michigan}
   \country{USA}
}

\author{Soujanya Poria}
\affiliation{%
  \institution{Singapore University of Technology and Design}
   \country{Singapore}
}




\renewcommand{\shortauthors}{Mehrish et al.}
\makeatletter
\let\@authorsaddresses\@empty
\makeatother

\begin{abstract}
The field of speech processing has undergone a transformative shift with the advent of deep learning. The use of multiple processing layers has enabled the creation of models capable of extracting intricate features from speech data. This development has paved the way for unparalleled advancements in speech recognition, text-to-speech synthesis, automatic speech recognition, and emotion recognition, propelling the performance of these tasks to unprecedented heights. The power of deep learning techniques has opened up new avenues for research and innovation in the field of speech processing, with far-reaching implications for a range of industries and applications. This review paper provides a comprehensive overview of the key deep learning models and their applications in speech-processing tasks. We begin by tracing the evolution of speech processing research, from early approaches, such as MFCC and HMM, to more recent advances in deep learning architectures, such as CNNs, RNNs, transformers, conformers, and diffusion models. We categorize the approaches and compare their strengths and weaknesses for solving speech-processing tasks. Furthermore, we extensively cover various speech-processing tasks, datasets, and benchmarks used in the literature and describe how different deep-learning networks have been utilized to tackle these tasks. Additionally, we discuss the challenges and future directions of deep learning in speech processing, including the need for more parameter-efficient, interpretable models and the potential of deep learning for multimodal speech processing. By examining the field's evolution, comparing and contrasting different approaches, and highlighting future directions and challenges, we hope to inspire further research in this exciting and rapidly advancing field.
\end{abstract}
\begin{teaserfigure} \includegraphics[width=\textwidth]{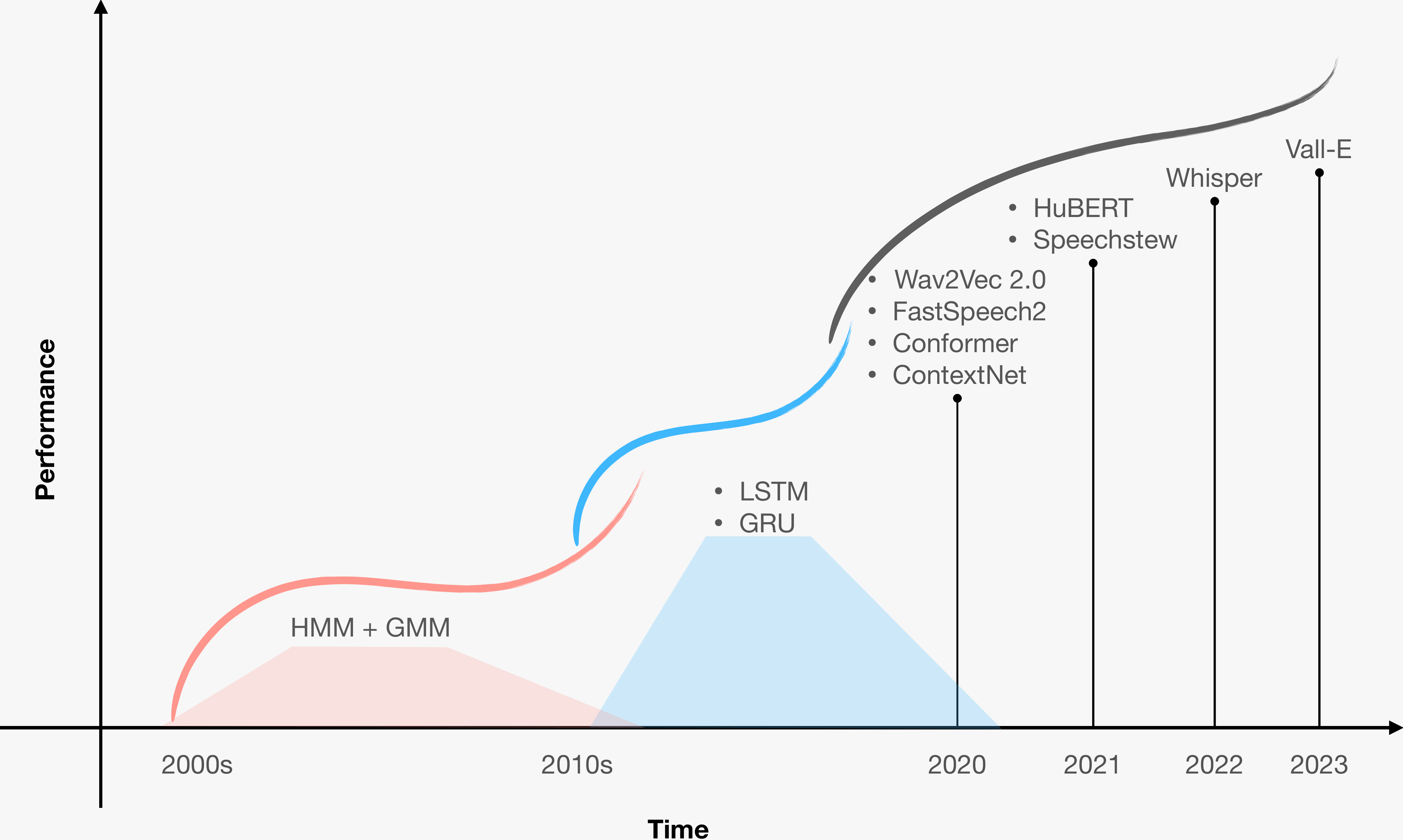}
    \caption{Evolution of speech processing models over the years.}
    \label{fig:evolution}
\end{teaserfigure}

\maketitle
\pagebreak
\tableofcontents
\newpage
\section{Introduction}
\label{sec:intro}

Humans employ language as a means to effectively convey their emotions and sentiments. Language encompasses a collection of words forming a vocabulary, accompanied by grammar, which dictates the appropriate usage of these words. It manifests in various forms, including written text, sign language, and spoken communication. Speech, specifically, entails the utilization of phonetic combinations of consonant and vowel sounds to articulate words from the vocabulary. Phonetics, in turn, pertains to the production and perception of sounds by individuals. Through speech, individuals are able to express themselves and convey meaning in their chosen language.

Speech processing is a field dedicated to the study and application of methods for analyzing and manipulating speech signals. It encompasses a range of tasks, including automatic speech recognition (ASR) \cite{yu2016automatic, nassif2019speech}, speaker recognition (SR) \cite{bai2021speaker}, and speech synthesis or text-to-speech \cite{ning2019review}. In recent years, speech processing has garnered increasing significance due to its diverse applications in areas such as telecommunications, healthcare, and entertainment. Notably, statistical modeling techniques, particularly Hidden Markov Models (HMMs), have played a pivotal role in advancing the field \cite{gales2008application, rabiner1989tutorial}. These models have paved the way for significant advancements and breakthroughs in speech processing research and development.

Over the past few years, the field of speech processing has been transformed by introducing powerful tools, including deep learning. \Cref{fig:evolution} illustrates the evolution of speech processing models over the years, the rapid development of deep learning architecture for speech processing reflects the growing complexity and diversity of the field. This technology has revolutionized the analysis and processing of speech signals using deep neural networks (DNNs), convolutional neural networks (CNNs), and recurrent neural networks (RNNs). These architectures have proven highly effective in various speech-processing applications, such as speech recognition, speaker recognition, and speech synthesis. This study comprehensively overviews the most critical and emerging deep-learning techniques and their potential applications in various speech-processing tasks.


Deep learning has revolutionized speech processing by its ability to automatically learn meaningful features from raw speech signals, eliminating the need for manual feature engineering. This breakthrough has led to significant advancements in speech processing performance, particularly in challenging scenarios involving noise, as well as diverse accents and dialects. By leveraging the power of deep neural networks, speech processing systems can now adapt and generalize more effectively, resulting in improved accuracy and robustness in various applications. The inherent capability of deep learning to extract intricate patterns and representations from speech data has opened up new possibilities for tackling real-world speech processing challenges.

Deep learning architectures have emerged as powerful tools in speech processing, offering remarkable improvements in various tasks. Pioneering studies, such as \cite{hinton2012deep}, have demonstrated the substantial gains achieved by deep neural networks (DNNs) in speech recognition accuracy compared to traditional HMM-based systems. Complementing this, research in \cite{abdel2014convolutional} showcased the effectiveness of convolutional neural networks (CNNs) for speech recognition. Moreover, recurrent neural networks (RNNs) have proven their efficacy in both speech recognition and synthesis, as highlighted in \cite{graves2013speech}. Recent advancements in deep learning have further enhanced speech processing systems, with attention mechanisms \cite{chorowski2015attention} and transformers \cite{vaswani2017attention} playing significant roles. Attention mechanisms enable the model to focus on salient sections of the input signal, while transformers facilitate modeling long-range dependencies within the signal. These developments have led to substantial improvements in the performance and versatility of speech processing systems, unlocking new possibilities for applications in diverse domains.

Although deep learning has made remarkable progress in speech processing, it still faces certain challenges that need to be addressed. These challenges include the requirement for substantial amounts of labeled data, the interpretability of the models, and their robustness to different environmental conditions. To provide a comprehensive understanding of the advancements in this domain, this paper presents an extensive overview of deep learning architectures employed in speech-processing applications. Speech processing encompasses the analysis, synthesis, and recognition of speech signals, and the integration of deep learning techniques has led to significant advancements in these areas. By examining the current state-of-the-art approaches, this paper aims to shed light on the potential of deep learning for tackling the existing challenges and further advancing speech processing research.

The paper provides a comprehensive exploration of deep-learning architectures in the field of speech processing. It begins by establishing the background, encompassing the definition of speech signals, speech features, and traditional non-neural models. Subsequently, the focus shifts towards an in-depth examination of various deep-learning architectures specifically tailored for speech processing, including RNNs, CNNs, Transformers, GNNs, and diffusion models. Recognizing the significance of representation learning techniques in this domain, the survey paper dedicates a dedicated section to their exploration.

Moving forward, the paper delves into an extensive range of speech processing tasks where deep learning has demonstrated substantial advancements. These tasks encompass critical areas such as speech recognition, speech synthesis, speaker recognition, speech-to-speech translation, and speech synthesis. By thoroughly analyzing the fundamentals, model architectures, and specific tasks within the field, the paper then progresses to discuss advanced transfer learning techniques, including domain adaptation, meta-learning, and parameter-efficient transfer learning.

Finally, in the conclusion, the paper reflects on the current state of the field and identifies potential future directions. By considering emerging trends and novel approaches, the paper aims to shed light on the evolving landscape of deep learning in speech processing and provide insights into promising avenues for further research and development.

\paragraph{Why this paper?} Deep learning has become a powerful tool in speech processing because it automatically learns high-level representations of speech signals from raw audio data. As a result, significant advancements have been made in various speech-processing tasks, including speech recognition, speaker identification, speech synthesis, and more. These tasks are essential in various applications, such as human-computer interaction, speech-based search, and assistive technology for people with speech impairments. For example, virtual assistants like Siri and Alexa use speech recognition technology, while audiobooks and in-car navigation systems rely on text-to-speech systems.

Given the wide range of applications and the rapidly evolving nature of deep learning, a comprehensive review paper that surveys the current state-of-the-art techniques and their applications in speech processing is necessary. Such a paper can help researchers and practitioners stay up-to-date with the latest developments and trends and provide insights into potential areas for future research. However, to the best of our knowledge, no current work covers a broad spectrum of speech-processing tasks.

A review paper on deep learning for speech processing can also be a valuable resource for beginners interested in learning about the field. It can provide an overview of the fundamental concepts and techniques used in deep learning for speech processing and help them gain a deeper understanding of the field. While some survey papers focus on specific speech-processing tasks such as speech recognition, a broad survey would cover a wide range of other tasks such as speaker recognition speech synthesis, and more. A broad survey would highlight the commonalities and differences between these tasks and provide a comprehensive view of the advancements made in the field.
\section{Background}
\label{sec:representation}
Before moving on to deep neural architectures, we discuss basic terms used in speech processing, low-level representations of speech signals, and traditional models used in the field.
\subsection{Speech Signals}
Signal processing is a fundamental discipline that encompasses the study of quantities that exhibit variations in space or time. In the realm of signal processing, a quantity exhibiting spatial or temporal variations is commonly referred to as a signal. Specifically, sound signals are defined as variations in air pressure. Consequently, a speech signal is identified as a type of sound signal, namely pressure variations, generated by humans to facilitate spoken communication. Transducers play a vital role in converting these signals from one form, such as air pressure, to another form, typically an electrical signal.

In signal processing, a signal that repetitively manifests after a fixed duration, known as a period, is classified as periodic. The reciprocal of this period represents the frequency of the signal. The waveform of a periodic signal defines its shape and concurrently determines its timbre, which pertains to the subjective perception of sound quality by humans. To facilitate the processing of speech, speech signals are commonly digitized. This entails converting them into a series of numerical values by measuring the signal's amplitude at consistent time intervals. The sampling rate, defined by the number of samples collected per second, determines the granularity of this digitization process.

\subsection{Speech Features}
Speech features are numerical representations of speech signals that are used for analysis, recognition, and synthesis. {Broadly, speech signals can be classified into two categories: time-domain features and frequency-domain features.} 

\textbf{Time-domain} features are derived directly from the amplitude of the speech signal over time. These are simple to compute and often used in real-time speech-processing applications. Some common time-domain features include:
\begin{itemize}
    \item Energy: Energy is a quantitative measure of the amplitude characteristics of a speech signal over time. It is computed by squaring each sample in the signal and summing them within a specific time window. This captures the overall strength and dynamics of the signal, revealing temporal variations in intensity. The energy measure provides insights into segments with higher or lower amplitudes, aiding in speech recognition, audio segmentation, and speaker diarization. It also helps identify events and transitions indicative of changes in vocal activity. By quantifying amplitude variations, energy analysis contributes to a comprehensive understanding of speech signals and their acoustic properties.
    \item Zero-crossing rate: The zero-crossing rate indicates how frequently the speech signal crosses the zero-axis within a defined time frame. It is computed by counting the number of polarity changes in the signal during a specific window.
    \item Pitch: Pitch refers to the perceived tonal quality in a speaker's voice, which is determined by analyzing the fundamental frequency of the speech signal. The fundamental frequency can be estimated through the application of pitch detection algorithms \cite{rabiner1976comparative} or by utilizing autocorrelation techniques \cite{tan2003pitch}. 
    \item  Linear predictive coding (LPC):Linear Predictive Coding (LPC) is a powerful technique that represents the speech signal as a linear combination of past samples, employing an autoregressive model. The estimation of model parameters is accomplished through methods like the Levinson-Durbin algorithm \cite{castiglioni2005levinson}. The obtained coefficients serve as a valuable feature representation for various speech-processing tasks.
\end{itemize}

\textbf{Frequency-domain} features are derived from the signal represented in the frequency domain also known as its spectrum. {A spectrum captures the distribution of energy as a function of frequency. Spectrograms are two-dimensional visual representations capturing the variations in a signal's spectrum over time. When compared against time-domain features, it is generally more complex to compute frequency-domain features as they tend to involve time-frequency transform operations such as Fourier transform.} 
\begin{itemize}
    \item  Mel-spectrogram: A Mel spectrogram, also known as a Mel-frequency spectrogram or Melspectrogram, is a representation of the short-term power spectrum of a sound signal. It is widely used in audio signal processing and speech recognition tasks. It is obtained by converting the power spectrum of a speech signal into a mel-scale, which is a perceptual scale of pitches based on the human auditory system's response to different frequencies. The mel-scale divides the frequency range into a set of mel-frequency bands, with higher resolution in the lower frequencies and coarser resolution in the higher frequencies. This scale is designed to mimic the non-linear frequency perception of human hearing. To compute the Melspectrogram, the speech signal is typically divided into short overlapping frames. For each frame, the Fast Fourier Transform (FFT) is applied to obtain the power spectrum. The power spectrum is then transformed into the mel-scale using a filterbank that converts the power values at different frequencies to their corresponding mel-frequency bands. Finally, the logarithm of the mel-scale power values is computed, resulting in the Melspectrogram.
    
    Melspectrogram provides a time-frequency representation of the audio signal, where the time dimension corresponds to the frame index, and the frequency dimension represents the mel-frequency bands. It captures both the spectral content and temporal dynamics of the signal, making it useful for tasks such as speech recognition, music analysis, and sound classification. By using the Melspectrogram, the representation of the audio signal is transformed to a more perceptually meaningful domain, which can enhance the performance of various audio processing algorithms. It is particularly beneficial in scenarios where capturing the spectral patterns and frequency content of the signal is important for the analysis or classification task at hand.
    \item Mel-frequency cepstral coefficients (MFCCs): Mel-frequency cepstral coefficients (MFCCs) are a feature representation widely utilized in various applications such as speech recognition, gesture recognition, speaker identification, and cetacean auditory perception systems. MFCCs capture the power spectrum of a sound over a short duration by utilizing a linear cosine transformation of a logarithmically-scaled power spectrum on a non-linear mel frequency scale. The MFCCs consist of a set of coefficients that collectively form a Mel-frequency cepstrum \footnote{https://en.wikipedia.org/wiki/Mel-frequency\_cepstrum}. With just 12 parameters related to the amplitude of frequencies, MFCCs provide an adequate number of frequency channels to analyze audio, while still maintaining a compact representation. The main objectives of MFCC extraction are to eliminate vocal fold excitation (F0) information related to pitch, ensure the independence of the extracted features, align with human perception of loudness and frequency, and capture the contextual dynamics of phones. The process of extracting MFCC features involves A/D conversion, pre-emphasis filtering, framing, windowing, Fourier transform, Mel filter bank application, logarithmic operation, discrete cosine transform (DCT), and liftering. By following these steps, MFCCs enable the extraction of informative audio features while avoiding redundancy and preserving the relevant characteristics of the sound signal.
\end{itemize}

Other types of speech features include formant frequencies, pitch contour, cepstral coefficients, wavelet coefficients, and spectral envelope. 
These features can be used for various speech-processing tasks, including speech recognition, speaker identification, emotion recognition, and speech synthesis.

In the field of speech processing, frequency-based representations such as Mel spectrogram and MFCC are widely used since they are more robust to noise as compared to temporal variations of the sound \cite{9955539}. Time-domain features can be useful when the task warrants this information (such as pauses, emotions, phoneme duration, and speech segments). It is noteworthy that the time-domain and frequency-domain features tend to capture different sets of information and thus can be used in conjunction to solve a task \cite{1165240,9053712,tang2021joint}.

\subsection{Traditional models for speech processing}

Traditional speech representation learning algorithms based on shallow models utilize basic non-parametric models for extracting features from speech signals. The primary objective of these models is to extract significant features from the speech signal through mathematical operations, such as Fourier transforms, wavelet transforms, and linear predictive coding (LPC). The extracted features serve as inputs to classification or regression models. The shallow models aim to extract meaningful features from the speech signal, enabling the classification or regression model to learn and make accurate predictions.
\begin{itemize}
    \item Gaussian Mixture Models (GMMs): Gaussian Mixture Models (GMMs) are powerful generative models employed to represent the probability distribution of a speech feature vector. They achieve this by combining multiple Gaussian distributions with different weights. GMMs have found widespread applications in speaker identification \cite{kinnunen2005real} and speech recognition tasks \cite{reynolds2003channel}. Specifically, in speaker identification, GMMs are utilized to capture the distribution of speaker-specific features, enabling the recognition of individuals based on their unique characteristics. Conversely, in speech recognition, GMMs are employed to model the acoustic properties of speech sounds, facilitating accurate recognition of spoken words and phrases. GMMs play a crucial role in these domains, enabling robust and efficient analysis of speech-related data.
    
    \item Support Vector Machines (SVMs): Support Vector Machines (SVMs) are a widely adopted class of supervised learning algorithms extensively utilized for various speech classification tasks \cite{smith2001speech}. They are particularly effective in domains like speaker recognition  \cite{hatch2006within,solomonoff2004channel,solomonoff2005advances} and phoneme recognition \cite{campbell2003phonetic}. SVMs excel in their ability to identify optimal hyperplanes that effectively separate different classes in the feature space. By leveraging this optimal separation, SVMs enable accurate classification and recognition of speech patterns. As a result, SVMs have become a fundamental tool in the field of speech analysis and play a vital role in enhancing the performance of speech-related classification tasks.
    
    \item Hidden Markov Models (HMMs): Hidden Markov Models (HMMs) have gained significant popularity as a powerful tool for performing various speech recognition tasks, particularly ASR \cite{gales2008application, rabiner1989tutorial}. In ASR, HMMs are employed to model the probability distribution of speech sounds by incorporating a sequential arrangement of hidden states along with corresponding observations. The training of HMMs is commonly carried out using the Baum-Welch algorithm, a variant of the Expectation Maximization algorithm, which enables effective parameter estimation and model optimization\footnote{Wikipedia:~Baum-Welch~algorithm: \url{http://en.wikipedia.org/wiki/Baum\%e2\%80\%93Welch\_algorithm}}.
    
    By leveraging HMMs in speech recognition, it becomes possible to predict the most likely sequence of speech sounds given an input speech signal. This enables accurate and efficient recognition of spoken language, making HMMs a crucial component in advancing speech recognition technology. Their flexibility and ability to model temporal dependencies contribute to their widespread use in ASR and various other speech-related applications, further enhancing our understanding and utilization of spoken language.

    \item The K-nearest neighbors (KNN) algorithm is a simple yet effective classification approach utilized in a wide range of speech-related applications, including speaker recognition \cite{sadjadi2014nearest} and language recognition. The core principle of KNN involves identifying the K-nearest neighbors of a given input feature vector within the training data and assigning it to the class that appears most frequently among those neighbors. This algorithm has gained significant popularity due to its practicality and intuitive nature, making it a reliable choice for classifying speech data in numerous real-world scenarios. By leveraging the proximity-based classification, KNN provides a straightforward yet powerful method for accurately categorizing speech samples based on their similarities to the training data. Its versatility and ease of implementation contribute to its widespread adoption in various speech-related domains, facilitating advancements in speaker recognition, language identification, and other applications in the field of speech processing.

    \item Decision trees: Decision trees are widely employed in speech classification tasks as a class of supervised learning algorithms. Their operation involves recursively partitioning the feature space into smaller regions, guided by the values of the features. Within each partition, a decision rule is established to assign the input feature vector to a specific class. The strength of decision trees lies in their ability to capture complex decision boundaries by hierarchically dividing the feature space. By analyzing the values of the input features at each node, decision trees efficiently navigate the classification process. This approach not only provides interpretability, but also facilitates the identification of key features contributing to the classification outcome. Through their recursive partitioning mechanism, decision trees offer a flexible and versatile framework for speech classification. They excel in scenarios where the decision rules are based on discernible thresholds or ranges of feature values. The simplicity and transparency of decision trees make them a valuable tool for understanding and solving speech-related classification tasks.
\end{itemize}


To summarize, conventional speech representation learning algorithms based on shallow models entail feature extraction from the speech signal, which is subsequently used as input for classification or regression models. These algorithms have found extensive applications in speech processing tasks like speech recognition, speaker identification, and speech synthesis. However, they have been progressively superseded by more advanced representation learning algorithms, particularly deep neural networks, due to their enhanced capabilities.

\section{Deep Learning Architectures and Their Applications in Speech Processing Tasks}

Deep learning architectures have revolutionized the field of speech processing by demonstrating remarkable performance across various tasks. With their ability to automatically learn hierarchical representations from raw speech data, deep learning models have surpassed traditional approaches in areas such as speech recognition, speaker identification, and speech synthesis. These architectures have been instrumental in capturing intricate patterns, uncovering latent features, and extracting valuable information from vast amounts of speech data. In this section, we delve into the applications of deep learning architectures in speech processing tasks, exploring their potential, advancements, and the impact they have had on the field. By examining the key components and techniques employed in these architectures, we aim to provide insights into the current state-of-the-art in deep learning for speech processing and shed light on the exciting prospects it holds for future advancements in the field.

\subsection{Recurrent Neural Networks (RNNs)}
\label{sec:rnn}
{It is natural to consider Recurrent Neural Networks for various speech processing tasks since the input speech signal is inherently a dynamic process \cite{salehinejad2017recent}. RNNs can model a given time-varying (sequential) patterns that were otherwise hard to capture by standard feedforward neural architectures. Initially, RNNs were used in conjunction with HMMs where the sequential data is first modeled by HMMs while localized classification is done by the neural network. However, such a hybrid model tends to inherit limitations of HMMs, for instance, HMM requires task-specific knowledge and independence constraints for observed states \cite{bourlard1994connectionist}. To overcome the limitations inherited by the hybrid approach, end-to-end systems completely based on RNNs became popular for sequence transduction tasks such as speech recognition and text\cite{graves2012sequence, kawakami2008supervised}. Next, we discuss RNN and it's variants:}

 
\subsubsection{RNN Models}

\paragraph{Vanilla RNN}

{Give input sequence of T states $(x_{1}, \ldots ,x_{T})$ with $x_i \in \mathbb{R}^d$, the output state at time $t$ can be computed as}
{
\begin{equation}
    h_{t} = \mathcal{H}(W_{hh}h_{t-1}+W_{xh}x_{t}+b_h)
\end{equation}
\begin{equation}\label{rnn_h}
    y_{t} = W_{hy}h_{t}+b_y
\end{equation}
}
{ where $W_{hh}, W_{hx}, W_{yh}$ are weight matrices and $b_h, b_y$ are bias vectors. $\mathcal{H}$ is non-linear activation functions such as Tanh, ReLU, and Sigmoid. RNNs are made of high dimensional hidden states, notice $h_t$ in the above equation, which makes it possible for them to model sequences and help overcome the limitation of feedforward neural networks. The state of the hidden layer is conditioned on the current input and the previous state, which makes the underlying operation recursive. Essentially, the hidden state $h_{t-1}$ works as a memory of past inputs $\{x_k\}_{k=1}^{t-1}$ that influence the current output $y_t$.}

\paragraph{Bidirectional RNNs}
{For numerous tasks in speech processing, it is more effective to process the whole utterance at once. For instance, in speech recognition, one-shot input transcription can be more robust than transcribing based on the partial (i.e. previous) context information \cite{graves2013speech}. The vanilla RNN has a limitation in such cases as they are unidirectional in nature, that is, output $y_t$ is obtained from $\{x_k\}_{k=1}^{t}$, and thus, agnostic of what comes after time $t$. Bidirectional RNNs (BRNNs) were proposed to overcome such shortcomings of RNNs \cite{schuster1997bidirectional}. BRRNs encode both future and past (input) context in separate hidden layers. The outputs of the two RNNs are then combined at each time step, typically by concatenating them together, to create a new, richer representation that includes both past and future context.}
{
\begin{align}
    \overrightarrow{h_{t}} &= \mathcal{H}(W_{\overrightarrow{hh}}\overrightarrow{h}_{t-1}+W_{\overrightarrow{xh}}x_{t}+b_{\overrightarrow{h}})\\
    \overleftarrow{h_{t}} &= \mathcal{H}(W_{\overleftarrow{hh}}\overleftarrow{h}_{t+1}+W_{\overleftarrow{xh}}x_{t}+b_{\overleftarrow{h}})\\
    y_{t} &= W_{\overrightarrow{hy}}\overrightarrow{h}_{t}+W_{\overleftarrow{hy}}\overleftarrow{h}_{t}+b_y
\end{align}}
{where high dimensional hidden states $\overrightarrow{h}_{t-1}$ and $\overleftarrow{h}_{t+1}$ are hidden states modeling the forward context from $1,2,\ldots,t-1$ and backward context from $T, T-1, \ldots, t+1$, respectively}.

\paragraph{Long Short-Term Memory}
{Vanilla RNNs are observed to face another limitation, that is, vanishing gradients that do not allow them to learn from long-range context information. To overcome this, a variant of RNN, named as LSTM, was specifically designed to address the vanishing gradient problem and enable the network to selectively retain (or forget) information over longer periods of time \cite{hochreiter1997long}. This attribute is achieved by maintaining separate purpose-built memory cells in the network: the long-term memory cell $c_t$ and the short-term memory cell $h_t$. In \Cref{rnn_h}, LSTM redefines the operator $\mathcal{H}$ in terms of forget gate $f_t$, input gate $i_t$, and output gate $o_t$,}
{
\begin{align}
    i_t &= \sigma(W_{xi}x_t + W_{hi}h_{t-1}+W_{ci}c_{t-1}+b_i),\\
    f_t &= \sigma(W_{xf}x_t + W_{hf}h_{t-1}+W_{cf}c_{t-1}+b_f),\\
    c_t &= f_t\odot c_{t-1} + i_t\odot \tanh{(W_{xc}x_t+W_{hc}h_{t-1}+b_c)},\\
    o_t &= \sigma(W_{xo}x_t + W_{ho}h_{t-1}+W_{co}c_{t}+b_o),\\
    h_t &= o_t\odot \tanh{(c_t)},
\end{align}}
{where $\sigma(x)=1/({1+e^{-x}})$ is a logistic sigmoid activation function. $c_t$ is a fusion of the information from the previous state of the long-term memory $c_{t-1}$, the previous state of short-term memory $h_{t-1}$, and current input $x_t$. $W$ and $b$ are weight matrices and biases. $\odot$ is the element-wise vector multiplication or Hadamard operator. Bidirectional LSTMs (BLSTMs) can capture longer contexts in both forward and backward directions \cite{graves2012sequence}.}

\paragraph{Gated Recurrent Units}




{Gated Recurrent Units (GRU) aim to be a computationally-efficient approximate of LSTM by using only two gates (vs three in LSTM) and a single memory cell (vs two in LSTM). To control the flow of information over time, a GRU uses an update gate $z_t$ to decide how much of the new input to be added to the previous hidden state and a reset gate $r_t$ to decide how much of previous hidden state information to be forgotten.}
{
\begin{align}
    z_{t} &= \sigma(W_{xz}x_{t} + W_{hz}h_{t-1}),\\
    r_{t} &= \sigma(W_{xr}x_{t} + W_{hr}h_{t-1}), \\
    h_{t} &= (1-z_{t})\odot h_{t-1} + z_{t}\odot \tanh(W_{xh}x_{t} + W_{rh}(r_{t}\odot h_{t-1})),
\end{align}
}
{where $\odot$ is element-wise multiplication between the two vectors (Hadamard product).}


%

RNNs and their variants are widely used in various deep learning applications like speech recognition, synthesis, and natural language understanding. Although seq2seq based on recurrent architectures such as LSTM/GRU has made great strides in speech processing, they suffer from the drawback of slow training speed due to internal recurrence. Another drawback of the RNN family is their inability to leverage information from long distant time steps accurately.

\paragraph{Connectionist Temporal Classification}
Connectionist Temporal Classification (CTC) \cite{graves2012connectionist} is a scoring and output function commonly used to train LSTM networks for sequence-based problems with variable timing. CTC has been applied to several tasks, including phoneme recognition, ASR, and other sequence-based problems. One of the major benefits of CTC is its ability to handle unknown alignment between input and output, simplifying the training process. When used in ASR  \cite{9688009,9747887,9053165}, CTC eliminates the need for manual data labeling by assigning probability scores to the output given any input signal. This is particularly advantageous for tasks such as speech recognition and handwriting recognition, where the input and output can vary in size. CTC also solves the problem of having to specify the position of a character in the output, allowing for more efficient training of the neural network without post-processing the output. Finally, the CTC decoder can transform the neural network output into the final text without post-processing.

\subsubsection{Application}

The utilization of RNNs in popular products such as Google's voice search and Apple's Siri to process user input and predict the output has been well-documented \cite{he2017streaming,li2017acoustic}. RNNs are frequently utilized in speech recognition tasks, such as the prediction of phonetic segments from audio signals \cite{papastratis2021speech}. They excel in use cases where context plays a vital role in outcome prediction and are distinct from CNNs as they utilize feedback loops to process a data sequence that informs the final output \cite{papastratis2021speech}.

In recent times, there have been advancements in the architecture of RNNs, which have been primarily focused on developing end-to-end (E2E) models \cite{li2020towards,9746187} for ASR. These E2E models have replaced conventional hybrid models and have displayed substantial enhancements in speech recognition \cite{li2020towards,li2021better}. However, a significant challenge faced by E2E RNN models is the synchronization of the input speech sequence with the output label sequence \cite{graves2012sequence}. To tackle this issue, a loss function called CTC \cite{graves2012connectionist} is utilized for training RNN models, allowing for the repetition of labels to construct paths of the same length as the input speech sequence. An alternative method is to employ an Attention-based Encoder-Decoder (AED) model based on RNN architecture, which utilizes an attention mechanism to align the input speech sequence with the output label sequence. However, AED models tend to perform poorly on lengthy utterances.

The development of Bimodal Recurrent Neural Networks (BRNN) has led to significant advancements in the field of Audiovisual Speech Activity Detection (AV-SAD) \cite{tao2019end}. BRNNs have demonstrated immense potential in improving the performance of speech recognition systems, particularly in noisy environments, by combining information from various sources. By integrating separate RNNs for each modality, BRNNs can capture temporal dependencies within and across modalities. This leads to successful outcomes in speech-based systems, where integrating audio and visual modalities is crucial for accurate speech recognition. Compared to conventional audio-only systems, BRNN-based AV-SAD systems display superior performance, particularly in challenging acoustic conditions where audio-only systems might struggle.

To enhance the performance of continuous speech recognition, LSTM networks have been utilized in hybrid architectures alongside CNNs \cite{passricha2019hybrid}. The CNNs extract local features from speech frames that are then processed by LSTMs over time \cite{passricha2019hybrid}. LSTMs have also been employed for speech synthesis, where they have been shown to enhance the quality of statistical parametric speech synthesis \cite{passricha2019hybrid}.

Aside from their ASR and speech synthesis applications, LSTM networks have been utilized for speech post-filtering. To improve the quality of synthesized speech, researchers have proposed deep learning-based post-filters, with LSTMs demonstrating superior performance over other post-filter types \cite{coto2019improving}. Bidirectional LSTM (Bi-LSTM) is another variant of RNN that has been widely used for speech synthesis \cite{fan2014tts}. Several RNN-based analysis/synthesis models such as WaveNet \cite{oord2016wavenet}, SampleRNN \cite{mehri2016samplernn}, and Tacotron have been developed. These neural vocoder models can generate high-quality synthesized speech from acoustic features without requiring intermediate vocoding steps.
\subsection{Convolutional Neural Networks}
\label{sec:cnn}
Convolutional neural networks (CNNs) are a specialized class of deep neural architecture consisting of one or more pairs of alternating convolutional and pooling layers. A convolution layer applies filters that process small local parts of the input, where these filters are replicated along the whole input space. A pooling layer converts convolution layer activations to low resolution by taking the maximum filter activation within a specified 
window and shifting across the activation map. CNNs are variants of fully connected neural networks widely used for processing data with grid-like topology. For example, time-series data (1D grid) with samples at regular intervals or images (2D grid) with pixels constitute a grid-like structure.

As discussed in \Cref{sec:representation}, the speech spectrogram retains more information than hand-crafted features, including speaker characteristics such as vocal tract length differences across speakers, distinct speaking styles causing formant to undershoot or overshoot, etc. Also, explicitly expressed these characteristics in the frequency domain. The spectrogram representation shows very strong correlations in time and frequency. Due to these characteristics of the spectrogram, it is a suitable input for a CNN processing pipeline that requires preserving locality in both frequency and time axis. For speech signals, modeling local correlations with CNNs will be beneficial. The CNNs can also effectively extract the structural features from the spectrogram and reduce the complexity of the model through weight sharing. This section will discuss the architecture of 1D and 2D CNNs used in various speech-processing tasks.

\subsubsection{CNN Model Variants}
\paragraph{2D CNN}
\label{2dcnn}
{Since spectrograms are two-dimensional visual representations, one can leverage CNN architectures widely used for visual data processing (images and videos) by performing convolutions in two dimensions.} The mathematical equation for a 2D convolutional layer can be represented as:
\begin{equation}
y_{i,j}^{(k)}= \sigma\Big(\sum_{l=1}^{L} \sum_{m=1}^{M} x_{i+l-1,j+m-1}^{(l)} w_{l,m}^{(k)} + b^{(k)}\Big)
\end{equation}
Here, $x_{i,j}^{(l)}$ is the pixel value of the $l^{th}$ input channel at the spatial location $(i,j)$, $w_{l,m}^{(k)}$ is the weight of the $m^{th}$ filter at the $l^{th}$ channel producing the $k^{th}$ feature map, and $b^{(k)}$ is the bias term for the $k^{th}$ feature map.

The output feature map $y_{i,j}^{(k)}$ is obtained by convolving the input image with the filters and then applying an activation function $\sigma$ to introduce non-linearity. The convolution operation involves sliding the filter window over the input image, computing the dot product between the filter and the input pixels at each location, and producing a single output pixel.

However, there are some drawbacks to using a 2D CNN for speech processing. One of the main issues is that 2D convolutions are computationally expensive, especially for large inputs. This is because 2D convolutions involve many multiplications and additions, and the computational cost grows quickly with the input size.

To address this issue, a 1D CNN can be designed to operate directly on the speech signal without needing a spectrogram. 1D convolutions are much less computationally expensive than 2D convolutions because they only operate on one dimension of the input. This reduces the multiplications and additions required, making the network faster and more efficient. In addition, 1D feature maps require less memory during processing, which is especially important for real-time applications. A neural network's memory requirements are proportional to its feature maps' size. By using 1D convolutions, the size of the feature maps can be significantly reduced, which can improve the efficiency of the network and reduce the memory requirements.


\paragraph{1D CNN}
1D CNN is essentially a special case of 2D CNN where the height of the filter is equal to the height the spectogram. Thus, the filter only slides along the temporal dimension and the height of the resultant feature maps is one. As such, 1D convolutions are computationally less expensive and memory efficient~\cite{KIRANYAZ2021107398}, as compared to 2D CNNs. Several studies \cite{kiranyaz2015convolutional,Karita2019ImprovingTE,abdeljaber2017real} have shown that 1D CNNs are preferable to their 2D counterparts in certain applications. For example, \citet{alsabhan2023human} found that the performance of predicting emotions with a 2D CNN model was lower compared to a 1D CNN model.

1D convolution is useful in speech processing for several reasons:
\begin{itemize}
    \item 
    {Since, speech signals are sequences of amplitudes sampled over time, 1D convolution can be applied along temporal dimension to capture temporal variations in the signal.} 
    \item \emph{Robustness to distortion and noise:} Since, 1D convolution allows local feature extraction, the resultant features are often resilient to global distortions of the signal. For instance, a speaker might be interrupted in the middle of an utterance. Local features would still produce robust representations for those relevant spans, which is key to ASR, among many speech processing task. On the other hand, speech signals are often contaminated with noise, making extracting meaningful information difficult. 1D convolution followed by pooling layers can mitigate the impact of noise~\cite{hendrycks2018benchmarking}, improving speech recognition systems' accuracy.
\end{itemize}


The basic building block of a 1D CNN is the convolutional layer, which applies a set of filters to the input data. A convolutional layer employs a collection of adjustable parameters called filters to carry out convolution operations on the input data, resulting in a set of feature maps as the output, which represent the activation of each filter at each position in the input data. The size of the feature maps depends on the size of the input data, the size of the filters, and the number of filters used. The activation function used in a 1D CNN is typically a non-linear function, such as the rectified linear unit (ReLU) function. 

Given an input sequence $x$ of length $N$, a set of $K$ filters $W_k$ of length $M$, and a bias term $b_k$, the output feature map $y_k$ of the $k^{th}$ filter is given by 
\begin{equation}
\label{1dcnn}
    y_k[n]  = \ReLU(b_k + \sum_{m=0}^{M-1} W_k[m] * x[n-m])
\end{equation}
where $n$ ranges from $M-1$ to $N-1$, and $*$ denotes the convolution operation. After the convolutional layer, the output tensor is typically passed through a pooling layer, reducing the feature maps' size by down-sampling. The most commonly used pooling operation is the max-pooling, which keeps the maximum value from a sliding window across each feature map.

CNNs often replace previously popular methods like HMMs and GMM-UBM in various cases. Moreover, CNNs possess the ability to acquire features that remain robust despite variations in speech signals resulting from diverse speakers, accents, and background noise. This is made possible due to three key properties of CNNs: locality, weight sharing, and pooling. The locality property enhances resilience against non-white noise by enabling the computation of effective features from cleaner portions of the spectrum. Consequently, only a smaller subset of features is affected by the noise, allowing higher network layers a better opportunity to handle the noise by combining higher-level features computed for each frequency band. This improvement over standard fully connected neural networks, which process all input features in the lower layers, highlights the significance of locality. As a result, locality reduces the number of network weights that must be learned.

\subsubsection{Application}

CNNs have proven to be versatile tools for a range of speech-processing tasks. They have been successfully applied to speech recognition \cite{nassif2019speech, 6857341}, including in hybrid NN-HMM models for speech recognition, and can be used for multi-class classification of words \cite{6288864}. In addition, CNNs have been proposed for speaker recognition in an emotional speech, with a constrained CNN model presented in \cite{simic2022speaker}.

CNNs, both 1D and 2D, have emerged as the core building block for various speech processing models, including acoustic models \cite{schneider2019wav2vec,gulati2020conformer,kriman2020quartznet} in ASR systems. For instance, in 2021, researchers from Facebook AI proposed wav2vec2.0 \cite{schneider2019wav2vec}, a hybrid ASR system based on CNNs for learning representations of raw speech signals that were then fed into a transformer-based language model. The system achieved state-of-the-art results on several benchmark datasets.

Similarly, Google's VGGVox \cite{Nagrani17} used a CNN with VGG architecture to learn speaker embeddings from Mel spectrograms, achieving state-of-the-art results in speaker recognition. CNNs have also been widely used in developing state-of-the-art speech enhancement and text-to-speech architectures. For instance, 	
the architecture proposed in \cite{li2021real,tzinis2022remixit} for Deep Noise Suppression (DNS) \cite{reddy2020interspeech} challenge and Google's Tacotron2 \cite{shen2018natural} are examples of models that use CNNs as their core building blocks. In addition to traditional tasks like ASR and speaker identification, CNNs have also been applied to non-traditional speech processing tasks like emotion recognition \cite{kakuba2022deep}, Parkinson's disease detection \cite{johri2019parkinson}, language identification \cite{singh2021spoken} and sleep apnea detection \cite{simply2019diagnosis}. In all these tasks, CNN extracted features from speech signals and fed them into the task classification model.

\subsubsection{Temporal Convolution Neural Networks}
\label{sec:tcn}
Recurrent neural networks, including RNNs, LSTMs, and GRUs, have long been popular for deep-learning sequence modeling tasks. They are especially favored in the speech-processing domain. However, recent studies have revealed that certain CNN architectures can achieve state-of-the-art accuracy in tasks such as audio synthesis, word-level language modelling, and machine translation, as reported in \cite{kalchbrenner2016neural,kalchbrenner2014convolutional,dauphin2017language}. The advantage of convolutional neural networks is that they enable faster training by allowing parallel computation. They can avoid common issues associated with recurrent models, such as the vanishing or exploding gradient problem or the inability to retain long-term memory.

In a recent study by \citet{bai2018empirical}, they proposed a generic Temporal Convolutional Neural Network (TCNN) architecture that can be applied to various speech-related tasks. This architecture combines the best practices of modern CNNs and has demonstrated comparable performance to recurrent architectures such as LSTMs and GRUs. The TCN approach could revolutionize speech processing by providing an alternative to the widely used recurrent neural network models.
\subsubsection{TCNN Model Variants}
\begin{figure}
    \centering
    \includegraphics[width=0.6\columnwidth]{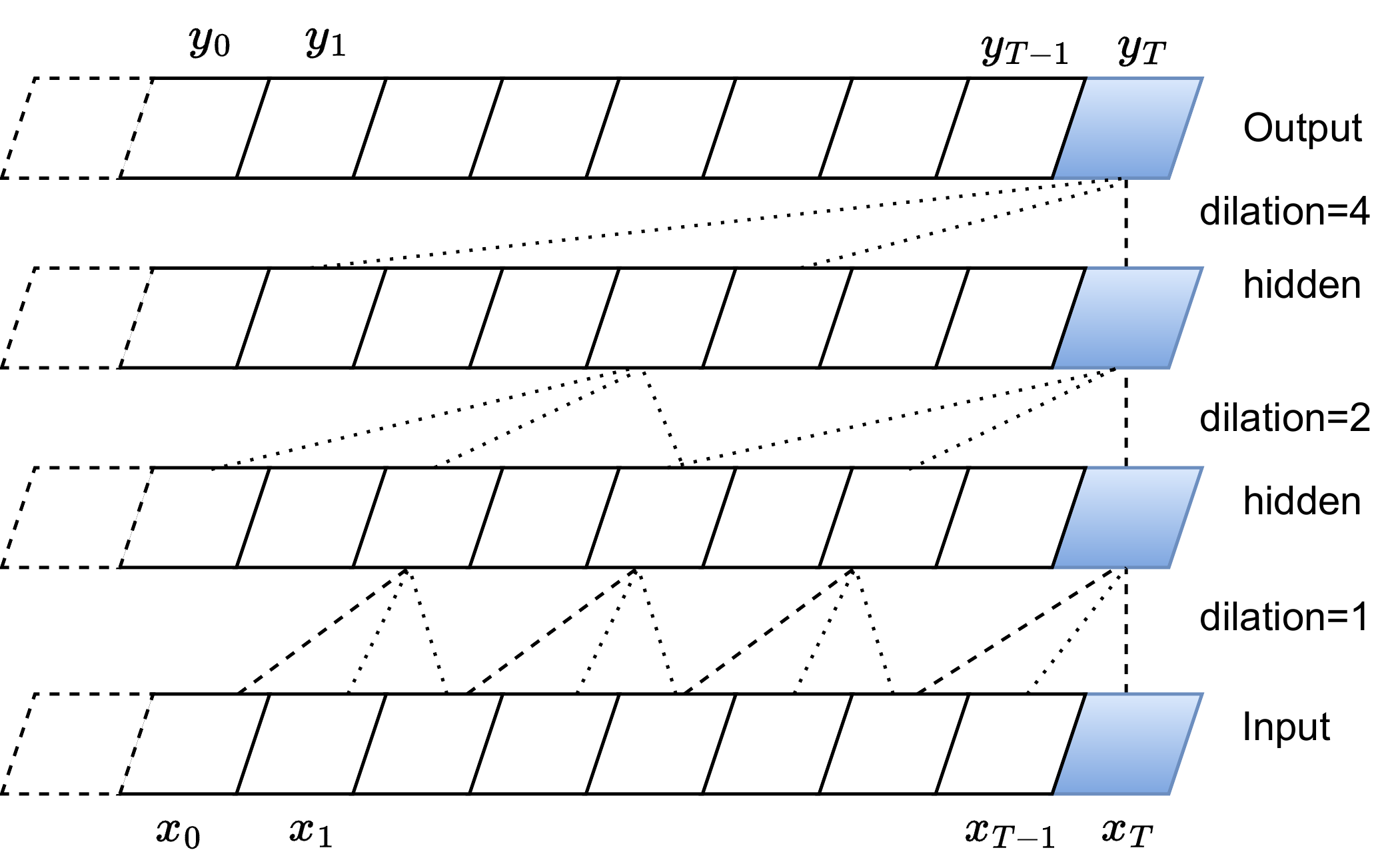}
    \caption{TCNNs leverage causal and dilated convolutions to model temporal dependencies in sequential data. Causal convolutions ensure that future information is not used during training, while dilated convolutions increase the receptive field without increasing computational complexity. This makes TCNNs an effective and efficient solution for a wide range of tasks, including speech recognition, action recognition, and music analysis.}
    \label{dilation}
\end{figure}

The architecture of TCNN is based upon two principles:(1) There is no information “leakage” from future to past;(2) the architecture can map an input sequence of any length to an output sequence of the same length, similar to RNN. TCN consists of dilated, causal 1D fully-convolutional layers with the same input and output lengths to satisfy the above conditions. In other words, TCNN is simply a 1D fully-convolutional network (FCN) with casual convolutions as shown in \Cref{dilation}.
\begin{itemize}
    \item \emph{Causal Convolution~\cite{oord2016wavenet}:} Causal convolution convolves the input at a specific time point $t$ solely with the temporally-prior elements.
    \item \emph{Dilated Convolution~\cite{Yu2015MultiScaleCA}:} By itself, causal convolution filters have a limited range of perception, meaning they can only consider a fixed number of elements in the past. Therefore, it is challenging to learn any dependency between temporally distant elements for longer sequences. Dilated convolution ameliorates this limitation by repeatedly applying dilating filters to expand the range of perception, as shown in \Cref{dilation}. The dilation is achieved by uniformly inserting zeros between the filter weights.
    
    Consider a 1-D sequence $x \in \mathbf{R}^{n}$ and a filter: $f: \{0,...,k-1\} \rightarrow \mathbf{R}$, the dilated convolution operation $F_d$ on an element $y$ of the sequence is defined as
    \begin{equation}
        F_d(y) = (x*_{d}f)(s) = \sum_{i=0}^{k-1}f(i).x_{y-d.i},
    \end{equation}
where $k$ is filter size, $d$ is dilation factor, and $y-d.i$ is the span along the past. The dilation step introduces a fixed step between every two adjacent filter taps. When $d=1$, a dilated convolution acts as a normal convolution. Whereas, for larger dilation, the filter acts on a wide but non-contiguous range of inputs. Therefore, dilation effectively expands the receptive field of the convolutional networks.
\end{itemize}

\subsubsection{Application}
Recent studies have shown that the TCNN architecture not only outperforms traditional recurrent networks like LSTMs and GRUs in terms of accuracy but also possesses a set of advantageous properties, including:

\begin{itemize}
\item Parallelism is a key advantage of TCNN over RNNs. In RNNs, time-step predictions depend on their predecessors' completion, which limits parallel computation. In contrast, TCNNs apply the same filter to each span in the input, allowing parallel application thereof. This feature enables more efficient processing of long input sequences compared to RNNs that process sequentially.

\item The receptive field size can be modified in various ways to enhance the performance of TCNNs. For example, incorporating additional dilated convolutional layers, employing larger dilation factors, or augmenting the filter size are all effective methods. Consequently, TCNNs offer superior management of the model's memory size and are highly adaptable to diverse domains.

\item When dealing with lengthy input sequences, LSTM and GRU models tend to consume a significant amount of memory to retain the intermediate outcomes for their numerous cell gates. On the other hand, TCNNs utilize shared filters throughout a layer, and the back-propagation route depends solely on the depth of the network. This makes TCNNs a more memory-efficient alternative to LSTMs and GRUs, especially in scenarios where memory constraints are a concern.
\end{itemize}

 TCNNs can perform real-time speech enhancement in the time domain \cite{pandey2019tcnn}. They have much fewer trainable parameters than earlier models, making them more efficient. TCNs have also been used for speech and music detection in radio broadcasts \cite{hung2022large,lemaire2019temporal}. 
 They have been used for single channel speech enhancement \cite{9601275,richter2020speech} and are trained as filter banks to extract features from waveform to improve the performance of ASR \cite{li2019single}.
 
\subsection{Transformers}
\label{sec:transformer}


While recurrence in RNNs (\cref{sec:rnn}) is a boon for neural networks to model sequential data, it is also a bane as the recurrence in time to update the hidden state intrinsically precludes parallelization. Additionally, although dedicated gated RNNs such as LSTM and GRU have helped to mitigate the vanishing gradient problem to some extent, it can still be a challenge to maintain long-term dependencies in RNNs. 

Proposed by \citet{vaswani2017attention}, Transformer solved a critical shortcoming of RNNs by allowing parallelization within the training sample, that is, facilitating the processing of the entire input sequence at once. Since then, the primary idea of using only the attention mechanism to construct an encoder and decoder has served as the basic recipe for many state-of-the-art architectures across the domains of machine learning. In this survey, we use \textbf{transformer} to denote architectures that are inspired by Transformer \cite{devlin2018bert, brown2020language, radford2019language, radford2018improving, han2022survey}. This section overviews the transformer's fundamental design proposed by \citet{vaswani2017attention} and its adaptations for different speech-related applications.

\subsubsection{Basic Architecture}
\begin{figure}
    \centering
    \includegraphics[width=0.8\columnwidth]{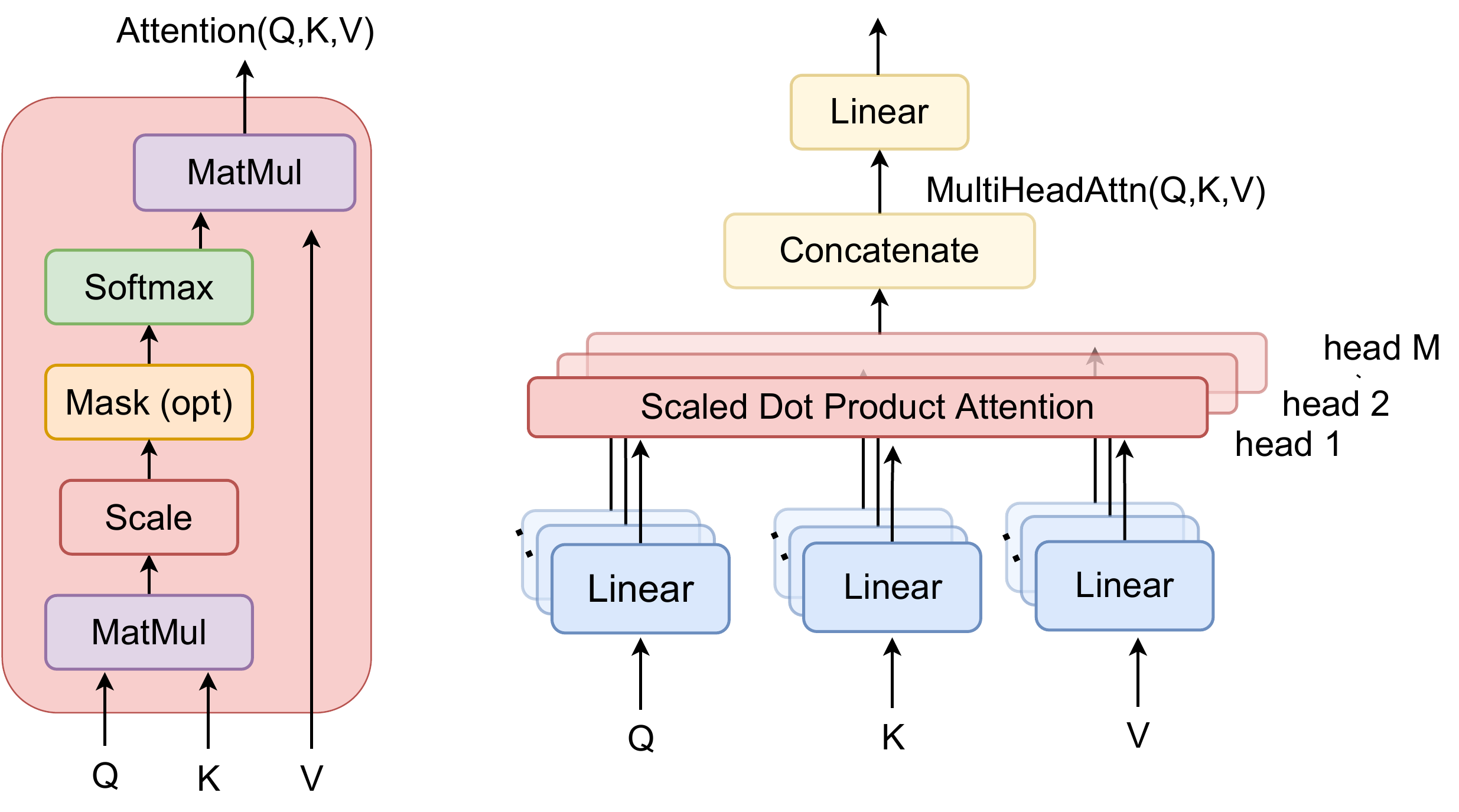}
    \caption{Illustrations of attention (left) and multi-headed attention (right).}
    \label{fig:Transformer}
\end{figure}

{Transformer architecture \cite{vaswani2017attention} comprises an attention-based encoder and decoder, with each module consisting of a stack of identical blocks. Each block in the encoder and decoder consists of two sub-layers: a multi-head attention (MHA) mechanism and a position-wise fully connected feedforward network as described in \Cref{fig:Transformer}. The MHA mechanism in the encoder allows each input element to attend to every other element in the sequence, enabling the model to capture long-range dependencies in the input sequence. The decoder typically uses a combination of MHA and encoder-decoder attention to attend to both the input sequence and the previously generated output elements. The feedforward network in each block of the Transformer provides non-linear transformations to the output of the attention mechanism. Next, we discuss operations involved in transformer layers, that is, multi-head attention and position-wise feedforward network:}


\paragraph{Attention in Transformers}
{Attention mechanism, first proposed by \citet{bahdanau2014neural}, has revolutionized sequence modeling and transduction models in various tasks of NLP, speech, and computer vision \cite{galassi2020attention, cho2015describing, wang2016survey, chaudhari2021attentive}. Broadly, it allows the model to focus on specific parts of the input or output sequence, without being limited by the distance between the elements. We can describe the attention mechanism as the mapping of a query vector and set of key-value vector pairs to an output. Precisely, the output vector is computed as a weighted summation of value vectors where the weight of a value vector is obtained by computing the compatibility between the query vector and key vector. Let, each query $Q$ and key $K$ are $d_k$ dimensional and value $V$ is $d_v$ dimensional. Specific to the Transformer, the compatibility function between a query and each key is computed as their dot product between scaled by $\sqrt{d_k}$. To obtain the weights on values, the scaled dot product values are passed through a softmax function:}

\begin{equation}
\label{attention}
    \Attn(\textbf{Q},\textbf{K},\textbf{V}) = \Softmax\left(\frac{\textbf{Q}\textbf{K}^{T}}{\sqrt{d_{k}}}\right)\textbf{V} 
\end{equation}

{Here multiple queries, keys, and value vectors, are packed together in matrix form respectively denoted by $\textbf{Q} \in {\mathbb{R}^{N\times d_{k}}}$, $\textbf{K} \in {\mathbb{R}^{M\times d_{k}}}$, and $\textbf{V} \in {\mathbb{R}^{M\times d_{v}}}$. \textit{N} and \textit{M} represent the lengths of queries and keys (or values). Scaling of dot product attention becomes critical to tackling the issue of small gradients with the increase in $d_k$ \cite{vaswani2017attention}.}

{
Instead of performing single attention in each transformer block, multiple attentions in lower-dimensional space have been observed to work better \cite{vaswani2017attention}. This observation gave rise to \textbf{Multi-Head Attention}: For $h$ heads and dimension of tokens in the model $d_m$, the $d_m$-dimensional query, key, and values are projected $h$ times to $d_k$, $d_k$, and $d_v$ dimensions using learnable linear projections\footnote{Projection weights are neither shared across heads nor query, key, and values.}. Each head performs attention operation as per \Cref{attention}. The $h$ $d_v$-dimensional are concatenated and projected back to $d_{m}$ using another projection matrix:

\begin{align}
    \MultiHead(\textbf{Q},\textbf{K},\textbf{V}) &= \Concat(\Head{}_{1},....\Head{}_{h})\textbf{W}^{O}, \\
    \text{with } \Head{}_{i} &=\Attn(\textbf{Q}\textbf{W}^{Q}_{i},\textbf{K}\textbf{W}^{K}_{i},\textbf{V}\textbf{W}^{V}_{i})
\end{align}

Where $\textbf{W}^{Q}, \textbf{W}^{K} \in {\mathbb{R}^{d_{model}\times d_{k}}},  \textbf{W}^{V} \in \mathbb{R}^{d_{model}\times d_{v}}, \textbf{W}^{O} \in {\mathbb{R}^{hd_{v}\times d_{model}}}$ are learnable projection matrices. Intuitively, multiple attention heads allow for attending to parts of the sequence differently (e.g., longer-term dependencies versus shorter-term dependencies). Intuitively, multiple attention heads allow for attending in different representational spaces jointly.
}

\paragraph{Position-wise FFN}
The position-wise FNN consists of two dense layers. It is referred to position-wise since the same two dense layers are used for each positioned item in the sequence and are equivalent to applying two $1\times1$ convolution layers.
\paragraph{Residual Connection and Normalization}
Residual connection and Layer Normalization are employed around each module for building a deep model. For example, each encoder block output can be defined as follows:
\begin{equation}
    H^{'} = \LayerNorm(\SelfAttention(X) + X)
\end{equation}
\begin{equation}
    H = \LayerNorm(\FFN(H^{'}) + H^{'})
\end{equation}
 $\SelfAttention(.)$ denotes attention module with $\textbf{Q} = \textbf{K} = \textbf{V} = \textbf{X}$, where $\textbf{X}$ is the output of the previous layer.

 Transformer-based architecture turned out to be better than many other architectures such as RNN, LSTM/GRU, etc. One of the major difficulties when applying a Transformer to
speech applications that it requires more complex configurations
(e.g., optimizer, network structure, data augmentation) than the conventional RNN-based models.  Speech signals are continuous-time signals with much higher dimensionality than text data. This high dimensionality poses significant computational challenges for the Transformer architecture, originally designed for sequential text data. Speech signals also have temporal dependencies, which means that the model needs to be able to process and learn from the entire signal rather than just a sequence of inputs. Also, speech signals are inherently variable and complex. The same sentence can be spoken differently and even by the same person at different times. This variability requires the model to be robust to differences in pitch, accent, and speed of speech.

\subsubsection{Application}
Recent advancements in NLP which lead to a paradigm shift in the field are highly attributed to the foundation models that are primarily a part of the transformers category, with self-attention being a key ingredient \cite{bommasani2021opportunities}. The recent models have demonstrated human-level performance in several professional and academic benchmarks. For instance, GPT4 scored within the top 10\% of test takers on a simulated version of the Uniform Bar Examination \cite{OpenAI2023GPT4TR}. While speech processing has not yet seen a shift in paradigm as in NLP owing to the capabilities of foundational models, even so, transformers have significantly contributed to advancement in the field including but not limited to the following tasks: automatic speech recognition, speech translation, speech synthesis, and speech enhancement, most of which we discuss in detail in \Cref{speech_processing_tasks}.

RNNs and Transformers are two widely adopted neural network architectures employed in the domain of Natural Language Processing (NLP) and speech processing. While RNNs process input words sequentially and preserve a hidden state vector over time, Transformers analyze the entire sentence in parallel and incorporate an internal attention mechanism. This unique feature makes Transformers more efficient than RNNs \cite{karita2019comparative}. Moreover, Transformers employ an attention mechanism that evaluates the relevance of other input tokens in encoding a specific token. This is particularly advantageous in machine translation, as it allows the Transformer to incorporate contextual information, thereby enhancing translation accuracy \cite{karita2019comparative}. To achieve this, Transformers combine word vector embeddings and positional encodings, which are subsequently subjected to a sequence of encoders and decoders. These fundamental differences between RNNs and Transformers establish the latter as a promising option for various natural language processing tasks \cite{karita2019comparative}.

A comparative study on transformer vs. RNN   \cite{karita2019comparative} in speech applications found that transformer neural networks achieve state-of-the-art performance in neural machine translation and other natural language processing applications \cite{karita2019comparative}. The study compared and analysed transformer and conventional RNNs in a total of 15 ASR, one multilingual ASR, one ST, and two TTS applications. The study found that transformer neural networks outperformed RNNs in most applications tested. Another survey of transformer-based models in speech processing found that transformers have an advantage in comprehending speech, as they analyse the entire sentence simultaneously, whereas RNNs process input words one by one.

Transformers have been successfully applied in end-to-end speech processing, including automatic speech recognition (ASR), speech translation (ST), and text-to-speech (TTS) \cite{li2019neural}. In 2018, the Speech-Transformer was introduced as a no-recurrence sequence-to-sequence model for speech recognition. To reduce the dimension difference between input and output sequences, the model's architecture was modified by adding convolutional neural network (CNN) layers before feeding the features to the transformer. In a later study \cite{nakatani2019improving}, the authors proposed a method to improve the performance of end-to-end speech recognition models based on transformers. They integrated the connectionist temporal classification (CTC) with the transformer-based model to achieve better accuracy and used language models to incorporate additional context and mitigate recognition errors.

In addition to speech recognition, the transformer model has shown promising results in TTS applications. The transformer based TTS model generates mel-spectrograms, followed by a WaveNet vocoder to output the final audio results \cite{li2019neural}. Several neural network-based TTS models, such as Tacotron 2, DeepVoice 3, and transformer TTS, have outperformed traditional concatenative and statistical parametric approaches in terms of speech quality \cite{li2019neural, shen2018natural, ping2017deep}.

One of the strengths of Transformer-based architectures for neural speech synthesis is their high efficiency while considering the global context \cite{gulati2020conformer,shi2020weak}. The Transformer TTS model has shown advantages in training and inference efficiency over RNN-based models such as Tacotron 2 \cite{shen2018natural}. The efficiency of the Transformer TTS network can speed up the training about 4.25 times \cite{li2019neural}. Moreover, Multi-Speech, a multi-speaker TTS model based on the Transformer \cite{li2019neural}, has demonstrated the effectiveness of synthesizing a more robust and better quality multi-speaker voice than naive Transformer-based TTS.

In contrast to the strengths of Transformer-based architectures in neural speech synthesis, large language models based on Transformers such as BERT \cite{devlin2018bert}, GPT \cite{radford2018improving}, XLNet \cite{yang2019xlnet}, and T5 \cite{raffel2020exploring} have limitations when it comes to speech processing. One of the issues is that these models require discrete tokens as input, necessitating using a tokenizer or a speech recognition system, introducing errors and noise. Furthermore, pre-training on large-scale text corpora can lead to domain mismatch problems when processing speech data. To address these limitations, dedicated frameworks have been developed for learning speech representations using transformers, including wav2vec \cite{schneider2019wav2vec}, data2vec \cite{baevski2022data2vec}, Whisper \cite{radford2022robust}, VALL-E \cite{wang2023neural}, Unispeech \cite{wang2021unispeech}, SpeechT5 \cite{ao2021speecht5} etc. We discuss some of them as follows.
\begin{itemize}
    \item Speech representation learning frameworks, such as wav2vec, have enabled significant advancements in speech processing tasks. One recent framework, w2v-BERT \cite{wang2019bridging}, combines contrastive learning and MLM to achieve self-supervised speech pre-training on discrete tokens. Fine-tuning wav2vec models with limited labeled data has also been demonstrated to achieve state-of-the-art results in speech recognition tasks \cite{baevski2019vq}. Moreover, XLS-R \cite{babu2021xls}, another model based on wav2vec 2.0, has shown state-of-the-art results in various tasks, domains, data regimes, and languages, by leveraging multilingual data augmentation and contrastive learning techniques on a large scale. These models learn universal speech representations that can be transferred across languages and domains, thus representing a significant advancement in speech representation learning.
    
    \item  Transformers have been increasingly popular in the development of frameworks for learning representations from multi-modal data, such as speech, images, and text. Among these frameworks, Data2vec \cite{baevski2022data2vec} is a self-supervised training approach that aims to learn joint representations to capture cross-modal correlations and transfer knowledge across modalities. It has outperformed other unsupervised methods for learning multi-modal representations in benchmark datasets. However, for tasks that require domain-specific models, such as speech recognition or speaker identification, domain-specific models may be more effective, particularly when dealing with data in specific domains or languages. The self-supervised training approach of Data2vec enables cost-effective and scalable learning of representations without requiring labeled data, making it a promising framework for various multi-modal learning applications.
    
    \item The field of speech recognition has undergone a revolutionary change with the advent of the Whisper model \cite{radford2022robust}. This innovative solution has proven to be highly versatile, providing exceptional accuracy for various speech-related tasks, even in challenging environments. The Whisper model achieves its outstanding performance through a minimalist approach to data pre-processing and weak supervision, which allows it to deliver state-of-the-art results in speech processing. The model is capable of performing multilingual speech recognition, translation, and language identification, thanks to its training on a diverse audio dataset. Its multitasking model can cater to various speech-related tasks, such as transcription, voice assistants, education, entertainment, and accessibility. One of the unique features of Whisper is its minimalist approach to data pre-processing, which eliminates the need for significant standardization and simplifies the speech recognition pipeline. The resulting models generalize well to standard benchmarks and deliver competitive performance without fine-tuning, demonstrating the potential of advanced machine learning techniques in speech processing.
    \item Text-to-speech synthesis has been a topic of interest for many years, and recent advancements have led to the development of new models such as VALL-E \cite{wang2023neural}. VALL-E is a novel text-to-speech synthesis model that has gained significant attention due to its unique approach to the task. Unlike traditional TTS systems, VALL-E treats the task as a conditional language modelling problem and leverages a large amount of semi-supervised data to train a generalized TTS system. It can generate high-quality personalized speech with a 3-second acoustic prompt from an unseen speaker and provides diverse outputs with the same input text. VALL-E also preserves the acoustic environment and the speaker's emotions about the acoustic prompt, without requiring additional structure engineering, pre-designed acoustic features, or fine-tuning. Furthermore, VALL-E X \cite{zhang2023speak} is an extension of VALL-E that enables cross-lingual speech synthesis, representing a significant advancement in TTS technology.
\end{itemize}
\begin{figure*}[h]
    \centering
    \includegraphics[width=\textwidth]{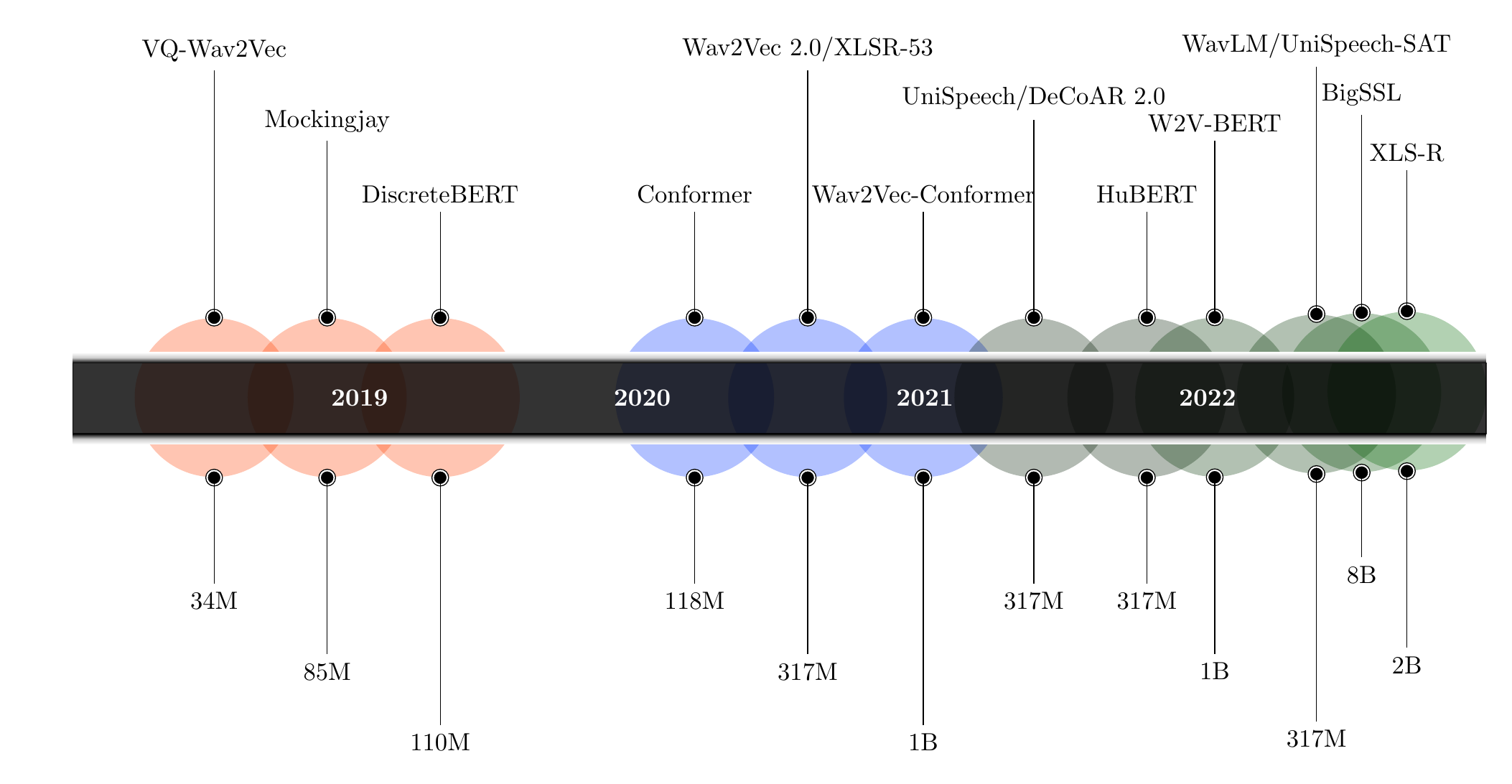}
    \caption{Timeline highlighting notable large Transformer models developed for speech processing, along with their corresponding parameter sizes.}
    \label{fig:timeline}
\end{figure*}

The timeline highlights the development of large transformer based models for speech processing is shown in \Cref{fig:timeline}. The size of the models has grown exponentially, with significant breakthroughs achieved in speech recognition, synthesis, and translation. These large models have set new performance benchmarks in the field of speech processing, but also pose significant computational and data requirements for training and inference.

\subsection{Conformer}
\label{sec:conformer}
The Transformer architecture, which utilizes a self-attention mechanism, has successfully replaced recurrent operations in previous architectures. Over the past few years, various Transformer variants have been proposed \cite{gulati2020conformer}. Architectures combining Transformers and CNNs have recently shown promising results on speech-processing tasks \cite{9414087}. To efficiently model both local and global dependencies of an audio sequence, several attempts have been made to combine CNNs and Transformers. One such architecture proposed by the authors is the Conformer \cite{gulati2020conformer}, a convolution-augmented transformer for speech recognition. Conformer outperforms RNNs, previous Transformers, and CNN-based models, achieving state-of-the-art performance in speech recognition. The Conformer model consists of several building blocks, including convolutional layers, self-attention layers, and feedforward layers. The architecture of the Conformer model can be summarized as follows:
\begin{itemize}
\item Input Layer: The Conformer model inputs a sequence of audio features, such as MFCCs or Mel spectrograms.

\item Convolutional Layers: Local features are extracted from the audio signal by processing the input sequence through convolutional layers.

\item Self-Attention Layers: The Conformer model incorporates self-attention layers following the convolutional layers. Self-attention is a mechanism that enables the model to focus on various sections of the input sequence while making predictions. This is especially advantageous for speech recognition because it facilitates capturing long-term dependencies in the audio signal.

\item Feedforward Layers: After the self-attention layers, the Conformer model applies a sequence of feedforward layers intended to process the output of the self-attention layers further and ready it for the ultimate prediction.

\item Output Layer: Finally, the output from the feedforward layers undergoes a softmax activation function to generate the final prediction, typically representing a sequence of character labels or phonemes.

\end{itemize}

The conformer model has emerged as a promising neural network architecture for various speech-related research tasks, including but not limited to speech recognition, speaker recognition, and language identification. In a recent study by \citet{gulati2020conformer}, the conformer model was demonstrated to outperform previous state-of-the-art models, particularly in speech recognition significantly. This highlights the potential of the conformer model as a key tool for advancing speech-related research.
\subsubsection{Application}
The Conformer model stands out among other speech recognition models due to its ability to efficiently model both local and global dependencies of an audio sequence. This is crucial for speech recognition, language translation, and audio classification \cite{assemblyai,gulati2020conformer,nvidiadocs}. The model achieves this through self-attention and convolution modules, combining the strengths of CNNs and Transformers. While CNNs capture local information in audio sequences, the self-attention mechanism captures global dependencies \cite{nvidiadocs}. The Conformer model has achieved remarkable performance in speech recognition tasks, setting benchmarks on datasets such as LibriSpeech and AISHELL-1. 

Despite these successes, speech synthesis and recognition challenges persist, including difficulties generating natural-sounding speech in non-English languages and real-time speech generation. To address these limitations, Wang et al. \cite{zhang2020pushing} proposed a novel approach that combines noisy student training with SpecAugment and large Conformer models pre-trained on the Libri-Light dataset using the wav2vec 2.0 pre-training method. This approach achieved state-of-the-art word error rates on the LibriSpeech dataset. Recently, \citet{wang2022conformer} developed Conformer-LHUC, an extension of the Conformer model that employs learning hidden unit contribution (LHUC) for speaker adaptation. Conformer-LHUC has demonstrated exceptional performance in elderly speech recognition and shows promise for the clinical diagnosis and treatment of Alzheimer's disease.

Several enhancements have been made to the Conformer-based model to address high word error rates without a language model, as documented in \cite{liu2022improvement}. \citet{wu2022deep} proposed a deep sparse Conformer to improve its long-sequence representation capabilities. Furthermore, \citet{burchi2023audio} have recently enhanced the noise robustness of the Efficient Conformer architecture by processing both audio and visual modalities. In addition, models based on Conformer, such as Transducers \cite{kim2021generalizing}, have been adopted for real-time speech recognition \cite{papastratis2021speech} due to their ability to process audio data much more quickly than conventional recurrent neural network (RNN) models. 
\subsection{Sequence to Sequence Models}
\label{sec:s2s}

The sequence-to-sequence (seq2seq) model in speech processing is popularly used for ASR, ST, and TTS tasks. The general architecture of the seq2seq model involves an encoder-decoder network that learns to map an input sequence to an output sequence of varying lengths. In the case of ASR, the input sequence is the speech signal, which is processed by the encoder network to produce a fixed-length feature vector representation of the input signal. The decoder network inputs this feature vector and produces the corresponding text sequence. This can be achieved through a stack of RNNs \cite{prabhavalkar2017comparison}, Transformer \cite{8462506} or Conformer \cite{gulati2020conformer} in the encoder and decoder networks.

The sequence-to-sequence model has emerged as a potent tool in speech translation. It can train end-to-end to efficiently map speech spectrograms in one language to their corresponding spectrograms in another. The notable advantage of this approach is eliminating the need for an intermediate text representation, resulting in improved efficiency. Additionally, the Seq2seq models have been successfully implemented in speech generation tasks, where they reverse the ASR approach. In such applications, the input text sequence serves as the input, with the encoder network creating a feature vector representation of the input text. The decoder network then leverages this representation to generate the desired speech signal.

\begin{figure}
    \centering
    \includegraphics[width=0.6\columnwidth]{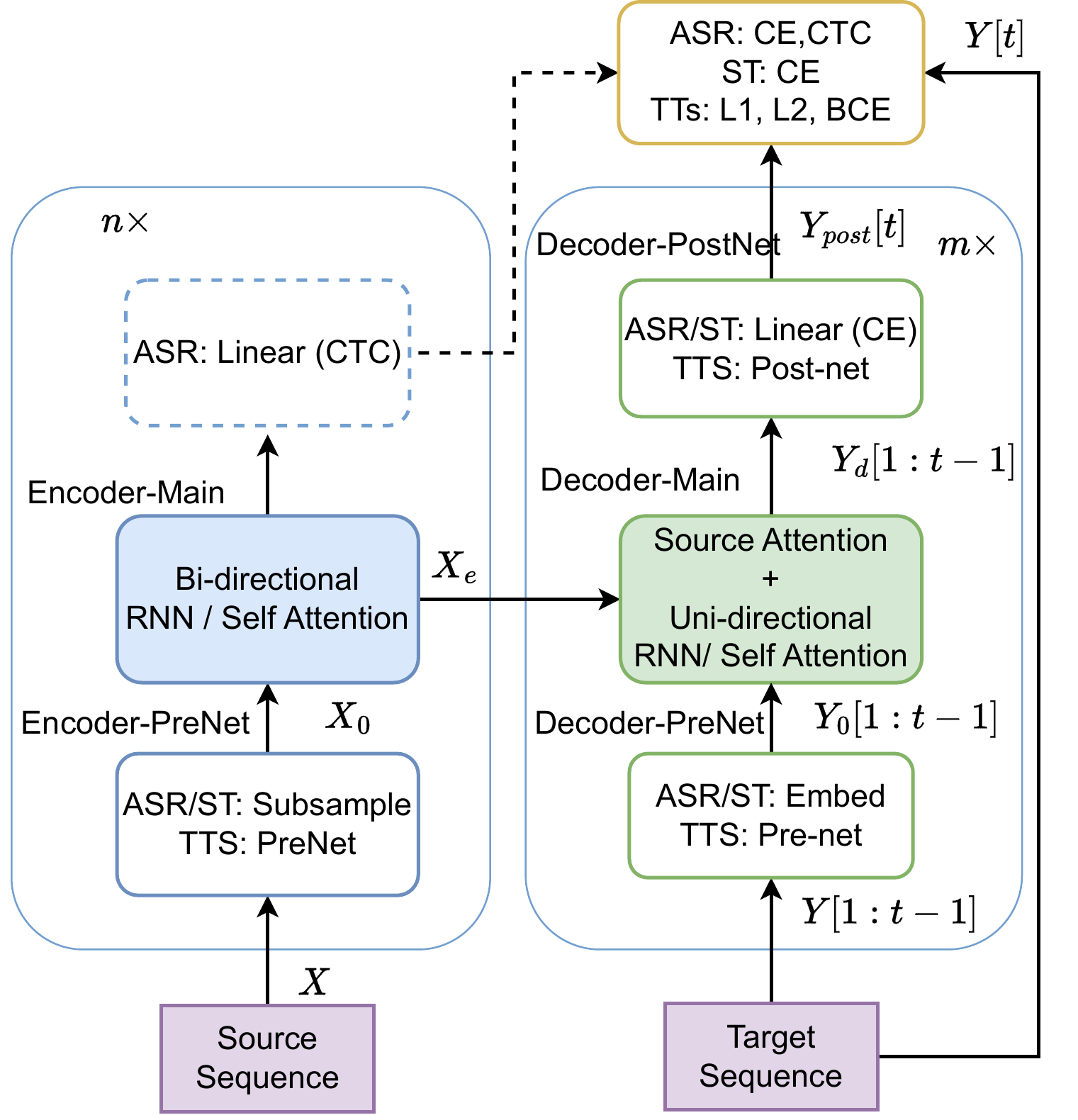}
    \caption{Unified formulation for Sequence-to-Sequence architecture in speech applications \cite{karita2019comparative}. $X$ and $Y$ are source and target sequences respectively.}
    \label{fig:s2s}
\end{figure}

\citet{karita2019comparative} conducted an extensive study comparing the performance of transformer and traditional RNN models on 15 different benchmarks for Automatic Speech Recognition (ASR), including a multilingual ASR benchmark, a Speech Translation (ST) benchmark, and two Text-to-Speech (TTS) benchmarks. In addition, they proposed a shared Sequence-to-Sequence (S2S) architecture for AST, TTS, and ST tasks, which is depicted in \Cref{fig:s2s}.

\begin{itemize}
    \item Encoder
    \begin{equation}
    \begin{aligned}
        X_{0} &= \EncoderPreNet(X), \\
        X_{e} &= \EncoderMain(X_{0})
    \end{aligned}
    \end{equation}
    where $X$ is the sequence of speech features (e.g. Mel spectrogram) for AST and ST and phoneme or character sequence for TTS.
    \item Decoder
        \begin{equation}
        \begin{aligned}
        Y_{0}[1:t-1] &= \DecoderPreNet(Y[1:t-1]),\\
        Y_{d}[t] &= \DecoderMain(X_{e},Y_{0}[1:t-1]),\\
        Y_{post}[1:t] &= \DecoderPostNet(Y_{d}[1:t]),
        \end{aligned}
    \end{equation}
    During the training stage, input to the decoder is ground truth target sequence $Y[1:t-1]$. The Decoder-Main module is utilized to produce a subsequent target frame. This is accomplished by utilizing the encoded sequence $X_{e}$ and the prefix of the target prefix $Y_{0}[1: t-1]$. The decoder is mostly unidirectional for sequence generation and often uses an attention mechanism \cite{bahdanau2014neural} to produce the output. 
\end{itemize}

Seq2seq models have been widely used in speech processing, initially based on RNNs. However, RNNs face the challenge of processing long sequences, which can lead to the loss of the initial context by the end of the sequence \cite{karita2019comparative}. To overcome this limitation, the transformer architecture has emerged, leveraging self-attention mechanisms to handle sequential data. The transformer has shown remarkable performance in tasks such as ASR, ST, and speech synthesis. As a result, the use of RNN-based seq2seq models has declined in favour of the transformer-based approach. 

\subsubsection{Application}
Seq2seq models have been used for speech processing tasks such as voice conversion \cite{8683282,huang2019voice}, speech synthesis \cite{wang2017tacotron,wang2019deep,okamoto2019real,9003956,huang2019voice}, and speech recognition. The field of ASR has seen significant progress, with several advanced techniques emerging as popular options. These include the CTC approach, which has been further developed and improved upon through recent advancements \cite{graves2014towards}, as well as attention-based approaches that have also gained traction \cite{chorowski2015attention}. The growing interest in these techniques has increased the use of seq2seq models in the speech community.

\begin{itemize}
    \item Attention-based Approaches: The attention mechanism is a crucial component of sequence-to-sequence models, allowing them to effectively weigh input acoustic features during decoding \cite{bahdanau2014neural,luong2015effective}. Attention-based Seq2seq models utilize previously generated output tokens and the complete input sequence to factorize the joint probability of the target sequence into individual time steps. The attention mechanism is conditioned on the current decoder states and runs over the encoder output representations to incorporate information from the input sequence into the decoder output. Incorporating attention mechanisms in Seq2Seq models has resulted in an impressive performance in various speech processing tasks, such as speech recognition \cite{nankaku2021neural,prabhavalkar2017comparison,tuske2019advancing,weng2018improving}, text-to-speech \cite{shen2018natural,8682353,9053915}, and voice conversion \cite{8683282,huang2019voice}. These models have demonstrated competitiveness with traditional state-of-the-art approaches. Additionally, attention-based Seq2Seq models have been used for confidence estimation tasks in speech recognition, where confidence scores generated by a speech recognizer can assess transcription quality \cite{li2021confidence}. Furthermore, these models have been explored for few-shot learning, which has the potential to simplify the training and deployment of speech recognition systems \cite{higy2018few}.
    \item Connectionist Temporal Classification: While attention-based methods create a soft alignment between input and target sequences, approaches that utilize CTC loss aim to maximize log conditional likelihood by considering all possible monotonic alignments between them. These CTC-based Seq2Seq models have delivered competitive results across various ASR benchmarks \cite{higuchi2022bert,majumdar2021citrinet,synnaeve2020end,gulati2020conformer} and have been extended to other speech-processing tasks such as voice conversion \cite{zhang2019sequence,9362095,liu2021any}, speech synthesis \cite{zhang2019sequence} etc. Recent studies have concentrated on enhancing the performance of Seq2Seq models by combining CTC with attention-based mechanisms, resulting in promising outcomes. This combination remains a subject of active investigation in the speech-processing domain.
\end{itemize}
\subsection{Reinforcement Learning}
\label{sec:rl}

Reinforcement learning (RL) is a machine learning paradigm that trains an agent to perform discrete actions in an environment and receive rewards or punishments based on its interactions. The agent aims to learn a policy that maximizes its long-term reward. In recent years, RL has become increasingly popular and has been applied to various domains, including robotics, game playing, and natural language processing. RL has been utilized in speech recognition, speaker diarization, and speech enhancement tasks in the speech field. One of the significant benefits of using RL for speech tasks is its ability to learn directly from raw audio data, eliminating the need for hand-engineered features. This can result in better performance compared to traditional methods that rely on feature extraction. By capturing intricate patterns and relationships in the audio data, RL-based speech systems have the potential to enhance accuracy and robustness.

\subsubsection{Basic Models}
The utilization of deep reinforcement learning (DRL) in speech processing involves the environment (a set of states $S$), agent, actions ($A$), and reward ($r$). The semantics of these components depends on the task at hand. For instance, in ASR tasks, the environment can be composed of speech features, the action can be the choices of phonemes, and the reward could be the correctness of those phonemes given the input. Audio signals are one-dimensional time-series signals that undergo pre-processing and feature extraction procedures. Pre-processing steps include noise suppression, silence removal, and channel equalization, improving audio signal quality and creating robust and efficient audio-based systems. Previous research has demonstrated that pre-processing improves the performance of deep learning-based audio systems \cite{latif2020speech}. 

Feature extraction is typically performed after pre-processing to convert the audio signal into meaningful and informative features while reducing their number. MFCCs and spectrograms are popular feature extraction choices in speech-based systems \cite{latif2020speech}. These features are then given to the DRL agent to perform various tasks depending on the application. For instance, consider the scenario where a human speaks to a DRL-trained machine, where the machine must act based on features derived from audio signals.


\begin{itemize}
\item \textit{Value-based DRL:}
Given the state of the environment ($s$), a value function $Q: S\times A \rightarrow \mathbb{R}$ is learned to estimate overall future reward $Q(s, a)$ should an action $a$ be taken. This value function is parameterized with deep networks like CNN, Transformers, etc. 

\item \textit{Policy-based DRL:} As opposed to value-based RL, policy-based RL methods learns a policy function $\pi: S \rightarrow A$ that chooses the best possible action ($a$) based on reward.
\item \textit{Model-based DRL:}
Unlike the previous two approaches, model-based RL learns the dynamics of the environment in terms of the state transition probabilities, i.e., a function $M: S\times A\times S \rightarrow \mathbb{R}$. Given such a model, policy, or value functions are optimized.
\end{itemize}

\subsubsection{Application}
In speech-related research, deep reinforcement learning can be used for several purposes, including:

\paragraph{Speech recognition and Emotion modeling} Deep reinforcement learning (DRL) can be used to train speech recognition systems \cite{kala2018reinforcement,rajapakshe2020deep,tjandra2018sequence,chung2020semi,9207023} to transcribe speech accurately. In this case, the system receives an audio input and outputs a text sequence corresponding to the spoken words. The environmental states might be learned from the input audio features. The actions might be the generated phonemes. The reward could be the similarity between the generated and gold phonemes, quantified in edit distance. Several works have also achieved promising results for non-native speech recognition \cite{radzikowski2019dual}

DRL pre-training has shown promise in reducing training time and enhancing performance in various Human-Computer Interaction (HCI) applications, including speech recognition \cite{rajapakshe2020deep}. Recently, researchers have suggested using a reinforcement learning algorithm to develop a Speech Enhancement (SE) system that effectively improves ASR systems.  However, ASR systems are often complicated and composed of non-differentiable units, such as acoustic and language models. Therefore, the ASR system's recognition outcomes should be employed to establish the objective function for optimizing the SE model.  Other than ASR, SE, some studies have also focused on SER using DRL algorithms \cite{lakomkin2018emorl,rajapakshe2022novel,kansizoglou2019active}


\paragraph{Speaker identification} Similarly, for speaker identification tasks, the actions can be the speaker's choices, and a binary reward can be the correctness of choice.

\paragraph{Speech synthesis and coding} Likewise, the states can be the input text, the actions can be the generated audio, and the reward could be the similarity between the gold and generated mel-spectrogram.

Deep reinforcement learning has several advantages over traditional machine learning techniques. It can learn from raw data without needing hand-engineered features, making it more flexible and adaptable. It can also learn from feedback, making it more robust and able to handle noisy environments.

However, deep reinforcement learning also has some challenges that must be addressed. It requires a lot of data to train and can be computationally expensive. It also requires careful selection of the reward function to ensure that the system learns the desired behavior.

\subsection{Graph Neural Network}
\label{sec:gnn}
Over the past few years, the field of Graph Neural Networks (GNNs) has witnessed a remarkable expansion as a widely adopted approach for analysing and learning from data on graphs. GNNs have demonstrated their potential in various domains, including computer science, physics, mathematics, chemistry, and biology, by delivering successful outcomes. Furthermore, in recent times, the speech-processing domain has also witnessed the growth of GNNs.

\begin{figure*}
    \centering
    \includegraphics[width=\textwidth]{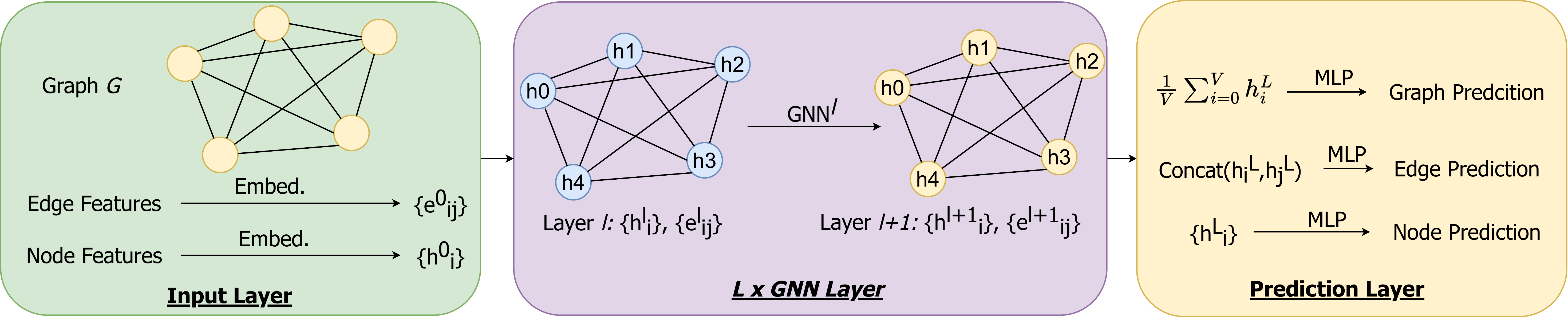}
    \caption{A standard experimental pipeline for GCNs, which embeds the graph node and embeds the graph node
edge features, performs several GNN layers to compute convolutional features, and finally
predicts a task-specific MLP layer.}
    \label{fig:GNN}
\end{figure*}

\subsubsection{Basic Models}
Speech processing involves analysing and processing audio signals, and GNNs can be useful in this context when we represent the audio data as a graph. In this answer, we will explain the architecture of GNNs for speech processing. The standard GNN pipeline is shown in \Cref{fig:GNN}, according to the application the GNN layer can consist of Graph Convolutional Layers \cite{zhang2019graph}, Graph Attention Layers \cite{velickovic2017graph}, or Graph Transformer \cite{yun2019graph}. 

\paragraph{Graph Representation of Speech Data}
The first step in using GNNs for speech processing is representing the speech data as a graph. One way to do this is to represent the speech signal as a sequence of frames, each representing a short audio signal segment. We can then represent each frame as a node in the graph, with edges connecting adjacent frames.

\paragraph{Graph Convolutional Layers}
Once the speech data is represented as a graph, we can use graph convolutional layers to learn representations of the graph nodes. Graph convolutional layers are similar to traditional ones, but instead of operating on a grid-like structure, they operate on graphs. These layers learn to aggregate information from neighboring nodes to update the features of each node.

\paragraph{Graph Attention Layers}
Graph attention layers can be combined with graph convolutional layers to give more importance to certain nodes in the graph. Graph attention layers learn to assign weights to neighbor nodes based on their features, which can help capture important patterns in speech data. Several works have used graph attention layers for neural speech synthesis \cite{liu2021graphspeech} or speaker verification \cite{jung2021graph} and diarization \cite{kwon2022multi}.

\paragraph{Recurrent Layers}
Recurrent layers can be used in GNNs for speech processing to capture temporal dependencies between adjacent frames in the audio signal. Recurrent layers allow the network to maintain an internal state that carries information from previous time steps, which can be useful for modeling the dynamics of speech signals.

\paragraph{Output Layers}
The output layer of a GNN for speech processing can be a classification layer that predicts a label for the speech data (e.g., phoneme or word) or a regression layer that predicts a continuous value (e.g., pitch or loudness). The output layer can be a traditional fully connected layer or a graph pooling layer that aggregates information from all the nodes in the graph.
\subsubsection{Application}

The advantages of using GNNs for speech processing tasks include their ability to represent the dependencies and interrelationships between various entities, which is suitable for speech processing tasks such as speaker diarization \cite{singh2023supervised,9054176,9688271}, speaker verification \cite{9414057,9746257}, speech synthesis \cite{9053355,liu2021graphspeech,sun2021graphpb}, or speech separation \cite{wang2023time,von2021graph}, which require the analysis of complex data representations. GNNs retain a state representing information from their neighborhood with arbitrary depth, unlike standard neural networks. GNNs can be used to model the relationship between phonemes and words. GNNs can learn to recognize words in spoken language by treating the phoneme sequence as a graph. GNNs can also be used to model the relationship between different acoustic features, such as pitch, duration, and amplitude, in speech signals, improving speech recognition accuracy.

GNNs have shown promising results in multichannel speech enhancement, where they are used for extracting clean speech from noisy mixtures captured by multiple microphones \cite{tzirakis2021multi}. The authors of a recent study \cite{nguyen2022multi} propose a novel approach to multichannel speech enhancement by combining Graph Convolutional Networks (GCNs) with spatial filtering techniques such as the Minimum Variance Distortionless Response (MVDR) beamformer. The algorithm aims to extract speech and noise from noisy signals by computing the Power Spectral Density (PSD) matrices of the noise and the speech signal of interest and then obtaining optimal weights for the beam former using a frequency-time mask. The proposed method combines the MVDR beam former with a super-Gaussian joint maximum a posteriori (SGJMAP) based SE gain function and a GCN-based separation network. The SGJMAP-based SE gain function is used to enhance the speech signals, while the GCN-based separation network is used to separate the speech from the noise further.

\subsection{Diffusion Probabilistic Model}
\label{sec:dpm}
\begin{figure*}[h]
    \centering
    \includegraphics[width=\textwidth]{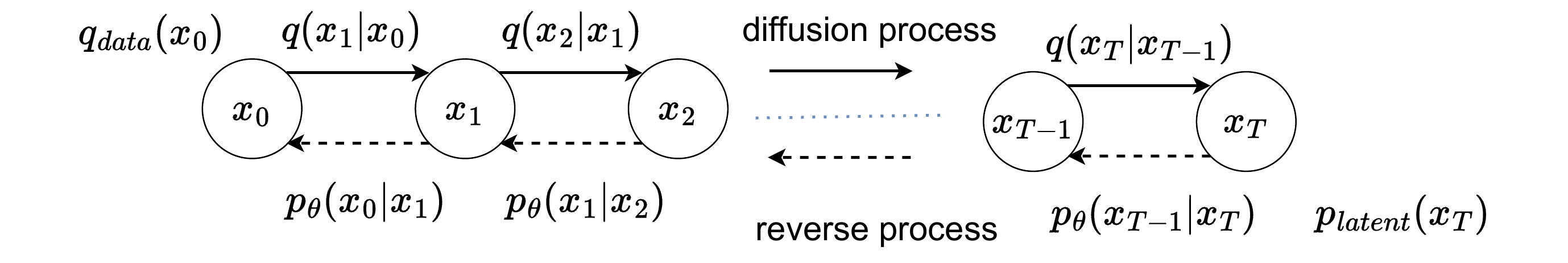}
    \caption{The Diffusion Probabilistic Model is a generative model that progressively transforms a noise distribution into the target data distribution through a series of diffusion steps, where the noise level decreases as the process continues. The model is trained by maximizing the likelihood of the data distribution and can be used for tasks such as speech synthesis, enhancement, and denoising.}
    \label{fig:diffusion}
\end{figure*}

Diffusion probabilistic models, inspired by non-equilibrium thermodynamics \cite{ho2020denoising, sohl2015deep}, have proven to be highly effective for generating high-quality images and audio. These models create a Markov chain of diffusion steps ($x_t \sim q(x_t|x_{t-1})$) from the original data ($x_{0}$) to the latent variable $x_{T}\sim \mathcal{N}(\mathbf{0}, \mathbf{I})$ by gradually adding pre-scheduled noise to the data. The reverse diffusion process then reconstructs the desired data samples ($x_{0}$) from the noise $x_{T}$, as shown in \cref{fig:diffusion}. Unlike VAE or flow models, diffusion models keep the dimensionality of the latent variables fixed. While mostly used for image and audio synthesis, diffusion models have potential applications in speech-processing tasks, such as speech synthesis and enhancement. This section offers a comprehensive overview of the fundamental principles of diffusion models and explores their potential uses in the speech domain.

\paragraph{Forward diffusion process}
Given a clean speech data $x_{0}\sim q_{data}(x_{0})$, 
\begin{equation}
    q(x_{1},...,x_{T}|x_{0}) = \prod_{t=1}^{T} q(x_{t}|x_{t-1}).
\end{equation}
At every time step $t$, $q(x_{t}|x_{t-1}):=\mathcal{N}(x_{t};\sqrt{1-\beta_{t}}x_{t-1},\beta_{t}\mathbf{I})$ where $\{\beta_{t} \in (0,1)\}_{t=1}^{T}$. As the forward process progresses, the data sample $x_{0}$ losses its distinguishable features, and as $T \to \infty$, $x_T$ approaches a standard Gaussian distribution.

\paragraph{Reverse diffusion process}
The reverse diffusion process is defined by a Markov chain from $x_{T}\sim \mathcal{N}(\mathbf{0}, \mathbf{I})$ to $x_{0}$ and parameterized by $\theta$:
\begin{equation}
    p_{\theta}(x_{0},...,x_{T-1}|x_{T}) = \prod_{t=1}^{T} p_{\theta}(x_{t-1}|x_{t})
\end{equation}
where $x_T \sim \mathcal{N}(0,I)$ and the transition probability $p_{\theta}(x_{t-1}|x_{t})$ is learnt through noise-estimation. This process eliminates the Gaussian noise added in the forward diffusion process.

\subsubsection{Application}
Diffusion models have emerged as a leading approach for generating high-quality speech in recent years \cite{chen2020wavegrad,kong2020diffwave,popov2021grad,popov2021diffusion,jeong2021diff,huang2022fastdiff}. These non-autoregressive models transform white noise signals into structured waveforms via a Markov chain with a fixed number of steps. One such model, FastDiff, has achieved impressive results in high-quality speech synthesis \cite{huang2022fastdiff}. By leveraging a stack of time-aware diffusion processes, FastDiff can generate high-quality speech samples 58 times faster than real-time on a V100 GPU, making it practical for speech synthesis deployment for the first time. It also outperforms other competing methods in end-to-end text-to-speech synthesis. Another powerful diffusion probabilistic model proposed for audio synthesis is DiffWave \cite{kong2020diffwave}. It is non-autoregressive and generates high-fidelity audio for different waveform generation tasks, such as neural vocoding conditioned on mel spectrogram, class-conditional generation, and unconditional generation. DiffWave delivers speech quality on par with the strong WaveNet vocoder \cite{oord2016wavenet} while synthesizing audio much faster.

Diffusion models have shown great promise in speech processing, particularly in speech enhancement \cite{9689602,serra2022universal,qiu2022srtnet,9746901}. Recent advances in diffusion probabilistic models have led to the development of a new speech enhancement algorithm that incorporates the characteristics of the noisy speech signal into the diffusion and reverses processes \cite{lu2022conditional}. This new algorithm is a generalized form of the probabilistic diffusion model, known as the conditional diffusion probabilistic model. During its reverse process, it can adapt to non-Gaussian real noises in the estimated speech signal. In addition, \citet{qiu2022srtnet} propose SRTNet, a novel method for speech enhancement that uses the diffusion model as a module for stochastic refinement. The proposed method comprises a joint network of deterministic and stochastic modules, forming the “enhance-and-refine” paradigm. The paper also includes a theoretical demonstration of the proposed method’s feasibility and presents experimental results to support its effectiveness.

\section{Speech Representation Learning}
The process of speech representation learning is essential for extracting pertinent and practical characteristics from speech signals, which can be utilized for various downstream tasks such as speaker identification, speech recognition, and emotion recognition. While traditional methods for engineering features have been extensively used, recent advancements in deep-learning-based techniques utilizing supervised or unsupervised learning have shown remarkable potential in this field. Nonetheless, a novel approach founded on self-supervised representation learning has surfaced, aiming to unveil the inherent structure of speech data and acquire representations that capture the underlying structure of the data. This approach surpasses traditional feature engineering methods and can significantly increase the accuracy to a considerable extent and effectiveness of downstream tasks. The primary objective of this new paradigm is to uncover informative and meaningful features from speech signals and outperform existing approaches. Therefore, this approach is considered a promising direction for future research in speech representation learning.

This section provides a comprehensive overview of the evolution of speech representation learning with neural networks. We will examine various techniques and architectures developed over the years, including the emergence of unsupervised representation learning methods like autoencoders, generative adversarial networks (GANs), and self-supervised representation learning frameworks. We will also examine the difficulties and constraints associated with these techniques, such as data scarcity, domain adaptation, and the interpretability of learned representations. Through a comprehensive analysis of the advantages and limitations of different representation learning approaches, we aim to provide insights into how to harness their power to improve the accuracy and robustness of speech processing systems.

\subsection{Supervised Learning}
{In supervised representation learning, the model is trained using annotated datasets to learn a mapping between input data and output labels. The set of parameters that define the mapping function is optimized during training to minimize the difference between the predicted and true output labels in the training data. The goal of supervised representation learning is to enable the model to learn a useful representation or features of the input data that can be used to accurately predict the output label for new, unseen data. For instance, supervised representation learning in speech processing using CNNs learn speech features from spectrograms. CNNs can identify patterns in spectrograms relevant to speech recognition, such as those corresponding to different phonemes or words. Unlike CNNs, which typically require spectrogram input, RNNs can directly take in the raw speech signals as input and learn to extract features or representations that are relevant for speech recognition or other speech-processing tasks. Learning speaker representations typically involves minimizing a loss function. \citet{chung2020defence} compares their effectiveness for speaker recognition tasks, we distill it in \Cref{lossfunction} to present an overview of commonly used loss functions. Additionally, a new angular variant of the prototypical loss is introduced in their work. Results from extensive experimental validation on the VoxCeleb1 test set indicate that the GE2E and prototypical networks outperform other models in terms of performance.}

\subsubsection{Deep speaker representations}

Speaker representation is a critical aspect of speech processing, allowing machines to analyze and process various parts of a speaker's voice, including pitch, intonation, accent, and speaking style. In recent years, deep neural networks (DNNs) have shown great promise in learning robust features for speaker recognition. This section reviews deep learning-based techniques for speaker representation learning that have demonstrated significant improvements over traditional methods.

These deep speaker representations can be applied to a range of speaker-recognition tasks beyond verification and identification, including diarization \cite{wang2018speaker,zhang2019fully,larcher2021speaker}, voice conversion \cite{wu2020one,lin2021s2vc,chou2019one}, multi-speaker TTS \cite{saito2021perceptual,paul2021universal,xue2022ecapa}, speaker adaptation \cite{chorowski2019unsupervised} etc. To provide a comprehensive overview, we analyzed deep embeddings from the perspectives of input raw \cite{jung2019rawnet,ravanelli2018speaker} or mel-spectogram \cite{snyder2018x}, network architecture \cite{lin2020wav2spk,desplanques2020ecapa}, temporal pooling strategies \cite{monteiro2019combining}, and loss functions \cite{snell2017prototypical,chung2020defence,wang2018cosface}. In the following subsection, we introduce two representative deep embeddings: $d$-vector \cite{variani2014deep} and $x$-vector \cite{snyder2017deep,snyder2018x}. These embeddings have been widely adopted recently and have demonstrated state-of-the-art performance in various speaker-recognition tasks. By understanding the strengths and weaknesses of different deep learning-based techniques for speaker-representation learning, we can better leverage their power to improve the accuracy and robustness of speaker-recognition systems.

\begin{itemize}
    \item \textbf{\textit{d}}-vector technique, proposed by Variani et al. (2014) \cite{variani2014deep}, serves as a frame-level speaker embedding method, as illustrated in Figure \ref{fig:dvectors}. In this approach, during the training phase, each frame within a training utterance is labeled with the speaker's true identity. This transforms the training process into a classification task, where a maxout Deep Neural Network (DNN) classifies the frames based on the speaker's identity. The DNN employs softmax as the output layer to minimize the cross-entropy loss between the ground-truth frame labels and the network's output. During the testing phase, the $d$-vector technique extracts the output activation of each frame from the last hidden layer of the DNN, serving as the deep embedding feature for that frame. To generate a compact representation called the $d$-vector, the technique computes the average of the deep embedding features from all frames within an utterance. The underlying hypothesis is that the compact representation space developed using a development set can effectively generalize to unseen speakers during the testing phase \cite{variani2014deep}.

\begin{figure}
    \centering
    \includegraphics[width = 0.6\columnwidth]{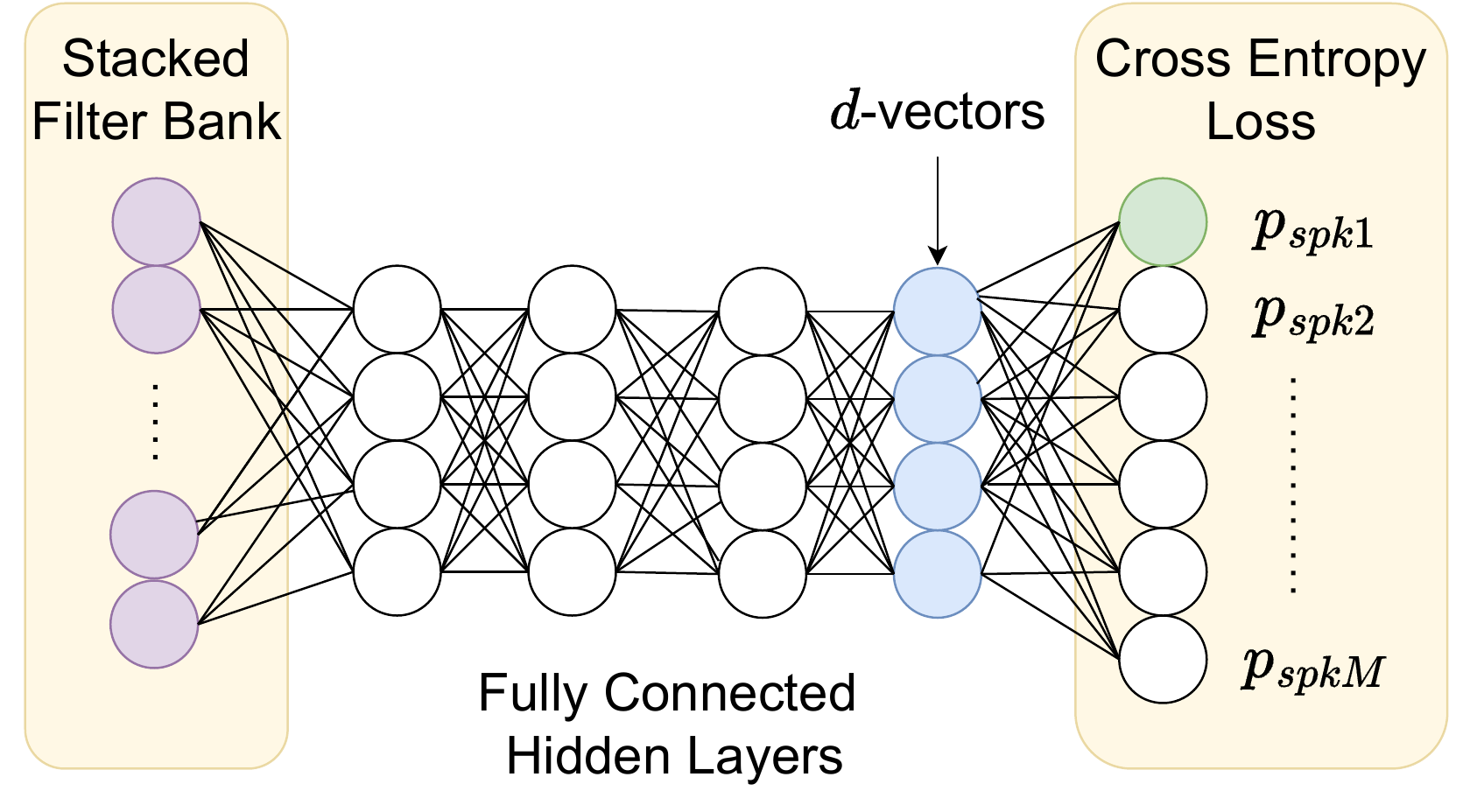}
    \caption{$d$-vector model architecture.}
    \label{fig:dvectors}
\end{figure}

\item \textbf{\textit{x}-vector} \cite{snyder2017deep,snyder2018x} is a segment-level speaker embedding and an advancement over the $d$-vector method as it incorporates additional modeling of temporal information and phonetic information in speech signals, resulting in improved performance compared to the $d$-vector. $x$-vector employs an aggregation process to move from frame-by-frame speaker labeling to utterance-level speaker labeling as highlighted in \Cref{fig:xvectors}. The network structure of the $x$-vector is depicted in a figure, which consists of time-delay layers for extracting frame-level speech embeddings, a statistical pooling layer for concatenating mean and standard deviation of embeddings as a segment-level feature, and a standard feedforward network for classifying the segment-level feature to its speaker. $x$-vector is the segment-level speaker embedding generated from the feedforward network's second-to-last hidden layer. The authors in \cite{yang2022data,9287426} have also discovered the significance of data augmentation in enhancing the performance of the $x$-vector.
\begin{figure}
    \centering
    \includegraphics[width = 0.6\columnwidth]{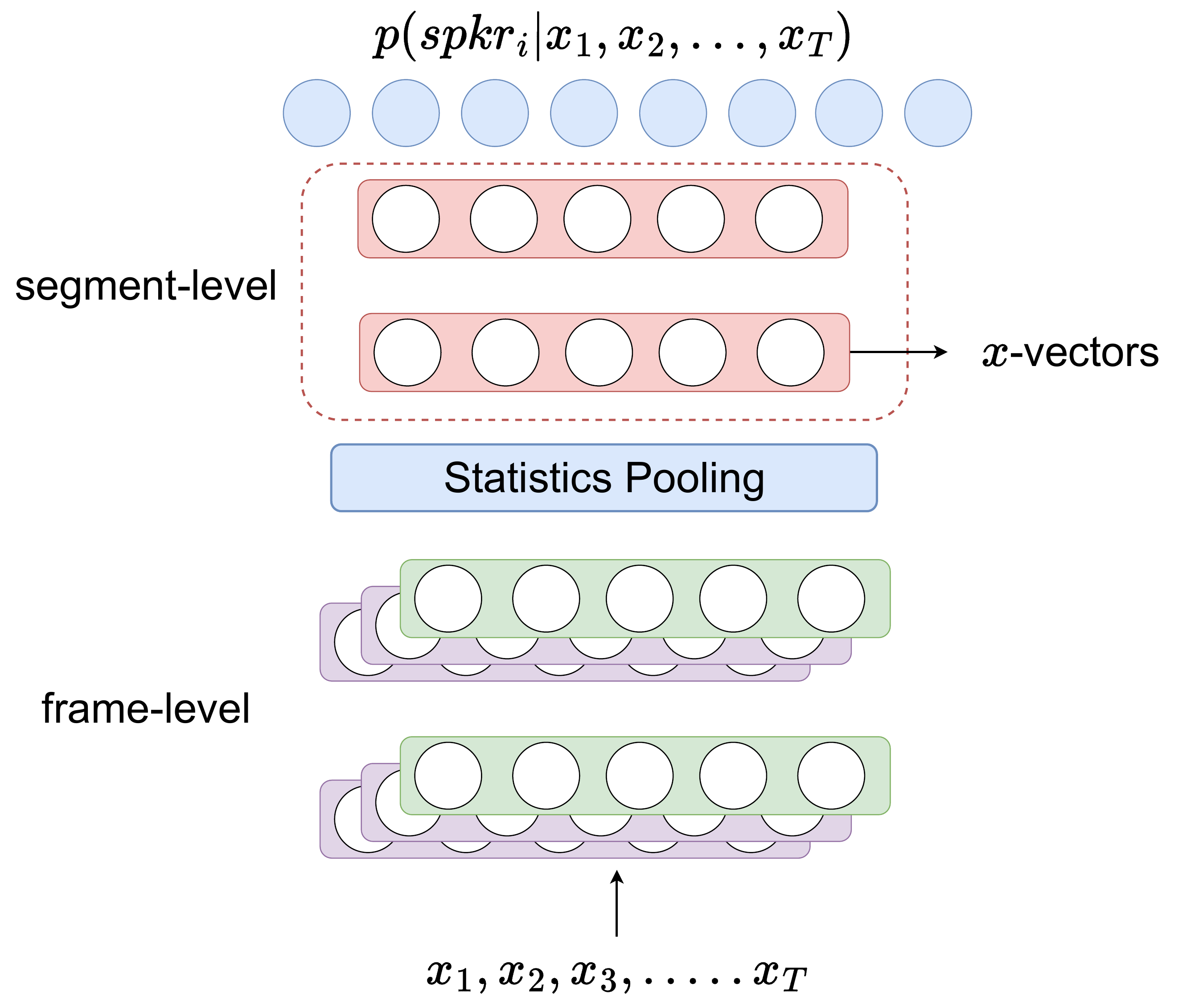}
    \caption{$x$-vector model architecture. $x_{1}$,$x_{2}$,....,$x_{T}$ are the spectral features such as Mel spectrograms of the speech utterance.}
    \label{fig:xvectors}
\end{figure}

\end{itemize}

\begin{table*}[h]
    \centering
    \caption{The table summarizes various loss functions used in training the speaker recognition models including their formulation \cite{chung2020defence}. }
    \label{tab:dataset}
    \resizebox{\columnwidth}{!}{%
        \begin{tabular}{@{}lll@{}}
            \toprule
            {Loss Function} &{Objective Type}& {Description}\\
            \midrule
            {Softmax}& Classification &$L_{S} = -\frac{1}{N}\sum_{i=1}^{N}\log\frac{\exp{W^{T}_{y_{i}}x_{i}+b_{y_{i}}}} {\sum_{j=1}^{C}\exp{W^{T}_{j}x_{i}+b_{y_{i}}}}$\\
            {AM-Softmax (CosFace) \cite{wang2018cosface}} & Classification &$L_{C} = -\frac{1}{N}\sum_{i=1}^{N}\log\frac{\exp{s(\cos{\theta_{y_{i},i}}-m)}}{\exp{s(\cos{\theta_{y_{i},i}}-m)}+ \sum_{j\neq y_{i}}\exp{s(\cos{\theta_{j,i}})}}$\\
            {AAM-Softmax (ArcFace) \cite{deng2019arcface}} & Classification &$    L_{A} = -\frac{1}{N}\sum_{i=1}^{N}\log\frac{\exp{s(\cos{\theta_{y_{i},i}}-m)}}{\exp{s(\cos{\theta_{y_{i},i}}+m)}+ \sum_{j\neq y_{i}}\exp{s(\cos{\theta_{j,i}})}}$\\
            {Triplet \cite{schroff2015facenet}} & Metric learning \cite{zhang2017end} &$    L_{T}=\frac{1}{N}\sum_{j=1}^{N} \max(0,||x_{j,0}-x_{j,1}||^{2}_{2},||x_{j,0}-x_{k\neq j,1}||^{2}_{2} + m)$\\
            {Prototypical \cite{snell2017prototypical}} & Metric learning \cite{snell2017prototypical} &$L_{P} = -\frac{1}{N}\sum_{j=1}^{N} \log\frac{\exp{\textbf{S}_{j,j}}}{\sum_{k=1}^{N}\exp{\textbf{S}_{j,k}}}$\\
            {Generalized end-to-end (GE2E) \cite{wan2018generalized}} & Metric learning \cite{wang2019adversarial} &$    L_{G} = -\frac{1}{N}\sum_{j,i} \log\frac{\exp{\textbf{S}_{j,i,j}}}{\sum_{k=1}^{N}\exp{\textbf{S}_{j,i,k}}}$\\
            {Angular Prototypical} & Metric learning &$    L_{AP} = -\frac{1}{N}\sum_{j,i} \log\frac{\exp{\textbf{S}_{j,i,j}}}{\sum_{k=1}^{N}\exp{\textbf{S}_{j,i,k}}}$
    \\
    \bottomrule
    \end{tabular}%
    }
    \label{lossfunction}
\end{table*}

\subsection{Unsupervised learning}
Unsupervised representation learning for speech processing has gained significant emphasis over the past few years. Similar to visual modality in CV and text modality in NLP, speech i.e. audio modality introduces unique challenges. Unsupervised speech representation learning is concerned with learning useful speech representations without using annotated data. Usually, the model is first pre-trained on the task where plenty of data is available. The model is then fined tuned or used to extract input representations for a small model, specifically targeting tasks with limited data.

One approach to addressing the unique challenges of unsupervised speech representation learning is to use probabilistic latent variable models (PLVM), which assume an unknown generative process produces the data and enables the learning of rich structural representations and reasoning about observed and unobserved factors of variation in complex datasets such as speech within a probabilistic framework. PLVM specified a joint distribution $p(x,z)$ over unobserved stochastic latent variable \textit{z} and observed variables \textit{x}. By factorizing the joint distribution into modular components, it becomes possible to learn rich structural representations and reason about observed and unobserved factors of variation in complex datasets such as speech within a probabilistic framework. 
\begin{figure}
    \centering
    \includegraphics[width=0.6\columnwidth]{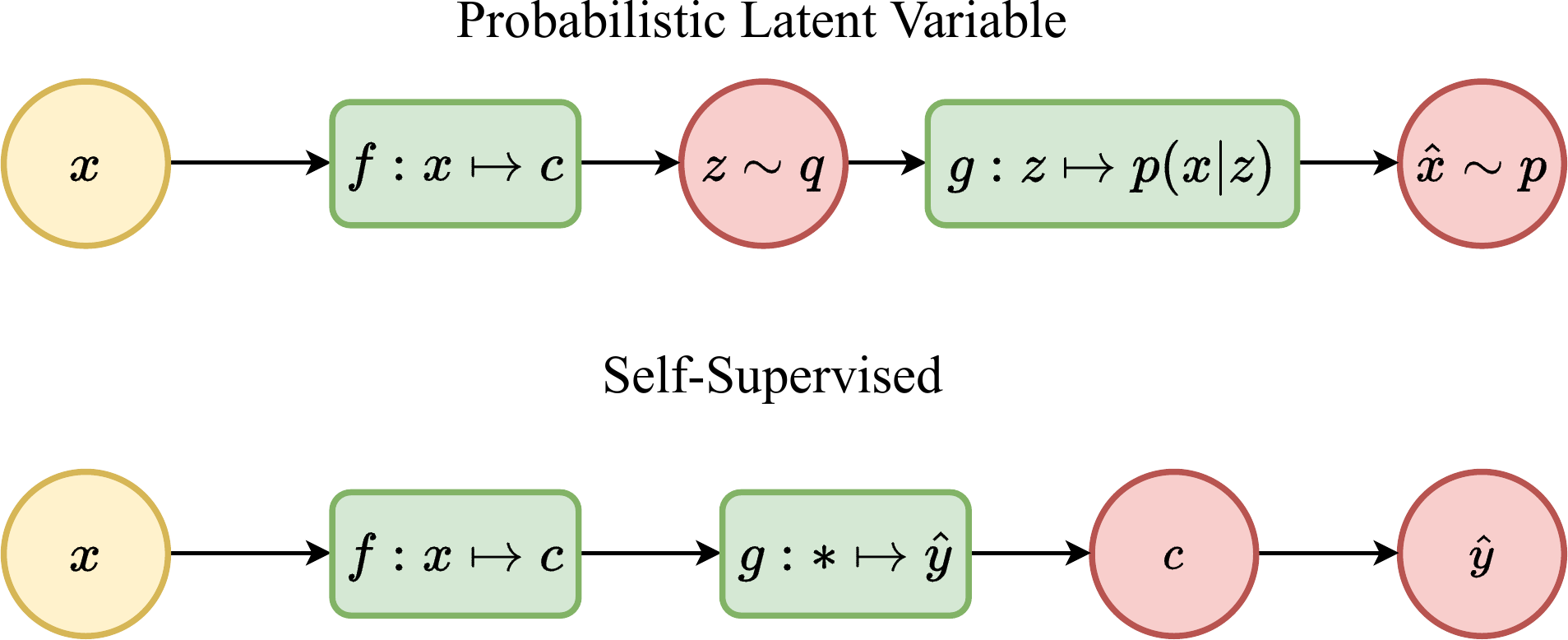}
    \caption{Overview of difference between probabilistic latent variable models and self-supervised learning. In latent variable models learn the functions $f(.)$ and $g(.)$ learn the parameters of distribution $p$ and $q$. The latent variable $z$ is used for representing learning.}
    \label{fig:unsupervised}
\end{figure}
    The likelihood of a PLVM given a data \textit{x} can be written as
    \begin{equation}
        p(x) = \int p(x|z) p(z) dz.
    \end{equation}

Probabilistic latent variable models provide a powerful way to learn a representation that captures the underlying relationships between observed and unobserved variables, without requiring explicit supervision or labels. These models involve unobserved latent variables that must be inferred from the observed data, typically using probabilistic inference techniques such as Markov Chain Monte Carlo (MCMC) methods. In the context of representation learning, Variational autoencoders (VAE) are commonly used with latent variable models for various speech processing tasks, leveraging the power of probabilistic modeling to capture complex patterns in speech data.

\subsection{Semi-supervised Learning}
Semi-supervised learning can be viewed as a process of optimizing a model using both labeled and unlabeled data. The set of labeled data points, denoted by $X_L$, contains $N_L$ items, where each item is represented as $(x_i, y_i)$ with $y_i$ being the label of $x_i$. On the other hand, the set of unlabeled data points, denoted by $X_U$, consists of $N_U$ items, represented as $x_{N_L+1}, x_{N_L+2}, ..., x_{N_L+N_U}$.

In semi-supervised learning, the objective is to train a model $f_{\theta}$ with parameters $\theta$ that can minimize the expected loss over the entire dataset. The loss function $L(y, f_{\theta}(x))$ is used to quantify the deviation between the model's prediction $f_{\theta}(x)$ and the ground truth label $y$. The expected loss can be mathematically expressed as:

\begin{equation}
L(y, f_{\theta}(x)) = E_{(x,y) \sim p_{data}(x,y)}[L(y,f_{\theta}(x))]
\end{equation}
where $p_{data}(x,y)$ is the underlying data distribution.In semi-supervised learning, the loss function is typically decomposed into two parts: a supervised loss term that is only defined on the labeled data, and an unsupervised loss term that is defined on both labeled and unlabelled data. The supervised loss term is calculated as follows:

\begin{equation}
    \mathcal{L}_{sup} = \frac{1}{N_{L}} \sum_{(x,y) \in X_{L}} L(y, f_{\theta}(x))
\end{equation}

The unsupervised loss term leverages the unlabelled data to encourage the model to learn meaningful representations that capture the underlying structure of the data. One common approach is to use a regularization term that encourages the model to produce similar outputs for similar input data. This can be achieved by minimizing the distance between the output of the model for two similar input data points. One such regularization term is the entropy minimization term, which can be expressed as:
\begin{equation}
    \mathcal{L}_{unsup} =  \frac{1}{N_{U}} \sum_{(x_{i}) \in X_{U}} \sum_{j=1}^{|y|} p_{\theta}(y_{j},x_{i}) \log p_{\theta} (y_{j},x_{i})
\end{equation}
where $p_{\theta}(y_j|x_i)$ is the predicted probability of the $j$-th label for the unlabelled data point $x_i$. Finally the overall objective function for semi-supervised learning can be expressed as $\mathcal{L} = \mathcal{L}_{sup} + \alpha \mathcal{L}_{unsup} $, $\alpha$ is a hyperparameter that controls the weight of the unsupervised loss term. The goal is to find the optimal parameters $\theta$ that minimize this objective function. Semi-supervised learning involves learning a model from both labelled and unlabelled data by minimizing a combination of supervised and unsupervised loss terms. By leveraging the additional unlabelled data, semi-supervised learning can improve the generalization and performance of the model in downstream tasks.

Semi-supervised learning techniques are increasingly being employed to enhance the performance of DNNs across a range of downstream tasks in speech processing, including ASR, TTS, etc. The primary objective of such approaches is to leverage large unlabelled datasets to augment the performance of supervised tasks that rely on labelled datasets. The recent advancements in speech recognition have led to a growing interest in the integration of semi-supervised learning methods to improve the performance of ASR and TTS systems \cite{zhang2020pushing,baskar2019semi,9795080,kahn2020self,xu2021self,9207023}. This approach is particularly beneficial in scenarios where labelled data is scarce or expensive to acquire. In fact, for many languages around the globe, labelled data for training ASR models are often inadequate, making it challenging to achieve optimal results. Thus, using a semi-supervised learning model trained on abundant resource data can offer a viable solution that can be readily extended to low-resource languages.

Semi-supervised learning has emerged as a valuable tool for addressing the challenges of insufficient annotations and poor generalization \cite{hady2013semi}. Research in various domains, including image quality assessment \cite{liu2019exploiting}, has demonstrated that leveraging both labelled and unlabelled data through semi-supervised learning can lead to improved performance and generalization. In the domain of speech quality assessment, several studies \cite{serra2021sesqa} have exploited the generalization capabilities of semi-supervised learning to enhance performance.

Moreover, semi-supervised learning has gained significant attention in other areas of speech processing, such as end-to-end speech translation \cite{pino2020self}. By leveraging large amounts of unlabelled data, semi-supervised learning approaches have demonstrated promising results in improving the performance and robustness of speech translation models. This highlights the potential of semi-supervised learning to address the limitations of traditional supervised learning approaches in a variety of speech processing tasks.

\subsection{Self-supervised representation learning (SSRL)} 

Self-supervised representation learning (SSRL) is a machine learning approach that focuses on achieving robust and in-depth feature learning while minimizing reliance on extensively annotated datasets, thus reducing the annotation bottleneck \cite{ericsson2022self,lee2022self}. SSRL comprises various techniques that allow models to be trained without needing human-annotated labels \cite{ericsson2022self,lee2022self}. One of the key advantages of SSRL is its ability to operate on unlabelled datasets, which reduces the need for large annotated datasets \cite{ericsson2022self,lee2022self}. In recent years, self-supervised learning has progressed rapidly, with some methods approaching or surpassing the efficacy of fully supervised learning methods. Self-supervised learning methods typically involve pretext tasks that generate pseudo labels for discriminative model training without actual labeling. The difference between self-supervised representation learning and unsupervised representation is highlighted in \cref{fig:unsupervised}. In contrast to unsupervised representation learning, SSRL techniques are designed to generate these pseudo labels for model training. The ability of SSRL to achieve robust and in-depth feature learning without relying heavily on annotated datasets holds great promise for the continued development of machine learning techniques.

SSRL differs from supervised learning mainly in terms of its data requirements. While supervised learning relies on labeled data, where the model learns from input-output pairs, SSL generates its own labels from the input data, eliminating the need for labeled data \cite{lee2022self}. The SSL approach trains the model to predict a portion of the input data, which is then utilized as a label for the task at hand \cite{lee2022self}. Although SSRL is an unsupervised learning technique, it seeks to tackle tasks commonly associated with supervised learning without relying on labeled data \cite{lee2022self}.

\begin{figure}
    \centering
    \includegraphics[width=0.8\columnwidth]{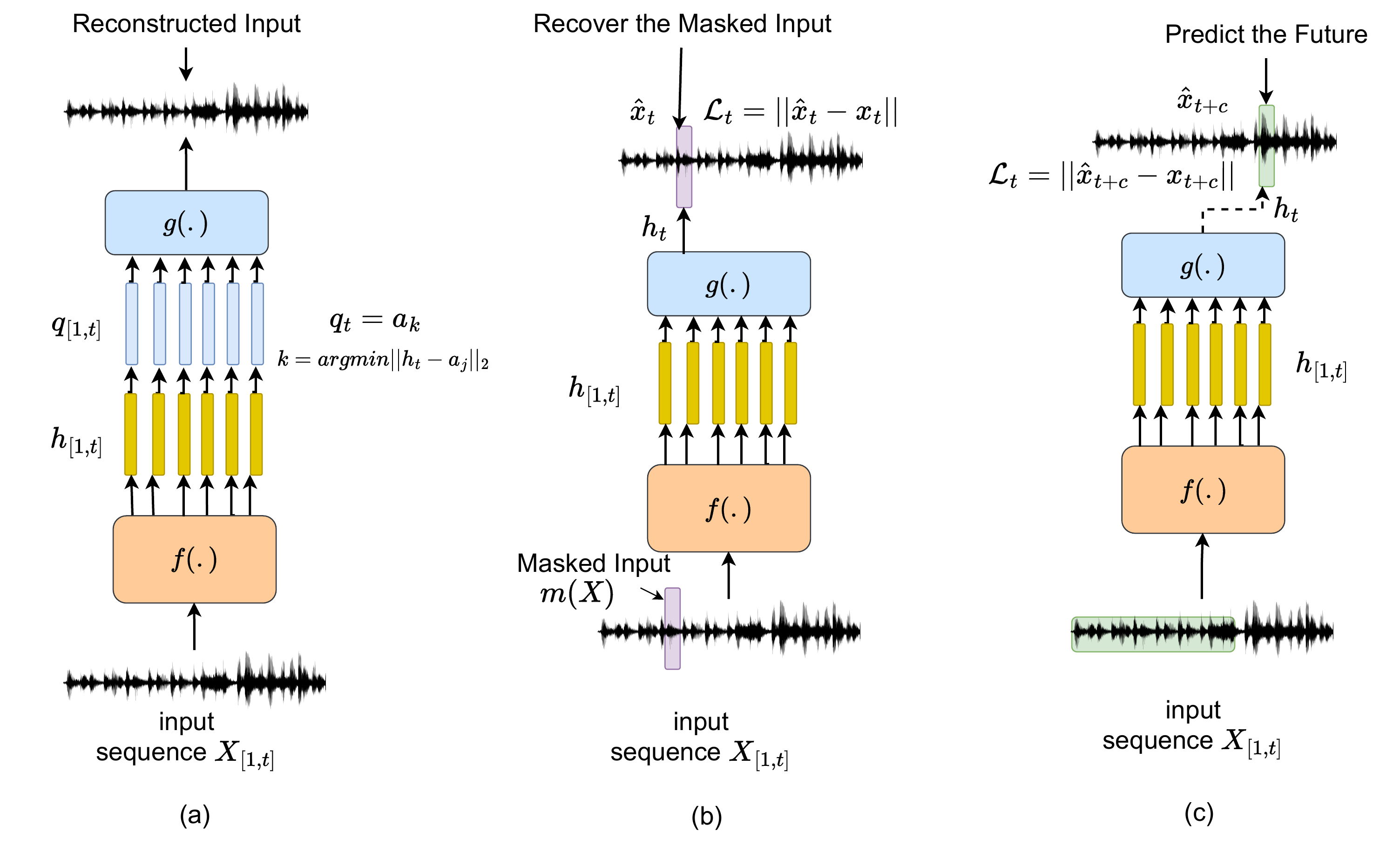}
    \caption{Generative approaches to self-supervised learning.}
    \label{fig:ssrl-gen}
\end{figure}

\subsubsection{Generative Models}
This method involves instructing a model to produce samples resembling the input data without explicitly learning the labels, creating valuable representations applicable to other tasks. The detailed architecture for generative models with three different variants is shown in \Cref{fig:ssrl-gen}. The earliest self-supervised method, predicting masked inputs using surrounding data, originated from the text field in 2013 with word2vec. The continuous bag of words (CBOW) concept of word2vec predicts a central word based on its neighbors, resembling ELMo and BERT's masked language modeling (MLM). These non-autoregressive generative approaches differ in their use of advanced structures, such as bidirectional LSTM (for ELMo) and transformer (for BERT), with recent models producing contextual embeddings. In the context of the speech, Mockingjay \cite{liu2020mockingjay} applied masking to all feature dimensions in the speech domain, whereas TERA \cite{liu2021tera} applied to mask only to a particular subset of feature dimensions. The summary of generative self-supervised approaches along with the data used for training the models are outlined in \Cref{SSRL:gen}. We further discuss different generative approaches as highlighted in \Cref{fig:ssrl-gen} as follows:

\begin{table*}[ht]
\centering
\caption{ Summary of \textit{generative self-supervised} approaches and proposed models for speech processing with associated metrics and training Data. \textbf{ASR}: Automatic Speech Recognition, \textbf{PR}: Phoneme Recognition. \textbf{PC}: Phoneme Classification, \textbf{SR}: Speaker Recognition, \textbf{LS}: LibriSpeech.}
\resizebox{0.95\textwidth}{!}{%
\begin{tabular}{cccccc}
\hline
\multirow{2}{*}{Model}      & \multirow{2}{*}{Reference} & \multirow{2}{*}{Task (Metric)} & \multirow{2}{*}{\begin{tabular}[c]{@{}c@{}}Pre-Training \\ Dataset (hours)\end{tabular}} & \multicolumn{2}{c}{Dataset}                                                                                                                                    \\ \cline{5-6} 
                            &                            &                                &                                                                                          & Training                                                                           & Test                                                                      \\ \hline
\multirow{2}{*}{Mockingjay} & \multirow{2}{*}{\cite{liu2020mockingjay}}    & PC                             & LS (360h)                                                                                & LS (360h)                                                                          & LS (test-clean)                                                           \\ \cline{3-6} 
                            &                            & SR                            & LS (360h)                                                                                & LS (100h)                                                                          & LS (100h)                                                                 \\ \hline
PASE                        & \cite{pascual2019learning}                     & ASR                            & LS (50 hr)                                                                               & DIRHA                                                                              & DIRHA                                                                     \\ \hline
PASE+                       & \cite{ravanelli2020multi}                    & ASR                            & LS (50 hr)                                                                               & \begin{tabular}[c]{@{}c@{}}DIRHA\\ CHiME-5\end{tabular}                            & \begin{tabular}[c]{@{}c@{}}DIRHA\\ CHiME-5\end{tabular}                   \\ \hline
DeCoAR                      & \cite{ling2020decoar}                     & ASR                            & \begin{tabular}[c]{@{}c@{}}LS (100h, 360h, 460 h,  960h)\\  WSJ si284\end{tabular}       & \begin{tabular}[c]{@{}c@{}}LS (100h, 360h, 460 h,  960h)\\  WSJ si284\end{tabular} & \begin{tabular}[c]{@{}c@{}}LS (test-clean)\\ LS (test-other)\end{tabular} \\ \hline
\end{tabular}}
\label{SSRL:gen}
\end{table*}

\begin{itemize}
    \item Auto-encoding Models: Auto-encoding Models have garnered significant attention in the domain of self-supervised learning, particularly Autoencoders (AEs) and Variational Autoencoders (VAEs). AEs consist of an encoder and a decoder that work together to reconstruct input while disregarding less important details, prioritizing the extraction of meaningful features. VAEs, a probabilistic variant of AEs, have found wide-ranging applications in the field of speech modeling. Furthermore, the vector-quantized variational autoencoder (VQ-VAE) \cite{van2017neural} has been developed as an extended generative model. The VQ-VAE introduces parameterization of the posterior distribution to represent discrete latent representations. Remarkably, the VQ-VAE has demonstrated notable success in generative spoken language modeling. By combining a discrete latent space with self-supervised learning, its performance is further improved.
        
    \item Autoregressive models: Autoregressive generative self-supervised learning uses autoregressive prediction coding technique \cite{chung2019unsupervised}  to model the probability distribution of a sequence of data points. This approach aims to predict the next data point in a sequence based on the previous data points. Autoregressive models typically use RNNs or a transformer architecture as a basic model. 
    
    The authors in paper \cite{oord2016wavenet} introduce a generative model for raw audio called WaveNet, based on PixelCNN \cite{van2016conditional}. To enhance the model's ability to handle long-range temporal dependencies, the authors incorporate dilated causal convolutions \cite{oord2016wavenet}. They also utilize Gated Residual blocks and skip connections to improve the model's expressivity.
    
    \item Masked Reconstruction: The concept of masked reconstruction is influenced by the masked language model (MLM) task proposed in BERT \cite{devlin2018bert}. This task involves masking specific tokens in input sentences with learned masking tokens or other input tokens, and training the model to reconstruct these masked tokens from the non-masked ones. Recent research has explored similar pretext tasks for speech representation learning that help models develop contextualized representations capturing information from the entire input, like the DeCoAR model \cite{ling2020decoar}. This approach assists the model in comprehending input data better, leading to more precise and informative representations.
    
\end{itemize}

\subsubsection{Contrastive Models}
\begin{figure}
    \centering
    \includegraphics[width=0.6\columnwidth]{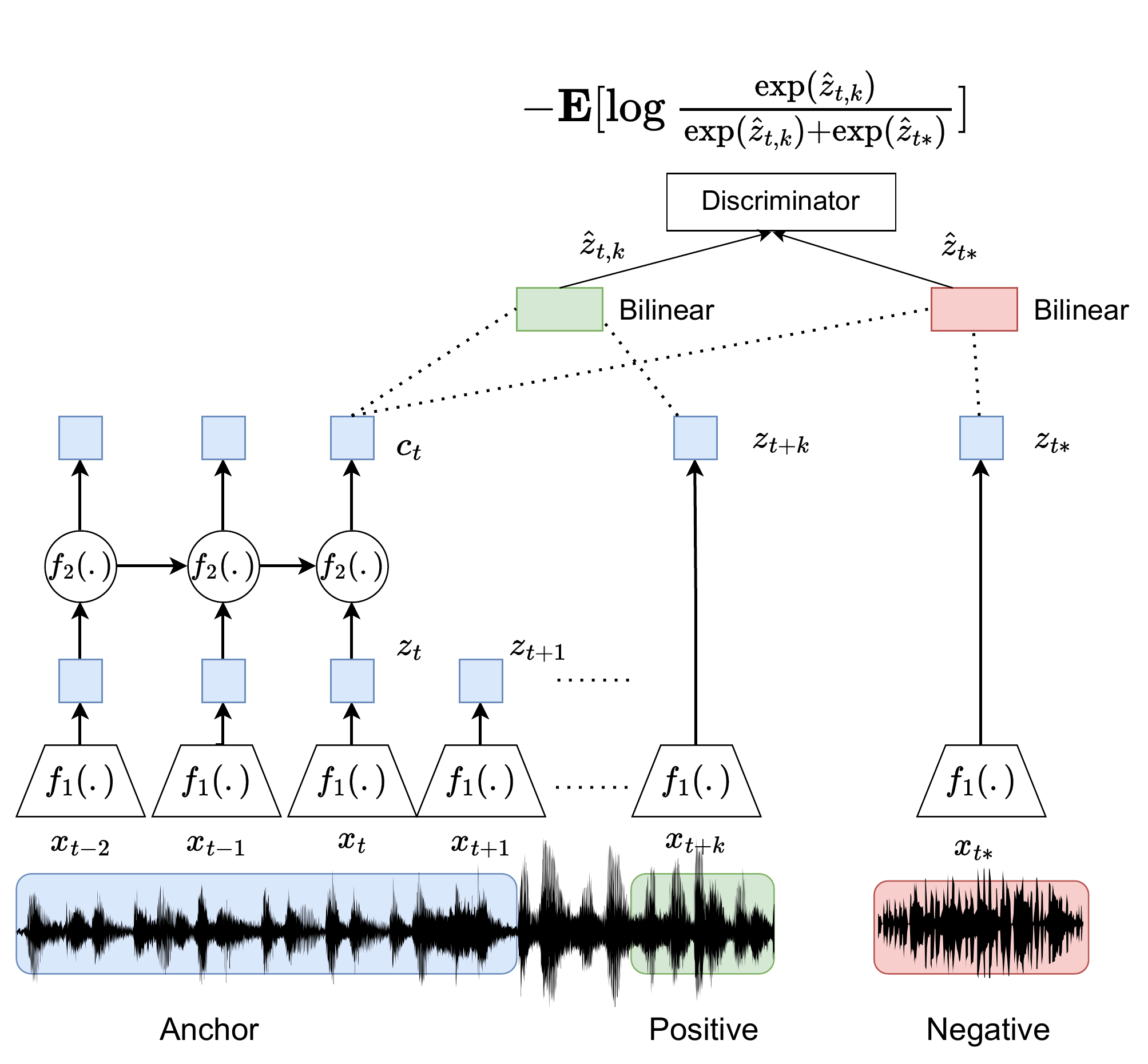}
    \caption{Contrastive Self-supervised learning: Contrastive Predictive Coding.}
    \label{fig:ssrl-con}
\end{figure}

The technique involves training a model to differentiate between similar and dissimilar pairs of data samples, which helps the model acquire valuable representations that can be utilized for various tasks, as shown on \Cref{fig:ssrl-con}. The fundamental principle of contrastive learning is to generate positive and negative pairs of training samples based on the comprehension of the data. The model must learn a function that assigns high similarity scores to two positive samples and low similarity scores to two negative samples. Therefore, generating appropriate samples is crucial for ensuring that the model comprehends the fundamental features and structures of the data. \Cref{SSRL:con} outlines popular contrastive self-supervised models used for different speech-processing tasks. We discuss Wav2Vec 2.0 since it has achieved state-of-the-art results in different downstream tasks.

\begin{itemize}
        \item Wav2Vec 2.0 \cite{baevski2020wav2vec} is a framework for self-supervised learning of speech representations that is one of the current state-of-the-art models for ASR \cite{baevski2020wav2vec}. The training of the model occurs in two stages. Initially, the model operates in a self-supervised mode during the first phase, where it uses unlabelled data and aims to achieve the best speech representation possible. The second phase is fine-tuning a particular dataset for a specific purpose. Wav2Vec 2.0 takes advantage of self-supervised training and uses convolutional layers to extract features from raw audio.
\end{itemize}

\begin{table*}[ht]
\centering
\caption{ Summary of \textit{contrastive self-supervised} approaches and proposed models for speech processing with associated metrics and training Data. \textbf{ASR}: Automatic Speech Recognition, \textbf{PR}: Phoneme Recognition. \textbf{PC}: Phoneme Classification, \textbf{SR}: Speaker Recognition, \textbf{LS}: LibriSpeech, \textbf{LL}: LibriLight, \textbf{WSJ}: Wall Street Journal. }
\resizebox{0.95\textwidth}{!}{%
\begin{tabular}{cccccc}
\hline
\multirow{2}{*}{Model}             & \multirow{2}{*}{Reference} & \multirow{2}{*}{Task} & \multirow{2}{*}{\begin{tabular}[c]{@{}c@{}}Pre-Training \\ Dataset (hours)\end{tabular}}                                    & \multicolumn{2}{c}{Dataset}                                                                                                                                                                                                                                             \\ \cline{5-6} 
                                   &                            &                                &                                                                                                                             & Training                                                                                                                     & Test                                                                                                                                     \\ \hline
\multirow{2}{*}{CPC}               & \multirow{2}{*}{\cite{oord2018representation}}    & PC                             & LS (100h)                                                                                                                   & LS (100h)                                                                                                                    & LS (100h)                                                                                                                                \\ \cline{3-6} 
                                   &                            & SR                             & LS (100h)                                                                                                                   & LS (100h)                                                                                                                    & LS (100h)                                                                                                                                \\ \hline
Modified CPC                       & \cite{riviere2020unsupervised}                     & PC                             & \begin{tabular}[c]{@{}c@{}}LS (100h, 360h)\\ Zerospeech2017(45h)\end{tabular}                                               & CV-Dataset                                                                                                                   & CV-Dataset                                                                                                                               \\ \hline
\multirow{2}{*}{Bidirectional CPC} & \multirow{2}{*}{\cite{kawakami2020learning}}    & ASR                            & \begin{tabular}[c]{@{}c@{}}WSJ (80h)\\ LS (960h)\\  TIMIT (5h)\\  SSA (1h)\\  TED3 (440h)\\  SwithBoard (310h)\end{tabular} & \begin{tabular}[c]{@{}c@{}}WSJ (80h)\\  LS (960h)\\  TIMIT (5h)\\  SSA (1h)\\  TED3 (440h)\\  SwithBoard (310h)\end{tabular} & \begin{tabular}[c]{@{}c@{}}WSJ (test92, test93)\\  LS (test-clean, test-other) \\ TED3 (dev, test)\\  SwithBoard (eval2000)\end{tabular} \\ \cline{3-6} 
                                   &                            & ASR-Multi                      & \begin{tabular}[c]{@{}c@{}}Audio Set (2500h)\\  AVSpeech (3100h)\\  CV-Dataset (430h\end{tabular}                           & \begin{tabular}[c]{@{}c@{}}Audio Set (2500h)\\  AVSpeech (3100h)\\  CV-Dataset (430h)\end{tabular}                           & \begin{tabular}[c]{@{}c@{}}OpenSLR \\ ALFFA\end{tabular}                                                                                 \\ \hline
\multirow{2}{*}{wav2vec}           & \multirow{2}{*}{\cite{schneider2019wav2vec}}    & ASR                            & \begin{tabular}[c]{@{}c@{}}LS 80/860h\\  LS 960h + WSJ (si284)\end{tabular}                                                 & WSJ (si284)                                                                                                                  & WSJ (eval92)                                                                                                                             \\ \cline{3-6} 
                                   &                            & PR                             & TIMIT                                                                                                                       & TIMIT                                                                                                                        & TIMIT                                                                                                                                    \\ \hline
\multirow{2}{*}{wav2vec 2.0}       & \multirow{2}{*}{\cite{baevski2020wav2vec}}    & ASR                            & \begin{tabular}[c]{@{}c@{}}LS (960h)\\ LL (60000h)\end{tabular}                                                             & LS (960h)                                                                                                                    & \begin{tabular}[c]{@{}c@{}}LS (test-clean)\\  LS (test-other)\end{tabular}                                                               \\ \cline{3-6} 
                                   &                            & PR                             & \begin{tabular}[c]{@{}c@{}}LS (960h)\\ LL (60000h)\end{tabular}                                                             & TIMIT                                                                                                                        & TIMIT                                                                                                                                    \\ \hline
\multirow{2}{*}{vq-wav2vec 2.0}    & \multirow{2}{*}{\cite{baevski2019vq}}    & ASR                            & LS (960h)                                                                                                                   & WSJ (si284)                                                                                                                  & WSJ (eval92)                                                                                                                             \\ \cline{3-6} 
                                   &                            & PR                             & LS (960h)                                                                                                                   & TIMIT                                                                                                                        & TIMIT                                                                                                                                    \\ \hline
wav2vec-C                          & \cite{sadhu2021wav2vec}                     & ASR                            & Alexa-10k                                                                                                                   & Alexa-eval                                                                                                                   & Alexa-eval                                                                                                                               \\ \hline
w2v-BERT                           & \cite{chung2021w2v}                     & ASR                            & LL (60000h)                                                                                                                 & LS (960h)                                                                                                                    & \begin{tabular}[c]{@{}c@{}}LS (test)\\  LS (test-other)\\  LS (dev) \\ LS (dev-other)\end{tabular}                                       \\ \hline
\multirow{2}{*}{Speech SimCLR}     & \multirow{2}{*}{\cite{jiang2020speech}}    & ASR                            & \begin{tabular}[c]{@{}c@{}}LS (960h) \\ WJS (si284)\\  TED2\end{tabular}                                                    & WJS (si284)                                                                                                                  & WJS (si284)                                                                                                                              \\ \cline{3-6} 
                                   &                            & PR                             & \begin{tabular}[c]{@{}c@{}}LS (960h) \\ WJS (si284) \\ TED2\end{tabular}                                                    & TIMIT                                                                                                                        & TIMIT                                                                                                                                    \\ \hline
UnSpeech                           & \cite{milde2018unspeech}                     & ASR-Mult                       & \begin{tabular}[c]{@{}c@{}}LL (60000h) \\ GigaSpeech (10000h)\\  VP (24000h)\end{tabular}                                   & SUPERB                                                                                                                       & SUPERB                                                                                                                                   \\ \hline
\end{tabular}}
\label{SSRL:con}
\end{table*}

In the speech field, researchers have explored different approaches to avoid overfitting, including augmentation techniques like Speech SimCLR \cite{jiang2020speech} and the use of positive and negative pairs through methods like Contrastive Predictive Coding (CPC) (\citet{ooster2019improving}), Wav2vec (v1, v2.0) (\citet{schneider2019wav2vec}), VQ-wav2vec (\citet{baevski2019vq}), and Discrete BERT \cite{baevski2019effectiveness}."
"In the graph field, researchers have developed approaches like Deep Graph Infomax (DGI) (Velickovic et al., 2019 \cite{velivckovic2018deep}) to learn representations that maximize the mutual information between local patches and global structures while minimizing mutual information between patches of corrupted graphs and the original graph's global representation.

\subsubsection{Predictive Models}
In training predictive models, the primary concept involves creating simpler objectives or targets to minimize the need for data generation. However, the most critical and difficult aspect is ensuring that the task's difficulty level is appropriate for the model to learn effectively. Predictive SSRL methods have been leveraged in ASR through transformer-based models to acquire meaningful representations \cite{baevski2019effectiveness,hsu2021hubert,liu2021tera} and have proven transformative in exploiting the growing abundance of data \cite{gao2023self}. \Cref{SSRL:pred} highlight popularly used SSRL methods along with the data used for training these models. In the following section we breifly discuss three popular predictive SSRL approaches used widely in various downstream tasks.
\begin{figure}
    \centering
    \includegraphics[width = 0.6\columnwidth]{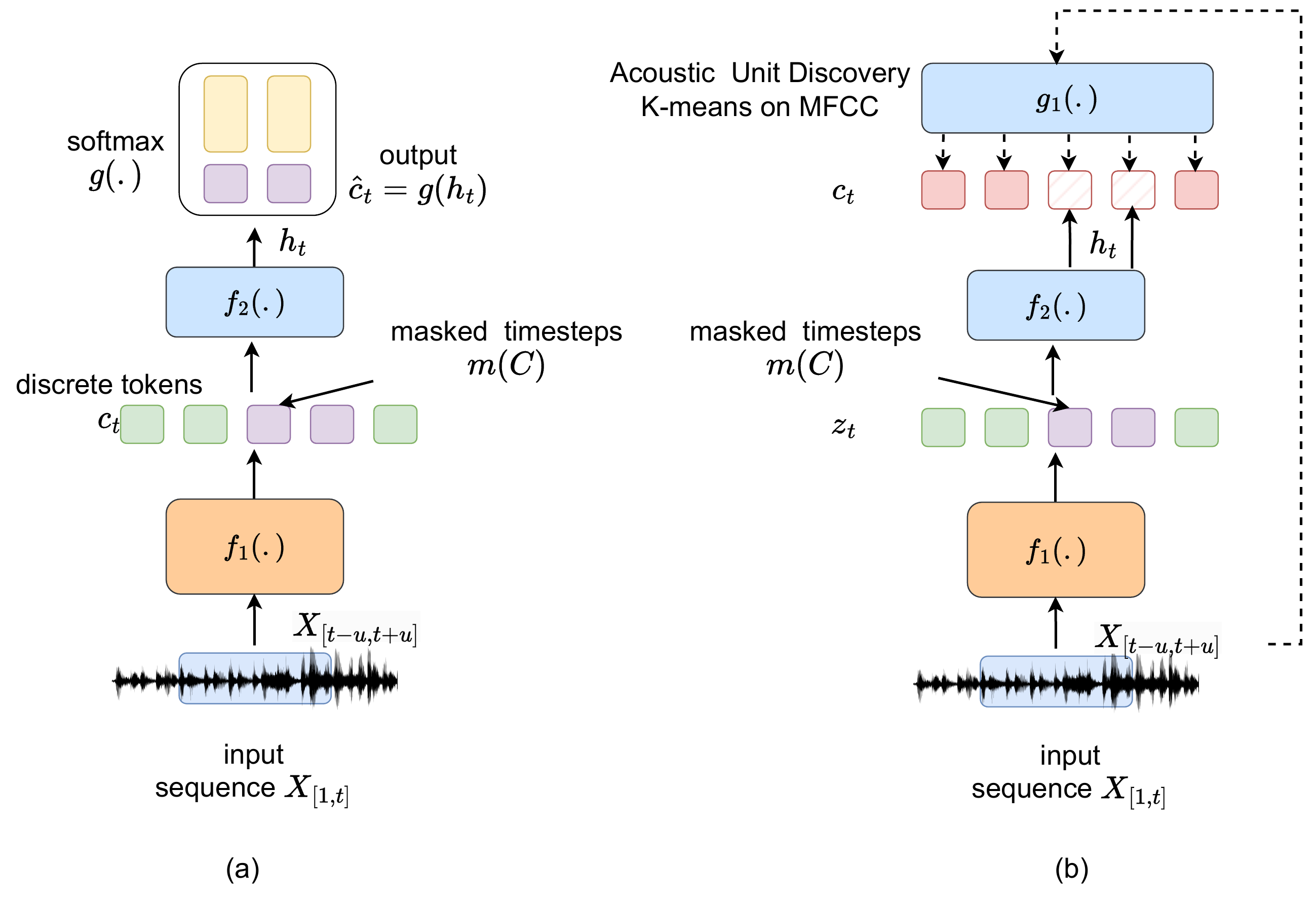}
    \caption{Predictive Self-supervised learning: (a) Discrete BERT (b) HuBERT.}
    \label{fig:ssrl-pred}
\end{figure}

\begin{itemize}
\item The direct application of BERT-type training to speech input presents challenges due to the unsegmented and unstructured nature of speech. To overcome this obstacle, a pioneering model known as Discrete BERT \cite{baevski2019effectiveness} has been developed. This model converts continuous speech input into a sequence of discrete codes, facilitating code representation learning. The discrete units are obtained from a pre-trained vq-wav2vec model \cite{baevski2019vq}, and they serve as both inputs and targets within a standard BERT model. The architecture of Discrete BERT, illustrated in \Cref{fig:ssrl-pred} (a), incorporates a softmax normalized output layer. During training, categorical cross-entropy loss is employed, with a masked perspective of the original speech input utilized for predicting code representations. Remarkably, the Discrete BERT model has exhibited impressive efficacy in self-supervised speech representation learning. Even with a mere 10-minute fine-tuning set, it achieved a Word Error Rate (WER) of 25\% on the standard test-other subset. This approach effectively tackles the challenge of directly applying BERT-type training to continuous speech input and holds substantial potential for significantly enhancing speech recognition accuracy
\begin{table*}[ht]
\centering
\caption{ Summary of \textit{ predictive self-supervised} approaches and proposed models for speech processing with associated metrics and training Data. \textbf{ASR}: Automatic Speech Recognition, \textbf{PR}: Phoneme Recognition. \textbf{PC}: Phoneme Classification, \textbf{SR}: Speaker Recognition, \textbf{LL}: LibriLight, \textbf{LS}: LibriSpeech.}
\resizebox{0.95\textwidth}{!}{%
\begin{tabular}{cccccc}
\hline
\multirow{2}{*}{Model}   & \multirow{2}{*}{Reference} & \multirow{2}{*}{Task (Metric)} & \multirow{2}{*}{\begin{tabular}[c]{@{}c@{}}Pre-Training \\ Dataset (hours)\end{tabular}}    & \multicolumn{2}{c}{Dataset}                                                                                                   \\ \cline{5-6} 
                         &                            &                                &                                                                                             & Training                 & Test                                                                                               \\ \hline
\multirow{2}{*}{BEST-RQ} & \multirow{2}{*}{\cite{chiu2022self}}    & ASR                            & LL (60000h)                                                                                 & LS (960h)                & \begin{tabular}[c]{@{}c@{}}LS (test)\\  LS (test-other)\\  LS (dev) \\ LS (dev-other)\end{tabular} \\ \cline{3-6} 
                         &                            & ASR-Multi                      & \begin{tabular}[c]{@{}c@{}}LL (60000h) \\ GigaSpeech (10000h)\\  VP (24000h)\end{tabular} & SUPERB                   & SUPERB                                                                                             \\ \hline
data2vec                 & \cite{baevski2022data2vec}                     & ASR                            & LS (960h)                                                                                   & LS (10m, 1h, 100h, 960h) & LS (960h)                                                                                          \\ \hline
Discrete BERT            & \cite{baevski2019effectiveness}                     & ASR                            & LS (960h)                                                                                   & LS (100h)                & \begin{tabular}[c]{@{}c@{}}LS (test)\\  LS (test-other)\end{tabular}                               \\ \hline
HuBERT                   & \cite{yoon2022hubert}                     & ASR                            & \begin{tabular}[c]{@{}c@{}}LS (960h)\\ LL (60000h)\end{tabular}                             & LS (960h)                & \begin{tabular}[c]{@{}c@{}}LS (test)\\  LS (test-other)\end{tabular}                               \\ \hline
WavLM                    & \cite{chen2022wavlm}                     & ASR                            & LL (60000h)                                                                                 & SUPERB                   & SUPERB                                                                                             \\ \hline
\end{tabular}}
\label{SSRL:pred}
\end{table*}

\item The HuBERT \cite{hsu2021hubert} and TERA  \cite{liu2021tera} models are two self-supervised approaches for speech representation learning. HuBERT uses an offline clustering step to align target labels with a BERT-like prediction loss, with the prediction loss applied only over the masked regions as outlined in \Cref{fig:ssrl-pred} (b). This encourages the model to learn a combined acoustic and language model over the continuous inputs. On the other hand, TERA is a self-supervised speech pre-training method that reconstructs acoustic frames from their altered counterparts using a stochastic policy to alter along various dimensions, including time, frequency, and tasks. These alterations help extract feature-based speech representations that can be fine-tuned as part of downstream models.

\end{itemize}

Microsoft has introduced UniSpeech-SAT \cite{chen2022unispeech} and WavLM \cite{chen2022wavlm} models, which follow the HuBERT framework. These models have been designed to enhance speaker representation and improve various downstream tasks. The key focus of these models is data augmentation during the pre-training stage, resulting in superior performance. WavLM model has exhibited outstanding effectiveness in diverse downstream tasks, such as automatic speech recognition, phoneme recognition, speaker identification, and emotion recognition. It is worth highlighting that this model currently holds the top position on the SUPERB leaderboard \cite{DBLP:journals/corr/abs-2105-01051}, which evaluates speech representations' performance in terms of reusability.

Self-supervised learning has emerged as a widely adopted and effective technique for speech processing tasks due to its ability to train models with large amounts of unlabeled data. A comprehensive overview of self-supervised approaches, evaluation metrics, and training data is provided in Table \ref{SSRL:pred} for speech recognition, speaker recognition, and speech enhancement. Researchers and practitioners can use this resource to select appropriate self-supervised methods and datasets to enhance their speech-processing systems. As self-supervised learning techniques continue to advance and refine, we can expect significant progress and advancements in speech processing.

\section{Speech Processing Tasks} \label{speech_processing_tasks}
In recent times, the field of speech processing has gained significant attention due to its rapid evolution and its crucial role in modern technological applications. This field involves the use of diverse techniques and algorithms to analyse and understand spoken language, ranging from basic speech recognition to more complex tasks such as spoken language understanding and speaker identification. Since speech is one of the most natural forms of communication, speech processing has become a critical component of many applications such as virtual assistants, call centres, and speech-to-text transcription. In this section, we provide a comprehensive overview of the various speech-processing tasks and the techniques used to achieve them, while also discussing the current challenges and limitations faced in this field and its potential for future development. 

The assessment of speech-processing models depends greatly on the calibre of datasets employed. By utilizing standardized datasets, researchers are enabled to objectively gauge the efficacy of varying approaches and identify scopes for advancement. The selection of evaluation metrics plays a critical role in this process, hinging on the task at hand and the desired outcome. Therefore, it is essential that researchers conduct a meticulous appraisal of different metrics to make informed decisions. This paper offers a thorough summary of frequently utilized datasets and metrics across diverse downstream tasks, as presented in \Cref{dataset} and, \Cref{objective}.

\renewcommand{\arraystretch}{1.5}
\begin{table*}[ht]
\centering
\caption{Comparative analysis of speech processing datasets: This table summarizes the essential features of different speech-processing datasets, including their typical applications in various speech-processing tasks. \textbf{ASR}: Automatic Speech Recognition, \textbf{PR}: Phoneme Recognition. \textbf{PC}: Phoneme Classification, \textbf{SR}: Speaker Recognition, \textbf{SV}: Speaker Verification, \textbf{SER}: Speech Emotion Recognition, \textbf{IC}: Intent Classification, \textbf{TTS}: Text-to-Speech, \textbf{VC}: Voice Conversion, \textbf{ST}: Speech Translation, \textbf{SS}: Speech Separation}
\resizebox{0.9\textwidth}{!}{%
\begin{tabular}{llllllllllllll}
\hline
Dataset                                               & Language       & Lenght (hours) & ASR                       & PR                        & PC                        & SR                        & SV                        & SER                                           & IC                        & TTS                       & VC                        & ST            & SS                                \\ \hline
TIMIT Acoustic-Phonetic Continuous Speech Corpus      & English        & 5.4            & \checkmark & \checkmark & \checkmark &                           &                           &                                               &                           &                           &                           &                                             &  \\ \hline
Lip Reading Sentences 2 (LRS2)                        & English        &                & \checkmark &                           &                           &                           &                           &                                               &                           &                           &                           &                                             &  \\ \hline
LibriSpeech (LS)                                      & English        & 1000           & \checkmark & \checkmark & \checkmark & \checkmark &                           &                                               &                           &                           &                           &                                              & \\ \hline
GigaSpeech                                            & English        & 10000          & \checkmark &                           &                           &                           &                           &                                               &                           &                           &                           &                                              & \\ \hline
Fleurs                                                & Multilingual   & 12             & \checkmark &                           &                           &                           &                           &                                               &                           &                           &                           &                                             & \\ \hline
LibriTTS                                              & English        & 585            &                           &                           &                           &                           &                           &                                               &                           & \checkmark & \checkmark &                                              & \\ \hline
L2ARCTIC                                              & English        & 11.2           &                           &                           &                           &                           &                           &                                               &                           & \checkmark & \checkmark &                                              & \\ \hline
CMUARCTIC                                             & English        & 20             &                           &                           &                           &                           &                           &                                               &                           & \checkmark & \checkmark &                                             &  \\ \hline
Wall Street Journal (WSJ)                             & English        &                & \checkmark & \checkmark & \checkmark &                           &                           &                                               &                           &                           &                           &                                             &  \\ \hline
VoxPopuli (VP)                                        & Multilingual   & 1800           & \checkmark &                           &                           &                           &                           &                                               &                           &                           &                           &                                              & \\ \hline
BABEL  (BBL)                                          & Multilingual   &                & \checkmark &                           &                           &                           &                           &                                               &                           &                           &                           &                                              & \\ \hline
Common Voice (CV-dataset)                             & Multilingual   & 9283           & \checkmark & \checkmark & \checkmark &                           &                           &                                               &                           &                           &                           &                                              & \\ \hline
CSTR VCTK                                             & English        &                &                           &                           &                           &                           &                           &                                               &                           &                           &                           &                                             &  \\ \hline
HUB 5                                                 & English        & 2000           & \checkmark &                           &                           &                           &                           &                                               &                           &                           &                           &                                              & \\ \hline
CHiME-5                                               & English        & 50.12          &                           &                           &                           &                           &                           &                                               &                           &                           &                           &                                              & \\ \hline
TED-LIUM 3 (TED 3)                                    & English        & 452            &                           &                           &                           &                           &                           &                                               &                           &                           &                           &                                              & \\ \hline
TED-LIUM 2 (TED 2)                                    & English        & 118            &                           &                           &                           &                           &                           &                                               &                           &                           &                           &                                              & \\ \hline
AISHELL-1                                             & Mandarin       & 520            & \checkmark &                           &                           &                           &                           &                                               &                           &                           &                           &                                              & \\ \hline
AISHELL-3                                             & Mandarin       & 85             &                           &                           &                           &                           &                           &                                               &                           & \checkmark & \checkmark &                                              & \\ \hline
AISHELL-4                                             & Mandarin       & 120            &                           &                           &                           &                           &                           &                                               &                           & \checkmark & \checkmark &                                              & \\ \hline
Arabic Speech Corpus                                  & Arabic         & 3.7            & \checkmark & \checkmark & \checkmark &                           &                           &                                               &                           &                           &                           &                                             &  \\ \hline
Persian Consonant Vowel Combination                   & Persian        & -              & \checkmark & \checkmark & \checkmark &                           &                           &                                               &                           &                           &                           &                                             &  \\ \hline
ALFFA                                                 & Multilingual   & 5.2-18.3       &                           &                           &                           &                           &                           &                                               &                           &                           &                           &                                              & \\ \hline
OpenSLR-multi                                         & Multilingual   & 4.4-265.9      &                           &                           &                           &                           &                           &                                               &                           &                           &                           &                                             &  \\ \hline
VCTK                                                  & English        & 44             &                           &                           &                           &                           &                           &                                               &                           &                           &                           &                                              & \\ \hline
VoxCeleb1/2                                           & English        &                &                           &                           &                           & \checkmark & \checkmark &                                               &                           &                           &                           &                                             &  \\ \hline
Fluent Speech Commands (FSC)                          & English        & 14.7           &                           &                           &                           &                           &                           &                                               & \checkmark &                           &                           &                                              & \\ \hline
Emotional Speech Dataset (ESD)                        & English        & 29             &                           &                           &                           &                           &                           & \checkmark                     &                           &                           &                           &                                              & \\ \hline
Interactive Emotional Dyadic Motion Capture (IEMOCAP) & English        & 12             &                           &                           & \multicolumn{1}{c}{}      & \multicolumn{1}{c}{}      & \multicolumn{1}{c}{}      & \multicolumn{1}{c}{\checkmark} & \multicolumn{1}{c}{}      & \multicolumn{1}{c}{}      & \multicolumn{1}{c}{}      & \multicolumn{1}{c}{}                         & \\ \hline
Multimodal EmotionLines Dataset ( MELD)               & English        &   -             &                           &                           & \multicolumn{1}{c}{}      & \multicolumn{1}{c}{}      & \multicolumn{1}{c}{}      & \multicolumn{1}{c}{\checkmark} & \multicolumn{1}{c}{}      & \multicolumn{1}{c}{}      & \multicolumn{1}{c}{}      & \multicolumn{1}{c}{}                        &  \\ \hline
LibraSeepch En-Fr                                     & English/French &    -            &                           &                           &                           &                           &                           &                                               &                           &                           &                           & \multicolumn{1}{c}{\checkmark} &\\ \hline
CoVoST-2                                              & Multilingual   & 2880           &                           &                           &                           &                           &                           &                                               &                           &                           &                           & \multicolumn{1}{c}{\checkmark} &\\ \hline
LibriLight (LL)                                              & English   & 60000           &       \multicolumn{1}{c}{\checkmark}                     &       \multicolumn{1}{c}{\checkmark}                     &                           &                           &                           &                                               &                           &                           &                           & &\\ \hline
\end{tabular}
}
\label{dataset}
\end{table*}

\begin{table*}[ht]
  \centering
  \caption{Comprehensive Evaluation Metrics for Speech Processing Tasks. This table provides a comprehensive overview of the evaluation metrics used to assess the performance of speech-based systems across various tasks such as ASR, speaker verification, and TTS. The table highlights the specific metrics employed for each task, along with the score range and commonly used datasets.}
  \label{tab:metrics}
  \resizebox{0.9\textwidth}{!}{%
  \begin{tabular}{|l|c|c|c|c|}
    \hline
    Tasks & Metric & Description & Score range & Evaluation dataset \\ \hline
    Automatic speech recognition & WER & Word Error Rate & 0-1 & TIMIT \\
     & CER & Character Error Rate & 0-1 & LibriSpeech \\ \hline
    Phoneme recognition & Accuracy & Classification accuracy & 0-1 & TIMIT \\ \hline
    Phoneme classification & F1-score & Harmonic mean of precision and recall & 0-1 & TIMIT \\ \hline
    Speaker recognition & EER & Equal Error Rate & 0-1 & VoxCeleb1 \\ \hline
    Speaker verification & FAR/FRR & False Acceptance Rate / False Rejection Rate & 0-1 & VoxCeleb1 \\ \hline
    Speech emotion recognition & Accuracy & Classification accuracy & 0-1 & IEMOCAP, ESD \\ \hline
    Intent classification & F1-score & Harmonic mean of precision and recall & 0-1 & ATIS, SNIPS \\ \hline
    Text-to-speech & MOS & Mean Opinion Score & 1-5 & LJSpeech, LibriTTS \\ \hline
    Voice conversion & MOS & Mean Opinion Score & 1-5 & VCC 2016 \\ \hline
    Speech translation & BLEU & Bilingual Evaluation Understudy & 0-1 & MuST-C \\ \hline
    Speech separation & SI-SDRi & Signal to Distortion Ratio & -20-30 & WSJ0-2mix \\ \hline
    Speech enhancement & PESQ & Perceptual Evaluation of Speech Quality & -0.5-4.5 & NOIZEUS \\ \hline
    Voice activity detection & F1-score & Harmonic mean of precision and recall & 0-1 & QUT-NOISE \\ \hline
  \end{tabular}%
  }
  \label{objective}
\end{table*}

\subsection{Automatic speech recognition (ASR) \& conversational multi-speaker AST}

\subsubsection{Task Description}
Automatic speech recognition (ASR) technology enables machines to convert spoken language into text or commands, serving as a cornerstone of human-machine communication and facilitating a wide range of applications such as speech-to-speech translation and information retrieval \cite{lu2020automatic}. ASR involves multiple intricate steps, starting with the extraction and analysis of acoustic features, including spectral and prosodic features, which are then employed to recognize spoken words. Next, an acoustic model matches the extracted features to phonetic units, while a language model predicts the most probable sequence of words based on the recognized phonetic units. Ultimately, the acoustic and language model outcomes are merged to produce the transcription of spoken words. Deep learning techniques have gained popularity in recent years, allowing for improved accuracy in ASR systems \cite{baevski2020wav2vec,radford2022robust}. This paper provides an overview of the key components involved in ASR and highlights the role of deep learning techniques in enhancing the technology's accuracy.

Most speech recognition systems that use deep learning aim to simplify the processing pipeline by training a single model to directly map speech signals to their corresponding text transcriptions. Unlike traditional ASR systems that require multiple components to extract and model features, such as HMMs and GMMs, end-to-end models do not rely on hand-designed components \cite{audhkhasi2019forget,li2022recent}. Instead, end-to-end ASR systems use DNNs to learn acoustic and linguistic representations directly from the input speech signals \cite{li2022recent}. One popular type of end-to-end model is the encoder-decoder model with attention. This model uses an encoder network to map input audio signals to hidden representations, and a decoder network to generate text transcriptions from the hidden representations. During the decoding process, the attention mechanism enables the decoder to selectively focus on different parts of the input signal \cite{li2022recent}.

End-to-end ASR models can be trained using various techniques such as CTC \cite{Karita2019ImprovingTE}, which is used to train models without explicit alignment between the input and output sequences, and RNNs, which are commonly used to model temporal dependencies in sequential data such as speech signals. Transfer learning-based approaches can also improve end-to-end ASR performance by leveraging pre-trained models or features \cite{liu2023towards,deng2022improving,sertolli2021representation}. While end-to-end ASR models have shown promising results in various applications, there is still room for improvement to achieve human-level performance \cite{liu2023towards,yoon2022hubert,kanda2022streaming,kanda2022transcribe,deng2022improving,fazel2021synthasr}. Nonetheless, deep learning-based end-to-end ASR architecture offers a promising and efficient approach to speech recognition that can simplify the processing pipeline and improve recognition accuracy.
\subsubsection{Dataset}
The development and evaluation of ASR systems are heavily dependent on the availability of large datasets. As a result, ASR is an active area of research, with numerous datasets used for this purpose. In this context, several popular datasets have gained prominence for use in ASR systems.
\begin{itemize}
    \item Common Voice: Mozilla's Common Voice project \cite{ardila2019common} is dedicated to producing an accessible, unrestricted collection of human speech for the purpose of training speech recognition systems. This ever-expanding dataset features contributions from more than $9,000$ speakers spanning $60$ different languages.

    \item LibriSpeech: LibriSpeech \cite{panayotov2015librispeech} is a corpus of approximately 1,000 hours of read English speech created from audiobooks in the public domain. It is widely used for speech recognition research and is notable for its high audio quality and clean transcription.
    
    \item VoxCeleb: VoxCeleb \cite{Nagrani17} is a large-scale dataset containing over 1 million short audio clips of celebrities speaking, which can be used for speech recognition and recognition research. It includes a diverse range of speakers from different backgrounds and professions.
    
    \item TIMIT: The TIMIT corpus \cite{garofolo1993timit} is a widely used speech dataset consisting of recordings consisting of 630 speakers representing eight major dialects of American English, each reading ten phonetically rich sentences. It has been used as a benchmark for speech recognition research since its creation in 1986.
    
    \item CHiME-5: The CHiME-5 dataset \cite{barker2018fifth} is a collection of recordings made in a domestic environment to simulate a real-world speech recognition scenario. It includes 6.5 hours of audio from multiple microphone arrays and is designed to test the performance of ASR systems in noisy and reverberant environments.
\end{itemize}

Other notable datasets include Google's Speech Commands Dataset \cite{warden2018speech}, the Wall Street Journal dataset\footnote{https://www.ldc.upenn.edu/}, and TED-LIUM \cite{rousseau2012ted}. 

\subsubsection{Models}
The use of RNN-based architecture in speech recognition has many advantages over traditional acoustic models. One of the most significant benefits is their ability to capture long-term temporal dependencies \cite{karita2019comparative} in speech data, enabling them to model the dynamic nature of speech signals. Additionally, RNNs can effectively process variable-length audio sequences, which is essential in speech recognition tasks where the duration of spoken words and phrases can vary widely. RNN-based models can efficiently identify and segment phonemes, detect and transcribe spoken words, and can be trained end-to-end, eliminating the need for intermediate steps. These features make RNN-based models particularly useful in real-time applications, such as speech recognition in mobile devices or smart homes \cite{dong2020rtmobile,he2019streaming}, where low latency and high accuracy are crucial.

In the past, RNNs were the go-to model for ASR. However, their limited ability to handle long-range dependencies prompted the adoption of the Transformer architecture. For example, in 2019, Google's Speech-to-Text API transitioned to a Transformer-based architecture that surpassed the previous RNN-based model, especially in noisy environments and for longer sentences, as reported in \cite{zhang2020transformer}. Additionally, Facebook AI Research introduced wav2vec 2.0, a self-supervised learning approach that leverages a Transformer-based architecture to perform unsupervised speech recognition. wav2vec 2.0 has significantly outperformed the previous RNN-based model and achieved state-of-the-art results on several benchmark datasets.

\begin{table*}[ht]
\caption{Table summarizing the performance of different ASR models in terms of WER\% on five different datasets (LibriSpeech test, LibriSpeech clean, TIMIT, Common Voice, WSJ eval92, and GigaSpeech) also highlighting the use of extra data during training. ZS stands for Zero-Shot Performance.}
\resizebox{\textwidth}{!}{%
\begin{tabular}{llcccllcc}
\hline
Model & Architecture & \begin{tabular}[c]{@{}c@{}}Extra\\
Training Data\end{tabular} & WER\% $\downarrow$ & WER\% $\downarrow$ & Model & Architecture & \begin{tabular}[c]{@{}c@{}}Extra\\
Training Data\end{tabular} & WER\% $\downarrow$ \\
\hline
\multicolumn{3}{c}{LibriSpeech test} & clean & \multicolumn{1}{c|}{others} & \multicolumn{4}{c}{TIMIT} \\
\hline Conformer + Wav2vec 2.0 \cite{zhang2020pushing} & Conformer + wav2vec2.0 & Y & 1.4 & \multicolumn{1}{c|}{2.6} & wav2vec 2.0 \cite{baevski2020wav2vec} & Transformer + CNN & Y & 8.3 \\
w2v-BERT XXL \cite{chung2021w2v} & CNN+Transformer & Y & 1.4 & \multicolumn{1}{c|}{2.5} & vq-wav2vec \cite{baevski2019vq} & Transformer + CNN & Y & 11.6 \\
SpeechStew (1B)\cite{chan2021speechstew} & Conformer & Y & 1.7 & \multicolumn{1}{c|}{3.3} & LSTM + Monophone Reg \cite{ravanelli2019pytorch} & LSTM & N & 14.5 \\
\cline{6-9} SpeechStew (100M) \cite{chan2021speechstew} & Conformer & N & 2.0 & \multicolumn{1}{c|}{4.0} & \multicolumn{4}{c}{Common Voice} \\
\cline{6-9} ContextNet + SpecAugment \cite{park2020improved} & LSTM+CNN & Y & 1.7 & \multicolumn{1}{c|}{3.4} & SpeechStew (1B) \cite{chan2021speechstew} & Conformer & N & 10.8 \\
Conformer (L) \cite{gulati2020conformer} & Conformer & N & 1.9 & \multicolumn{1}{c|}{4.1} & Whisper \cite{radford2022robust} & & N & 9.5 \\
\cline{6-9} ContextNet \cite{han2020contextnet} & Conformer + wav2vec2.0 & N & 1.9 & \multicolumn{1}{c|}{3.4} & \multicolumn{4}{c}{WSJ eval92} \\
\cline{6-9} Squeezeformer \cite{kim2022squeezeformer} & Conformer & N & 2.47 & \multicolumn{1}{c|}{5.97} & SpeechStew (100M) \cite{chan2021speechstew} & Conformer & N & 1.3 \\
LSTM Transducer \cite{zeyer2021librispeech} & LSTM & N & 2.23 & \multicolumn{1}{c|}{5.6} & tdnn+chain \cite{povey2016purely} & TDNN & N & 2.32 \\
\cline{6-9} Transformer Transducer \cite{liu2021improving} & Transformer & N & 2.0 & \multicolumn{1}{c|}{4.2} & \multicolumn{4}{c}{GigaSpeech} \\
\cline{6-9} Whisper \cite{radford2022robust} & & N & 2.7 (ZS) & \multicolumn{1}{c|}{5.6 (ZS)} & Conformer/Transformer-AED \cite{chen2021gigaspeech} & Conformer & N & 10.80 \\
\hline
\end{tabular}
}
\label{performance:ASR}
\end{table*}

Transformer for the ASR task is first proposed in \cite{8462506}, where authors include CNN layers before submitting preprocessed speech features to the input. By incorporating more CNN layers, it becomes feasible to diminish the gap between the sizes of the input and output sequences, given that the number of frames in audio exceeds the number of tokens in text. This results in a favorable impact on the training process. The change in the original architecture is minimal, and the model achieves a competitive word error rate (WER) of $10.9\%$ on the Wall Street Journal (WSK) speech recognition dataset (Table \ref{performance:ASR}). Despite its numerous advantages, Transformers in its pristine state has several issues when applied to ASR. RNN, with its overall training speed (i.e., convergence) and better WER because of effective joint training and decoding methods, is still the best option. 

The authors in \cite{8462506} propose the Speech Transformer, which has the advantage of faster iteration time, but slower convergence compared to RNN-based ASR. However, integrating the Speech Transformer with the naive language model (LM) is challenging. To address this issue, various improvements in the Speech Transformer architecture have been proposed in recent years. For example, \cite{Karita2019ImprovingTE} suggests incorporating the Connectionist Temporal Classification (CTC) loss into the Speech Transformer. CTC is a popular technique used in speech recognition to align input and output sequences of varying lengths and one-to-many or many-to-one mappings. It introduces a blank symbol representing gaps between output symbols and computes the loss function by summing probabilities across all possible paths. The loss function encourages the model to assign high probabilities to correct output symbols and low probabilities to incorrect output symbols and the blank symbol, allowing the model to predict sequences of varying lengths. The CTC loss is commonly used with RNNs such as LSTM and GRU, which are well-suited for sequential data. CTC loss is a powerful tool for training neural networks to perform sequence-to-sequence tasks where the input and output sequences have varying lengths and mappings between them are not one-to-one.

Various other improvements have also been proposed to enhance the performance of Speech Transformer architecture and integrate it with the naive language model, as the use of the transformer directly for ASR has not been effective in exploiting the correlation among the speech frames. The sequence order of speech, which the recurrent processing of input features can represent, is an important distinction. The degradation in performance for long sentences is reported using absolute positional embedding (AED) \cite{chorowski2015attention}. The problems associated with long sequences can become more acute for transformer \cite{zhou2019improving}. To address this issue, a transition was made from absolute positional encoding to relative positional embeddings  \cite{zhou2019improving}. Whereas authors in  \cite{tsunoo2019transformer} replace positional embeddings with pooling layers. In a considerably different approach, the authors in \cite{mohamed2019transformers} propose a novel way of combining positional embedding with speech features by replacing positional encoding with trainable convolution layers. This update further improves the stability of
optimization for large-scale learning of transformer networks. The above works confirmed the superiority of their techniques against sinusoidal positional encoding.

\begin{table}[t]
\centering
\caption{Comparison of performance between wav2vec2.0 Large and Whisper on different datasets. The zero-shot Whisper model consistently outperforms wav2vec2.0 Large on several datasets, indicating significant performance differences.}
\resizebox{0.6\columnwidth}{!}{%
\begin{tabular}{l|cc}
\hline
Dataset           & wav2vec2.0 Large & Whisper Large \\ \hline
Common Voice      & 29.9             & \textbf{9.0}  \\
Fleurs En         & 14.6             & \textbf{4.4}  \\
Tedlium           & 10.5             & \textbf{4.0}  \\
CHiME6            & 65.8             & \textbf{25.5} \\
VoxPopuli En      & 17.9             & \textbf{7.3}  \\
Switchboard       & 28.3             & \textbf{13.8} \\
CallHome          & 34.8             & \textbf{17.6} \\
LibriSpeech Clean & 2.7              & \textbf{2.7}  \\
LibriSpeech Other & 6.2              & \textbf{5.2}  \\ \hline
\end{tabular}}
\label{w2vandwhisper}
\end{table}

In 2016, Baidu introduced a hybrid ASR model called Deep Speech 2 \cite{amodei2016deep}  that uses both RNNs and Transformers. The model also uses CNNs to extract features from the audio signal, followed by a stack of RNNs to model the temporal dependencies and a Transformer-based decoder to generate the output sequence. This approach achieved state-of-the-art results on several benchmark datasets such as LibriSpeech, VoxForge, WSJeval92 etc. The transition of ASR models from RNNs to Transformers has significantly improved performance, especially for long sentences and noisy environments. 

The Transformer architecture has been widely adopted by different companies and research groups for their ASR models, and it is expected that more organizations will follow this trend in the upcoming years. One of the advanced speech models that leverage this architecture is the Universal Speech Model (USM) \cite{zhang2023google} developed by Google, which has been trained on over 12 million hours of speech and 28 billion sentences of text in more than 300 languages. With its 2 billion parameters, USM can recognize speech in both common languages like English and Mandarin and less-common languages. Other popular acoustic models for speech recognition include Quartznet \cite{kriman2020quartznet}, Citrinet \cite{majumdar2021citrinet}, and Conformer \cite{gulati2020conformer}. These models can be chosen and switched based on the specific use case and performance requirements of the speech recognition pipeline. For example, Conformer-based acoustic models are preferred for addressing robust ASR, as shown in a recent study. Another study found that Conformer-1\footnote{https://www.assemblyai.com/blog/conformer-1/} is more effective in handling real-world data and can produce up to $43\%$ fewer errors on noisy data than other popular ASR models. Additionally, fine-tuning pre-trained models such as BERT  \cite{devlin2018bert} and GPT \cite{radford2018improving} has been explored for ASR tasks, leading to state-of-the-art performance on benchmark datasets like LibriSpeech (refer to Table \ref{performance:ASR}). An open-source toolkit called Vosk\footnote{https://alphacephei.com/vosk/lm} provides pre-trained models for multiple languages optimized for real-time and efficient performance, making it suitable for applications that require such performance.

The field of speech recognition has made significant progress by adopting unsupervised pre-training techniques, such as those utilized by Wav2Vec 2.0 \cite{baevski2020wav2vec}. Another recent advancement in automatic speech recognition (ASR) is the whisper model, which has achieved human-level accuracy when transcribing the LibriSpeech dataset. These two cutting-edge frameworks, Wav2Vec 2.0 and whisper, currently represent the state-of-the-art in ASR. The whisper model is trained on an extensive supervised dataset, including over 680,000 hours of audio data collected from the web, which has made it more resilient to various accents, background noise, and technical jargon. The whisper model is also capable of transcribing and translating audio in multiple languages, making it a versatile tool. OpenAI has released inference models and code, laying the groundwork for the development of practical applications based on the whisper model.

In contrast to its predecessor, Wav2Vec 2.0 is a self-supervised learning framework that trains models on unlabeled audio data before fine-tuning them on specific datasets. It uses a contrastive predictive coding (CPC) loss function to learn speech representations directly from raw audio data, requiring less labeled data. The model's performance has been impressive, achieving state-of-the-art results on several ASR benchmarks. These advances in unsupervised pre-training techniques and the development of novel ASR frameworks like Whisper and Wav2Vec 2.0 have greatly improved the field of speech recognition, paving the way for new real-world applications. In summary, the \Cref{w2vandwhisper} highlights the varying effectiveness of wav2vec2.0 large and whisper models across different datasets.

\subsection{Neural Speech Synthesis}
\subsubsection{Task Description}
\begin{figure*}
    \centering
    \includegraphics[width=\textwidth]{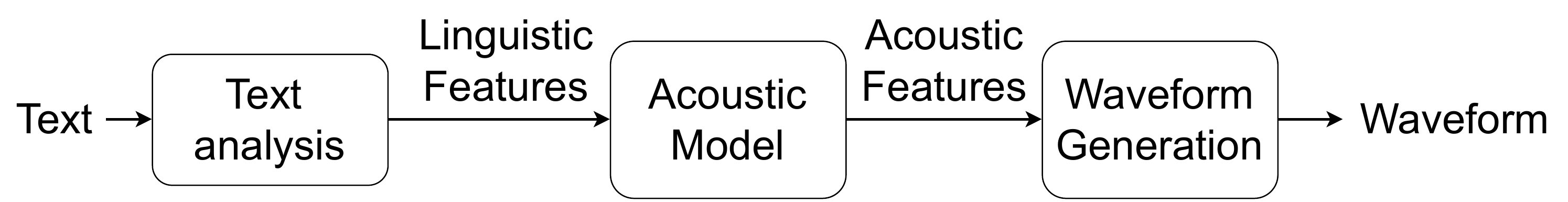}
    \caption{Neural Text-to-speech (TTS) pipeline: a diagram showing the main modules of a typical TTS system. The system takes text input and processes it through various stages to generate speech output. The text analysis module tokenizes the input text and generates linguistic features such as phonemes and prosody. The acoustic model module then converts these linguistic features into acoustic features, such as mel spectrograms, using a neural network. Finally, the waveform generation module synthesizes the speech waveform from the acoustic features using another neural network.}
    \label{fig:TTS}
\end{figure*}

Neural speech synthesis is a technology that utilizes artificial intelligence and deep learning techniques to create speech from text or other inputs. Its applications are widespread, including in healthcare, where it can be used to develop assistive technologies for those who are unable to communicate due to neurological impairments. To generate speech, deep neural networks like CNNs, RNNs, transformers, and diffusion models are trained using phonemes and the mel spectrum. The process involves several components, such as text analysis, acoustic models, and vocoders, as shown in \Cref{fig:TTS}. Acoustic models convert linguistic features into acoustic features, which are then used by the vocoder to synthesize the final speech signal. Various architectures, including neural vocoders based on GANs like HiFi-GAN \cite{kong2020hifi}, are used by the vocoder to generate speech. Neural speech synthesis also enables manipulation of voice, pitch, and speed of speech signals using frameworks such as Fastspeech2  \cite{ren2020fastspeech} and NANSY/NANSY++ \cite{choi2021neural,choi2022nansy++}. These frameworks use information bottleneck to disentangle analysis features for controllable synthesis. The research in neural speech synthesis can be classified into two prominent approaches: autoregressive and non-autoregressive models. Autoregressive models generate speech one element at a time, sequentially, while non-autoregressive models generate all the elements simultaneously, in parallel. \Cref{TTS:Landscape} outlines the different architecture proposed under each category.

The evaluation of synthesized speech is of paramount importance for assessing its quality and fidelity. It serves as a means to gauge the effectiveness of different speech synthesis techniques, algorithms, and parameterization methods. In this regard, the application of statistical tests has emerged as a valuable approach to objectively measure the similarity between synthesized speech and natural speech \cite{franco2019application}. These tests complement the traditional Mean Opinion Score (MOS) evaluations and provide quantitative insights into the performance of speech synthesis systems. Additionally, widely used objective metrics such as Mel Cepstral Distortion (MCD) and Word Error Rate (WER) contribute to the comprehensive evaluation of synthesized speech, enabling researchers and practitioners to identify areas for improvement and refine the synthesis process. By employing these objective metrics and statistical tests, the evaluation of synthesized speech becomes a rigorous and systematic process, enhancing the overall quality and fidelity of speech synthesis techniques.

\subsubsection{Datasets}
The field of neural speech synthesis is rapidly advancing and relies heavily on high-quality datasets for effective training and evaluation of models. One of the most frequently utilized datasets in this field is the LJ Speech \cite{ljspeech17}, which features about $24$ hours of recorded speech from a single female speaker reading passages from the public domain LJ Speech Corpus. This dataset is free and has corresponding transcripts, making it an excellent choice for text-to-speech synthesis tasks. Moreover, it has been used as a benchmark for numerous neural speech synthesis models, including Tacotron \cite{wang2017tacotron}, WaveNet \cite{oord2016wavenet}, and DeepVoice \cite{arik2017deep,gibiansky2017deep}.

Apart from the LJ Speech dataset, several other datasets are widely used in neural speech synthesis research. The CMU Arctic \cite{kominek2004cmu}  and L2 Arctic \cite{zhao2018l2} datasets contain recordings of English speakers with diverse accents reading passages designed to capture various phonetic and prosodic aspects of speech. The LibriSpeech \cite{panayotov2015librispeech}, VoxCeleb \cite{Nagrani17}, TIMIT Acoustic-Phonetic Continuous Speech Corpus \cite{garofolo1993timit}, and Common Voice Dataset \cite{ardila2019common} are other valuable datasets that offer ample opportunities for training and evaluating text-to-speech synthesis models.

\subsubsection{Models}

Neural network-based text-to-speech (TTS) systems have been proposed using neural networks as the basis for speech synthesis, particularly with the emergence of deep learning. In Statistical Parametric Speech Synthesis (SPSS), early neural models replaced HMMs for acoustic modeling. The first modern neural TTS model, WaveNet \cite{oord2016wavenet}, generated waveforms directly from linguistic features. Other models, such as DeepVoice 1/2 \cite{arik2017deep,gibiansky2017deep}, used neural network-based models to follow the three components of statistical parametric synthesis. End-to-end models, including Tacotron 1 \& 2 \cite{wang2017tacotron,shen2018natural}, Deep Voice 3, and FastSpeech 1 \& 2  \cite{ren2019fastspeech,ren2020fastspeech}, simplified text analysis modules and utilized mel-spectrograms to simplify acoustic features with character/phoneme sequences as input. Fully end-to-end TTS systems, such as ClariNet \cite{ping2018clarinet}, FastSpeech 2 \cite{ren2020fastspeech}, and EATS \cite{donahueend}, are capable of directly generating waveforms from text inputs. Compared to concatenative synthesis~\footnote{https://en.wikipedia.org/wiki/Concatenative\_synthesis} and statistical parametric synthesis, neural network-based speech synthesis offers several advantages including superior voice quality, naturalness, intelligibility, and reduced reliance on human preprocessing and feature development. Therefore, end-to-end TTS systems represent a promising direction for advancing the field of speech synthesis.

\begin{table*}[]
\centering
\caption{Exploring the Landscape of TTS and Vocoder Architectures: Autoregressive and Non-Autoregressive Models.}
\resizebox{\textwidth}{!}{%
\begin{tabular}{ccc}
\hline
Method                   & Text-To-Speech                                                                                                                                                                                                                                                                                            & Vocoder                                                                                                                                                                                          \\ \hline
Autoregressive Model     & \begin{tabular}[c]{@{}c@{}}Tacotron \cite{wang2017tacotron} ,Tacotron2 \cite{shen2018natural}, Deep Voice 1,2,3\\ Transformer-TTS \cite{li2019neural}, DurIAN \cite{yu2019durian}, Flowtron \cite{valle2020flowtron}\\ RobuTrans \cite{li2020robutrans}, DeviceTTS \cite{huang2020devicetts},Wave-Tacotron \cite{weiss2021wave}\\ Apple TTS \cite{achanta2021device}\end{tabular}                                                                                                                                        & \begin{tabular}[c]{@{}c@{}}WaveNet \cite{oord2016wavenet}, WaveRNN \cite{kalchbrenner2018efficient}, WaveGAN \cite{pena2021wave}\\ LPCNet \cite{valin2019lpcnet}, GAN-TTS \cite{binkowski2019high}, MultiBand-WaveRNN \cite{yu2019durian}\\ ImporvedLPCNet \cite{valin2022neural}, Bunched LPCNet2 \cite{park2022bunched}\end{tabular}                                                         \\ \hline
Non-Autoregressive Model & \begin{tabular}[c]{@{}c@{}}ParaNet \cite{peng2020non}, FastSpeech \cite{ren2019fastspeech}, JDI-T \cite{lim2020jdi}, EATS \cite{donahue2020end}\\ FastSpeech2 \cite{ren2020fastspeech}, FastPitch \cite{lancucki2021fastpitch}, Glow-TTS \cite{kim2020glow}\\ Flow-TTS \cite{miao2020flow}, SpeedySpeech \cite{vainer2020speedyspeech}\\ Parallel Tacotron \cite{elias2021parallel}, BVAE-TTS \cite{lee2021bidirectional} \\ Parallel Tacotron2 \cite{elias2021parallel}, Grad-TTS \cite{popov2021grad}, VITS \cite{kim2021conditional}\\ RAD-TTS \cite{shih2021rad}, WaveGrad2 \cite{chen2021wavegrad}, DelightfulTTS \cite{liu2021delightfultts} \\ PortaSpeech \cite{ren2021portaspeech}, DiffGAN-TTS \cite{liu2022diffgan}, JETS \cite{lim2022jets}\\ WavThruVec \cite{siuzdak2022wavthruvec}, FastDiff \cite{huang2022fastdiff}, CLONE \cite{liu2022controllable}\end{tabular} & \begin{tabular}[c]{@{}c@{}}Parallel-WaveNet \cite{oord2018parallel}, WaveGlow \cite{prenger2019waveglow}, Parallel-WaveGAN \cite{yamamoto2020parallel}\\ MelGAN \cite{kumar2019melgan}, MultiBand-MelGAN \cite{yang2021multi}, VocGAN \cite{yang2020vocgan}, WaveGrad \cite{chen2020wavegrad}\\  DiffWave \cite{kong2020diffwave}, HiFi-GAN \cite{kong2020hifi}, StyleMelGAN \cite{mustafa2021stylemelgan}, Fre-GAN \cite{kim2021fre}\\ iSTFTNet \cite{kaneko2022istftnet}, Avocodo \cite{bak2022avocodo} \end{tabular} \\ \hline
\end{tabular}} 
\label{TTS:Landscape}
\end{table*}

 Transformer models have become increasingly popular for generating mel-spectrograms in TTS systems \cite{ren2020fastspeech,li2019neural}. These models are preferred over RNN structures in end-to-end TTS systems because they improve training and inference efficiency \cite{ren2019fastspeech,li2019neural}. In a study conducted by \citet{li2019neural}, a multi-head attention mechanism replaced both RNN structures and the vanilla attention mechanism in Tacotron 2 \cite{shen2018natural}. This approach addressed the long-distance dependency problem and improved pluralization. Phoneme sequences were used as input to generate the mel-spectrogram, and speech samples were synthesized using WaveNet as a vocoder. Results showed that the transformer-based TTS approach was $4.25$ times faster than Tacotron 2 and achieved similar MOS (Mean Opinion Score) performance.

Aside from the work mentioned above, there are other studies that are based on the Tacotron architecture. For example, \citet{skerry2018towards} and \citet{wang2018style} proposed Tacotron-based models for prosody control. These models use a separate encoder to compute style information from reference audio that is not provided in the text. Another noteworthy work is the Global-style-Token (GST) \cite{wang2018style} which improves on style embeddings by adding an attention layer to capture a wider range of acoustic styles.

The FastSpeech \cite{ren2019fastspeech} algorithm aims to improve the inference speed of TTS systems. To achieve this, it utilizes a feedforward network based on 1D convolution and the self-attention mechanism in transformers to generate Mel-spectrograms in parallel. Additionally, it solves the issue of sequence length mismatch between the Mel-spectrogram sequence and its corresponding phoneme sequence by employing a length regulator based on a duration predictor. The FastSpeech model was evaluated on the LJSpeech dataset and demonstrated significantly faster Mel-spectrogram generation than the autoregressive transformer model while maintaining comparable performance. FastPitch builds on FastSpeech by conditioning the TTS model on fundamental frequency or pitch contour, which improves convergence and eliminates the need for knowledge distillation of Mel-spectrogram targets in FastSpeech.

FastSpeech 2 \cite{ren2020fastspeech} represents a transformer-based Text-to-Speech (TTS) system that addresses the limitations of its predecessor, FastSpeech, while effectively handling the challenging one-to-many mapping problem in TTS. It introduces the utilization of a broader range of speech information, including energy, pitch, and more accurate duration, as conditional inputs. Furthermore, FastSpeech 2 trains the system directly on a ground-truth target, enhancing the quality of the synthesized speech. Additionally, a simplified variant called FastSpeech 2s has been proposed in [61], eliminating the requirement for intermediate Mel-spectrograms and enabling the direct generation of speech from text during inference. Experimental evaluations conducted on the LJSpeech dataset demonstrated that both FastSpeech 2 and FastSpeech 2s offer a streamlined training pipeline, resulting in fast, robust, and controllable speech synthesis compared to FastSpeech.

Furthermore, in addition to the transformer-based TTS systems like FastSpeech 2 and FastSpeech 2s, researchers have also been exploring the potential of Variational Autoencoder (VAE) based TTS models \cite{lee2021bidirectional,hsuhierarchical,guo2022multi,kim2021conditional}. These models can learn a latent representation of speech signals from textual input and may be able to produce high-quality speech with less training data and greater control over the generated speech characteristics. For example, authors in \cite{kim2021conditional} used a conditional variational autoencoder (CVAE) to model the acoustic features of speech and an adversarial loss to improve the naturalness of the generated speech. This approach involved conditioning the CVAE on the linguistic features of the input text and using an adversarial loss to match the distribution of the generated speech to that of natural speech. Results from this method have shown promise in generating speech that exhibits natural prosody and intonation. 

WaveGrad \cite{chen2020wavegrad} and DiffWave \cite{kong2020diffwave} have emerged as significant contributions in the field, employing diffusion models to generate raw waveforms with exceptional performance. In contrast, GradTTS \cite{popov2021grad} and DiffTTS \cite{jeong2021diff} utilize diffusion models to generate mel features rather than raw waveforms. Addressing the intricate challenge of one-shot many-to-many voice conversion, DiffVC \cite{popov2021diffusion} introduces a novel solver based on stochastic differential equations. Expanding the scope of sound generation to include singing voice synthesis, DiffSinger \cite{liu2022diffsinger} introduces a shallow diffusion mechanism. Additionally, Diffsound \cite{yang2022diffsound} proposes a sound generation framework that incorporates text conditioning and employs a discrete diffusion model, effectively resolving concerns related to unidirectional bias and accumulated errors.

EdiTTS \cite{DBLP:journals/corr/abs-2110-02584} introduces a diffusion-based audio model that is specifically tailored for the text-to-speech task. Its innovative approach involves the utilization of the denoising reversal process to incorporate desired edits through coarse perturbations in the prior space. Similarly, Guided-TTS \cite{kim2022guided} and Guided-TTS2 \cite{kim2022guided2} stand as early text-to-speech models that have effectively harnessed diffusion models for sound generation. Furthermore, Levkovitch et al. \cite{levkovitch2022zero} have made notable contributions by combining a voice diffusion model with a spectrogram domain conditioning technique. This combined approach facilitates text-to-speech synthesis, even with previously unseen voices during the training phase, thereby enhancing the model's versatility and capabilities.

InferGrad \cite{chen2022infergrad} enhances the diffusion-based text-to-speech model by incorporating the inference process during training, particularly when a limited number of inference steps are available. This improvement results in faster and higher-quality sampling. SpecGrad \cite{koizumi2022specgrad} introduces adaptations to the time-varying spectral envelope of diffusion noise based on conditioning log-mel spectrograms, drawing inspiration from signal processing techniques. ItoTTS \cite{wu2021hat} presents a unified framework that combines text-to-speech and vocoder models, utilizing linear SDE (Stochastic Differential Equation) as its fundamental principle. ProDiff \cite{huang2022prodiff} proposes a progressive and efficient diffusion model specifically designed for generating high-quality text-to-speech synthesis. Unlike traditional diffusion models that require a large number of iterations, ProDiff parameterizes the model by predicting clean data and incorporates a teacher-synthesized mel-spectrogram as a target to minimize data discrepancies and improve the sharpness of predictions. Finally, Binaural Grad \cite{leng2022binauralgrad} explores the application of diffusion models in binaural audio synthesis, aiming to generate binaural audio from monaural audio sources. It accomplishes this through a two-stage diffusion-based framework.

\begin{figure}
    \centering
    \includegraphics[width=0.6\columnwidth]{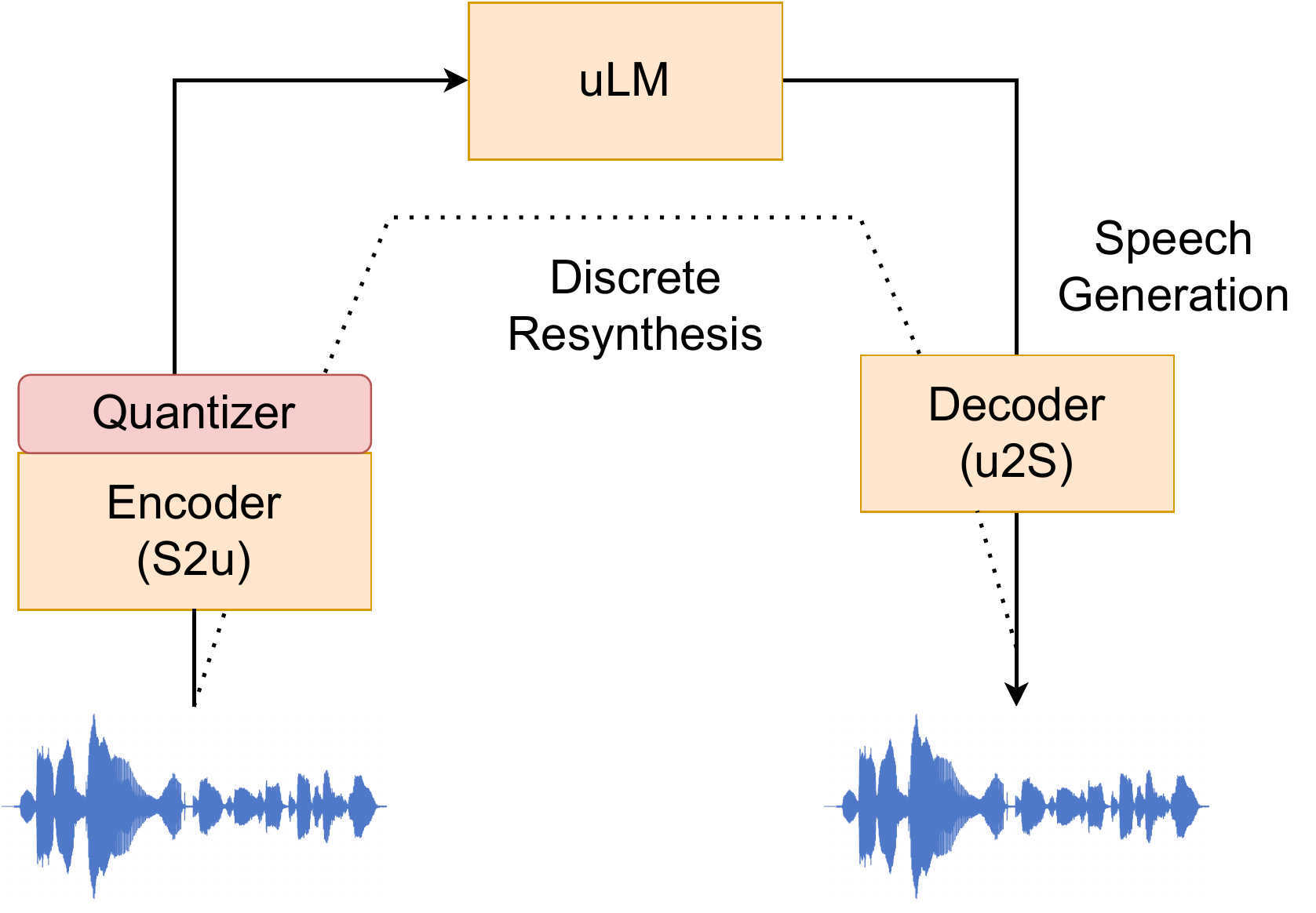}
    \caption{The architecture of the Generative Spoken Language Model GSLM introduced by Meta in \cite{lakhotia2021generative}. GSLM model operates through a three-part architecture.  Firstly, the encoder takes the speech waveform and transforms it into distinct units represented as S2u. Secondly, the decoder reverses this mapping by converting the units back to the original waveform, represented as u2S. Finally, the language model is unit-based and captures the distribution of unit sequences, which can be viewed as a form of pseudo-text. }
    \label{fig:GSLM}
\end{figure}
\subsubsection{Alignment}
Improving the alignment of text and speech in TTS architecture has been the focus of recent research \cite{kim2020glow,popov2021grad,ju2022trinitts,miao2021efficienttts,li2022styletts,shih2021rad,9746686,9747707,chen2021speech,ren2021portaspeech,bai20223,zhang2018forward,battenberg2020location,shen2020non}. Traditional TTS models require external aligners to provide attention alignments of phoneme-to-frame sequences, which can be complex and inefficient. Although autoregressive TTS models use an attention mechanism to learn these alignments online, these alignments tend to be brittle and often fail to generalize to long utterances and out-of-domain text, resulting in missing or repeating words.

In their study \cite{drexler2019explicit}, the authors presented a novel text encoder network that includes an additional objective function to explicitly align text and speech encodings. The text encoder architecture is straightforward, consisting of an embedding layer, followed by two bidirectional LSTM layers that maintain the input's resolution. The study utilized the same subword segmentation for the input text as for the ASR output targets. While RNN models with soft attention mechanisms have been proven to be highly effective in various tasks, including speech synthesis, their use in online settings results in quadratic time complexity due to the pass over the entire input sequence for generating each element in the output sequence. In \cite{raffel2017online}, the authors proposed an end-to-end differentiable method for learning monotonic alignments, enabling the computation of attention in linear time. Several enhancements, such as those proposed in \cite{chiu2017monotonic}, have been proposed in recent years to improve alignment in TTS models. Additionally, in \cite{badlani2022one}, the authors introduced a generic alignment learning framework that can be easily extended to various neural TTS models. 

The use of normalizing flow has been introduced to address output diversity issues in parallel TTS architectures. This technique is utilized to model the duration of speech, as evidenced by studies conducted in \cite{kim2020glow,shih2021rad,miao2021efficienttts}. One such flow-based generative model is Glow-TTS \cite{kim2020glow}, developed specifically for parallel TTS without the need for an external aligner. The model employs the generic Glow architecture previously used in computer vision and vocoder models to produce mel-spectrograms from text inputs, which are then converted to speech audio. Glow-TTS has demonstrated superior synthesis speed over the autoregressive model, Tacotron 2, while maintaining comparable speech quality.

Recently, a new TTS model called EfficientTTS \cite{miao2021efficienttts} has been introduced. This model outperforms previous models such as Tacotron 2 and Glow-TTS in terms of speech quality, training efficiency, and synthesis speed. The EfficientTTS model uses a multi-head attention mechanism to align input text and speech encodings, enabling it to generate high-quality speech with fewer parameters and faster synthesis speed. Overall, the introduction of normalizing flow and the development of models such as Glow-TTS and EfficientTTS have significantly improved the quality and efficiency of TTS systems.

To resolve output diversity issues in parallel TTS architectures, normalizing flow has been introduced to model the duration of speech \cite{kim2020glow,shih2021rad,miao2021efficienttts}. Glow-TTS \cite{kim2020glow} is a flow-based generative model for parallel TTS that does not require any external aligner12345. It is built on the generic Glow model that is previously used in computer vision and vocoder models3. Glow-TTS is designed to produce mel-spectrograms from text input, which can then be converted to speech audio4. It has been shown to achieve an order-of-magnitude speed-up over the autoregressive model, Tacotron 2, at synthesis with comparable speech quality. EfficientTTS is a recent study that proposed a new TTS model, which significantly outperformed models such as Tacotron 2 \cite{shen2018natural} and Glow-TTS \cite{kim2020glow} in terms of speech quality, training efficiency, and synthesis speed. The EfficientTTS \cite{miao2021efficienttts} model uses a multi-head attention mechanism to align the input text and speech encodings, enabling it to generate high-quality speech with fewer parameters and faster synthesis speed.


\subsubsection{Speech Resynthesis}
Speech resynthesis is the process of generating speech from a given input signal. The input signal can be in various forms, such as a digital recording, text, or other types of data. The aim of speech resynthesis is to create an output that closely resembles the original signal in terms of sound quality, prosody, and other acoustic characteristics. Speech resynthesis is an important research area with various applications, including speech enhancement \cite{tan2019learning,hsu2022revise,maiti2020speaker}, and voice conversion \cite{maimon2022speaking}. 
Recent advancements in speech resynthesis have revolutionized the field by incorporating self-supervised discrete representations to generate disentangled representations of speech content, prosodic information, and speaker identity. These techniques enable the generation of speech in a controlled and precise manner, as seen in \cite{lakhotia2021generative,polyak2021speech,qian2022contentvec,sicherman2023analysing}. The objective is to generate high-quality speech that maintains or degrades acoustic cues, such as phonotactics, syllabic rhythm, or intonation, from natural speech recordings.

Speech resynthesis is a vital research area with various applications, including speech enhancement and voice conversion, and recent advancements have revolutionized the field by incorporating self-supervised discrete representations. These techniques enable the generation of high-quality speech that maintains or degrades acoustic cues from natural speech recordings, and they have been used in the GSLM \cite{lakhotia2021generative}  architecture for acoustic modeling, speech recognition, and synthesis, as outlined in Figure \ref{fig:GSLM}. It comprises a discrete speech encoder, a generative language model, and a speech decoder, all trained without supervision. GSLM is the only prior work addressing the generative aspect of speech pre-training, which builds a text-free language model using discovered units.

\subsubsection{Voice Conversion}
Modifying a speaker's voice in a provided audio sample to that of another individual is called voice conversion, preserving linguistic content information. TTS and Voice conversion share a common objective of generating natural speech.
While models based on RNNs and CNNs have been successfully applied to voice conversion, the use of the transformer has shown promising results. Voice Transformer Network (VTN) \cite{huang2019voice} is a seq2seq voice conversion (VC) model based on the transformer architecture with TTS pre-training. Seq2seq VC models are attractive as they can convert prosody, and the VTN is a novel approach in this field that has been proven to be effective in converting speech from a source to a target without changing the linguistic content.

ASR and TTS-based voice conversion is a promising approach to voice conversion \cite{tian2019vocoder}. It involves using an ASR model to transcribe the source speech into the linguistic representation and then using a TTS model to synthesize the target speech with the desired voice characteristics \cite{polyak2019tts}. However, this approach overlooks the modeling of prosody, which plays an important role in speech naturalness and conversion similarity. To address this issue, researchers have proposed to directly predict prosody from the linguistic representation in a target-speaker-dependent manner \cite{zhang2020voice}. Other researchers have explored using a mix of ASR and TTS features to improve the quality of voice conversion \cite{huang2021prosody,zhao2021towards,chou2019one,zhang2019non}. 

CycleGAN \cite{kaneko2019cyclegan,kaneko2020cyclegan,kaneko2021maskcyclegan}, VAE \cite{choi2021neural,9053854,kameoka2019acvae}, and VAE with the generative adversarial network \cite{hsu2017voice} are other popular VC other popular approaches for non-parallel-voice conversion. CycleGAN-VC \cite{kaneko2019cyclegan} uses a cycle-consistent adversarial network to convert the source voice to the target voice and can generate high-quality speech without any extra data, modules, or alignment procedure. Several improvements and modifications are also proposed in recent years \cite{kaneko2020cyclegan,kaneko2021maskcyclegan,hsu2017voice}. VAE-based voice conversion is a promising approach that can generate high-quality speech with a small amount of training data \cite{choi2021neural,9053854,kameoka2019acvae}.

\subsubsection{Vocoders}
The field of audio synthesis has undergone significant advancements in recent years, with various approaches proposed to enhance the quality of synthesized audio. Prior studies have concentrated on improving discriminator architectures or incorporating auxiliary training losses. For instance, MelGAN introduced a multiscale discriminator that uses window-based discriminators at different scales and applies average pooling to downsample the raw waveform. It enforces the correspondence between the input Mel spectrogram and the synthesized waveform using an L1 feature matching loss from the discriminator. In contrast, GAN-TTS \cite{binkowski2019high} utilizes an ensemble of discriminators that operate on random windows of different sizes and enforce the mapping between the conditioner and the waveform adversarially using conditional discriminators. Another approach, parallel WaveGAN \cite{yamamoto2020parallel}, extends the single short-time Fourier transform loss to multi-resolution and employs it as an auxiliary loss for GAN training. Recently, some researchers have improved MelGAN by integrating the multi-resolution short-time Fourier transform loss. HiFi-GAN reuses the multi-scale discriminator from MelGAN and introduces the multi-period discriminator for high-fidelity synthesis. UnivNet employs a multi-resolution discriminator that takes multi-resolution spectrograms as input and can enhance the spectral structure of a synthesized waveform. In contrast, CARGAN integrates partial autoregression into the generator to enhance pitch and periodicity accuracy. The recent generative models for modeling raw audio can be categorized into the following groups.
\begin{itemize}
    \item  Autoregressive models: Although WaveNet is renowned for its exceptional ability to generate high-quality speech, including natural-sounding intonation and prosody, other neural vocoders have emerged as potential alternatives in recent years. For instance, LPCNet \cite{valin2019lpcnet} employs a combination of linear predictive coding (LPC) and deep neural networks (DNNs) to generate speech of similar quality while being computationally efficient and capable of producing low-bitrate speech. Similarly, SampleRNN \cite{mehri2016samplernn}, an unconditional end-to-end model, has demonstrated potential as it leverages a hierarchical RNN architecture and is trained end-to-end to generate raw speech of high quality.
    \item Generative Adversarial Network (GAN) vocoders: Numerous vocoders have been created that employ Generative Adversarial Networks (GANs) to generate speech of exceptional quality. These GAN-based vocoders, which include MelGAN  MelGAN \cite{kumar2019melgan}and HiFIGAN \cite{kong2020hifi}, are capable of producing high-fidelity raw audio by conditioning on mel spectrograms. Furthermore, they can synthesize audio at speeds several hundred times faster than real-time on a single GPU, as evidenced by research conducted in \cite{donahue2018adversarial,binkowskihigh,yamamoto2020parallel,kong2020hifi,kumar2019melgan}.
    \item Diffusion-based models: In recent years, there have been several novel architectures proposed that are based on diffusion. Two prominent examples of these are WaveGrad \cite{chenwavegrad} and DiffWave \cite{kong2020diffwave}. The WaveGrad model architecture builds upon prior works from score matching and diffusion probabilistic models, while the DiffWave model uses adaptive noise spectral shaping to adapt the diffusion noise. This adaptation, achieved through time-varying filtering, improves sound quality, particularly in high-frequency bands. Other examples of diffusion-based vocoders include InferGrad \cite{chen2022infergrad}, SpecGrad \cite{koizumi2022specgrad}, and Priorgrad \cite{leepriorgrad}. InfraGrad incorporates the inference process into training to reduce inference iterations while maintaining high quality. SpecGrad adapts the diffusion noise distribution to a given acoustic feature and uses adaptive noise spectral shaping to generate high-fidelity speech waveforms.
    \item Flow-based models: Parallel WaveNet, WaveGlow, etc. \cite{luong2021flowvocoder,prenger2019waveglow,kim2018flowavenet,ping2020waveflow,lee2020nanoflow} are based on normalizing flows and are capable of generating high-fidelity speech in real-time. While flow-based vocoders generally perform worse than autoregressive vocoders with regard to modeling the density of speech signals, recent research \cite{luong2021flowvocoder} has proposed new techniques to improve their performance.
\end{itemize} 

Universal neural vocoding is a challenging task that has achieved limited success to date. However, recent advances in speech synthesis have shown a promising trend toward improving zero-shot performance by scaling up model sizes. Despite its potential, this approach has yet to be extensively explored. Nonetheless, several approaches have been proposed to address the challenges of universal vocoding. For example, WaveRNN has been utilized in previous studies to achieve universal vocoding (\citet{lorenzo2018towards}; \citet{paul2020speaker}). Another approach \citet{jiao2021universal} developed involves constructing a universal vocoder using a flow-based model. Additionally, the GAN vocoder has emerged as a promising candidate for this task, as suggested by ~\citet{you2021gan}.
\subsubsection{Controllable Speech Synthesis}
Controllable Speech Synthesis \cite{9054556,9640518,9003829,9053732,ren2019fastspeech,wang2018style,8778667} is a rapidly evolving research area that focuses on generating natural-sounding speech with the ability to control various aspects of speech, including pitch, speed, and emotion.  Controllable Speech Synthesis is positioned in the emerging field of affective computing at the intersection of three disciplines: expressive speech analysis \cite{tits2019visualization}, natural language processing, and machine learning. This field aims to develop systems capable of recognizing, interpreting, and generating human-like emotional responses in interactions between humans and machines.

Expressive speech analysis is a critical component of this field. It provides mathematical tools to analyse speech signals and extract various acoustic features, including pitch, loudness, and duration, that convey emotions in speech. Natural language processing is also crucial to this field, as it helps to process the text input and extract the meaning and sentiment of the words. Finally, machine learning techniques are used to model and control the expressive features of the synthesized speech, enabling the systems to produce more expressive and controllable speech \cite{9053678,9420276,valle2020flowtron,kulkarni2020transfer,sorin2020principal,zhao2023emotion,pamisetty2023prosody,huang2022generspeech,lee2022hierspeech}.

In the last few years, notable advancements have been achieved in this field \cite{raitio2020controllable,kenter2019chive,habibie2022motion}, and several approaches have been proposed to enhance the quality of synthesized speech. For example, some studies propose using deep learning techniques to synthesize expressive speech and conditional generation models to control the prosodic features of speech \cite{raitio2020controllable,kenter2019chive}. Others propose using motion matching-based algorithms to synthesize gestures from speech  \cite{habibie2022motion}.

\subsubsection{Disentangling and Transferring}
The importance of disentangled representations for neural speech synthesis cannot be overstated, as it has been widely recognized in the literature that this approach can greatly improve the interpretability and expressiveness of speech synthesis models \cite{ma2019neural, hsu2019disentangling, qian2020unsupervised}. Disentangling multiple styles or prosody information during training is crucial to enhance the quality of expressive speech synthesis and control. Various disentangling techniques have been developed using adversarial and collaborative games, the VAE framework, bottleneck reconstructions, and frame-level noise modeling combined with adversarial training.

For instance, \citet{ma2019neural} have employed adversarial and collaborative games to enhance the disentanglement of content and style, resulting in improved controllability. \citet{hsu2019disentangling} have utilized the VAE framework with adversarial training to separate speaker information from noise. \citet{qian2020unsupervised} have introduced speech flow, which can disentangle rhythm, pitch, content, and timbre through three bottleneck reconstructions. In another work based on, adversarial training, \citet{zhang2021denoispeech} have proposed a method that disentangles noise from the speaker by modeling the noise at the frame level. 

 Developing high-quality speech synthesis models that can handle noisy data and generate accurate representations of speech is a challenging task. To tackle this issue, \citet{zhang2022hifidenoise} propose a novel approach involving multi-length adversarial training. This method allows for modeling different noise conditions and improves the accuracy of pitch prediction by incorporating discriminators on the mel-spectrogram. By replacing the traditional pitch predictor model with this approach, the authors demonstrate significant improvements in the fidelity of synthesized speech.

\subsubsection{Robustness}
Using neural TTS models can present issues with robustness, leading to low-quality audio samples for unseen or atypical text. In response, \citet{li2020robutrans} proposed RobuTrans \cite{li2020robutrans}, a robust transformer that converts input text to linguistic features before feeding it to the encoder. This model also includes modifications to the attention mechanism and position embedding, resulting in improved MOS scores compared to other TTS models. Another approach to enhancing robustness is the s-Transformer, introduced by \citet{wang2020s}, which models speech at the segment level, allowing it to capture long-term dependencies and use segment-level encoder-decoder attention. This technique performs similarly to the standard transformer, exhibiting robustness for extra-long sentences. Lastly, \citet{zheng2020improving} proposed an approach that combines a local recurrent neural network with the transformer to capture sequential and local information in sequences. Evaluation of a $20$-hour Mandarin speech corpus demonstrated that this model outperforms the transformer alone in performance.

In their recent paper \cite{yang2022norespeech}, the authors proposed a novel method for extracting dynamic prosody information from audio recordings, even in noisy environments. Their approach employs probabilistic denoising diffusion models and knowledge distillation to learn speaking style features from a teacher model, resulting in a highly accurate reproduction of prosody and timber. This model shows great potential in applications such as speech synthesis and recognition, where noise-robust prosody information is crucial. Other noteworthy advances in the development of robust TTS systems include the work by \cite{shih2021rad}, which focuses on a robust speech-text alignment module, as well as the use of normalizing flows for diverse speech synthesis.

\subsubsection{Low-Resource Neural Speech Synthesis}
High-quality paired text and speech data are crucial for building high-quality Text-to-Speech (TTS) systems \cite{gabrys2022voice}. Unfortunately, most languages are not supported by popular commercialized speech services due to the lack of sufficient training data \cite{xu2020lrspeech}. To overcome this challenge, researchers have developed TTS systems under low data resource scenarios using various techniques \cite{gabrys2022voice, xu2020lrspeech, elneima2022adversarial, tu2019end}.

Several techniques have been proposed by researchers to enhance the efficiency of low-resource/Zero-shot TTS systems. One of these is the use of semi-supervised speech synthesis methods that utilize unpaired training data to improve data efficiency, as suggested in a study by \citet{liu2022simple}. Another method involves cascading pre-trained models for ASR, MT, and TTS to increase data size from unlabelled speech, as proposed by \citet{nguyen2022improving}. In addition, researchers have employed crowdsourced acoustic data collection to develop TTS systems for low-resource languages, as shown in a study by \citet{butryna2020google}. \citet{huang2022generspeech} introduced a zero-shot style transfer approach for out-of-domain speech synthesis that generates speech samples exhibiting a new and distinctive style, such as speaker identity, emotion, and prosody.

\subsection{Speaker recognition}
\subsubsection{Task Description}

Speech signal consists of information on various characteristics of a speaker, such as origin, identity, gender, emotion, etc. This property of speech allows speech-based speaker profiling with a wide range of applications in forensics, recommendation systems, etc. The research on recognizing speakers is extensive and aims to solve two major tasks: speaker identification (what is the identity?) and speaker verification (is the speaker he/she claims to be?). Speaker recognition/verification tasks require extracting a fixed-length vector, called speaker embedding, from unconstrained utterances. These embeddings represent the speakers and can be used for identification or verification tasks. Recent state-of-the-art speaker-embedding-extractor models are based on DNNs and have shown superior performance on both speaker identification and verification tasks. 
\begin{itemize}
    \item \textbf{Speaker Recognition} (SR) relies on speaker identification as a key aspect, where an unknown speaker's speech sample is compared to speech models of known speakers to determine their identity. The primary aim of speaker identification is to distinguish an individual's identity from a group of known speakers. This process involves a detailed analysis of the speaker's voice characteristics such as pitch, tone, accent, and other pertinent features to establish their identity. Recent advancements in deep learning techniques have significantly enhanced speaker identification, leading to the creation of accurate, efficient, and end-to-end models. Various deep learning-based models such as CNNs, RNNs, and their combinations have demonstrated exceptional performance in several subtasks of speaker identification, including verification, identification, diarization, and robust recognition \cite{ravanelli2020multi,kawakami2020learning,kinoshita2020improving}.
    \item \textbf{Speaker Verification} (SV) is a process that involves confirming the identity of a speaker through their speech. It differs from speaker identification, which aims to identify unknown speakers by comparing their voices with that of registered speakers in a database. Speaker verification verifies whether a speaker is who they claim to be by comparing their voice with an available speaker template. Deep learning-based speaker verification relies on Speaker Representation based on embeddings, which involves learning low-dimensional vector representations from speech signals that capture speaker characteristics, such as pitch and speaking style, and can be used to compare different speech signals and determine their similarity.
\end{itemize}

\subsubsection{Dataset}
The VoxCeleb dataset (VoxCeleb 1 \& 2) is widely used in speaker recognition research, as mentioned in \cite{Nagrani17}. This dataset consists of speech data collected from publicly available media, employing a fully automated pipeline that incorporates computer vision techniques. The pipeline retrieves videos from YouTube and applies active speaker verification using a two-stream synchronization CNN. Speaker identity is further confirmed through CNN-based facial recognition. Another commonly employed dataset is TIMIT, which comprises recordings of phonetically balanced English sentences spoken by a diverse set of speakers. TIMIT is commonly used for evaluating speech recognition and speaker identification systems, as referenced in \cite{garofolo1993timit}.

 Other noteworthy datasets in the field include the SITW database \cite{mclaren2016speakers}, which provides hand-annotated speech samples for benchmarking text-independent speaker recognition technology, and the RSR2015 database \cite{larcher2012rsr2015}, which contains speech recordings acquired in a typical office environment using multiple mobile devices. Additionally, the RedDots project \cite{lee2015reddots} and VOICES corpus \cite{richey2018voices} offer unique collections of offline voice recordings in furnished rooms with background noise, while the CN-CELEB database \cite{fan2020cn} focuses on a specific person of interest extracted from bilibili.com using an automated pipeline followed by human verification. 
 
 The BookTubeSpeech dataset \cite{pham2020toward} was also collected using an automated pipeline from BookTube videos, and the Hi-MIA database \cite{qin2020hi} was designed specifically for far-field scenarios using multiple microphone arrays. The FFSVC20 challenge \cite{qin2020ffsvc} and DIHARD challenge \cite{ryant2018first} are speaker verification and diarization research initiatives focusing on far-field and robustness challenges, respectively. Finally, the LibriSpeech dataset \cite{panayotov2015librispeech}, originally intended for speech recognition, is also useful for speaker recognition tasks due to its included speaker identity labels.


\subsubsection{Models}
Speaker identification (SI) and verification (SV) are crucial research topics in the field of speech technology due to their significant importance in various applications such as security \cite{edu2020smart}, forensics \cite{koval2020practice}, biometric authentication \cite{hanifa2021review}, and speaker diarization \cite{xiao2021microsoft}. Speaker recognition has become more popular with technological advancements, including the Internet of Things (IoT), smart devices, voice assistants, smart homes, and humanoids. Therefore, a significant quantity of research has been conducted in this field, and many methods have been developed, making the state-of-the-art in this field quite mature and versatile. However, it has become increasingly challenging to provide an overview of the various methods due to the high number of studies in the field.

A neural network approach for speaker verification was first attempted by \citet{6854363} in 2014, utilizing four fully connected layers for speaker classification. Their approach has successfully verified speakers with short-duration utterances by obtaining the $d$-vector by averaging the output of the last hidden layer across frames. Although various attempts have been made to directly learn speaker representation from raw waveforms by other researchers (\citet{ravanelli2018speaker,jung2019rawnet}), other well-designed neural networks like CNNs and RNNs have been proposed for speaker verification tasks by \citet{ye2021deep}. Nevertheless, the field still requires more powerful deep neural networks for superior extraction of speaker features.

Speaker verification has seen notable advancements with the advent of more powerful deep neural networks. One such model is the $x$-vector-based system proposed by \citet{snyder2018x}, which has gained widespread popularity due to its remarkable performance. Since its introduction, the $x$-vector system has undergone significant architectural enhancements and optimized training procedures \cite{deng2019arcface}. The widely-used ResNet \cite{he2016deep} architecture has been incorporated into the system to improve its performance further. Adding residual connections between frame-level layers has been found to improve the embeddings \cite{garcia2020jhu,zeinali2019but}. This technique has also aided in faster convergence of the back-propagation algorithm and mitigated the vanishing gradient problem \cite{he2016deep}. \citet{tang2019deep} proposed further improvements to the $x$-vector system. They introduced a hybrid structure based on TDNN and LSTM to generate complementary speaker information at different levels. They also suggested a multi-level pooling strategy to collect the speaker information from global and local perspectives. These advancements have significantly improved speaker verification systems' performance and paved the way for further developments in the field.

\citet{desplanques2020ecapa} propose a state-of-the-art architecture for speaker verification utilizing a Time Delay Neural Network (TDNN) called ECAPA-TDNN. The paper presents a range of enhancements to the existing $x$-vector architecture that leverages recent developments in face verification and computer vision. Specifically, the authors suggest three major improvements. Firstly, they propose restructuring the initial frame layers into 1-dimensional Res2Net modules with impactful skip connections, which can better capture the relationships between different time frames. Secondly, they introduce Squeeze-and-Excitation blocks to the TDNN layers, which help highlight the most informative channels and improve feature discrimination. Lastly, the paper proposes channel attention propagation and aggregation to efficiently propagate attention weights through multiple TDNN layers, further enhancing the model's ability to discriminate between speakers.

Additionally, the paper presents a new approach that utilizes ECAPA-TDNN from the speaker recognition domain as the backbone network for a multiscale channel adaptive module. The proposed method achieves promising results, demonstrating the effectiveness of the proposed architecture in speaker verification. Overall, ECAPA-TDNN  offers a comprehensive solution to speaker verification by introducing several novel contributions that improve the existing $x$-vector architecture, which has been state-of-the-art in speaker verification for several years. The proposed approach also achieves promising results, suggesting that the proposed architecture can effectively tackle the challenges of speaker verification.

The attention mechanism is a powerful method for obtaining a more discriminative utterance-level feature by explicitly selecting frame-level representations that better represent speaker characteristics. Recently, the Transformer model with a self-attention mechanism has become effective in various application fields, including speaker verification. The Transformer architecture has been extensively explored for speaker verification. TESA \cite{mary2021s} is an architecture based on the Transformer's encoder, proposed as a replacement for conventional PLDA-based speaker verification to capture speaker characteristics better. TESA outperforms PLDA on the same dataset by utilizing the next sentence prediction task of BERT \cite{devlin2018bert}. \citet{zhu2021serialized} proposed a method to create fixed-dimensional speaker verification representation using a serialized multi-layer multi-head attention mechanism. Unlike other studies that redesign the inner structure of the attention module, their approach strictly follows the original Transformer, providing simple but effective modifications.

\subsection{Speaker Diarization}
\subsubsection{Task Description}
\begin{figure*}
    \centering
    \includegraphics[width=\columnwidth]{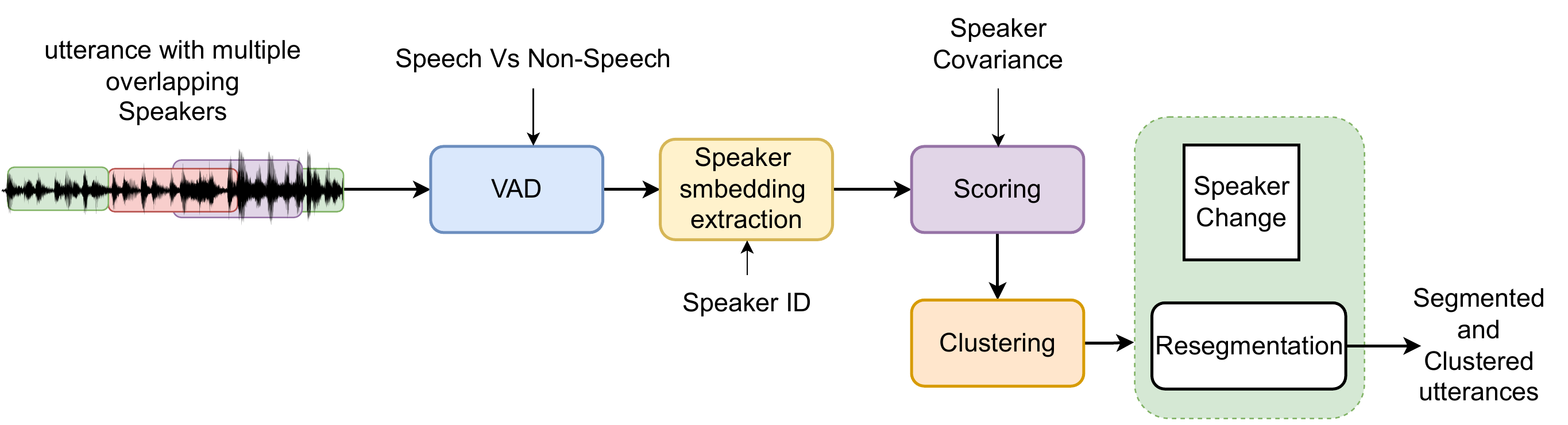}
    \caption{Speaker diarization system diagram showcasing the process of identifying and differentiating multiple speakers in an audio recording using various techniques such as VAD, segmentation, clustering and re-segmentation.}
    \label{fig:sd}
\end{figure*}

Speaker diarization is a critical component in the analysis of multi-speaker audio data, and it addresses the question of "who spoke when." The term "diarize" refers to the process of making a note or keeping a record of events, as per the English dictionary. A traditional speaker diarization system comprises several crucial components  that work together to achieve accurate and efficient speaker diarization. In this section, we will discuss the different components of a speaker diarization system (\Cref{fig:sd}) and their role in achieving accurate speaker diarization.

\begin{itemize}

\item \textit{Acoustic Features Extraction}: In the analysis of multi-speaker speech data, one critical component is the extraction of acoustic features \cite{anguera2012speaker,tranter2006overview}. This process involves extracting features such as pitch, energy, and MFCCs from the audio signal. These acoustic features play a crucial role in identifying different speakers by analyzing their unique characteristics.

\item \textit{Segmentation}: Segmentation is a crucial component in the analysis of multi-speaker audio data, where the audio signal is divided into smaller segments based on the silence periods between speakers \cite{anguera2012speaker,tranter2006overview}. This process helps in reducing the complexity of the problem and makes it easier to identify different speakers in smaller segments

\item \textit{Speaker Embedding Extraction}: This process involves obtaining a low-dimensional representation of each speaker's voice, which is commonly referred to as speaker embedding. This is achieved by passing the acoustic features extracted from the speech signal through a deep neural network, such as a CNN or RNN\cite{snyder2017deep}.

\item \textit{Clustering}: In this component, the extracted speaker embeddings are clustered based on similarity, and each cluster represents a different speaker \cite{anguera2012speaker,tranter2006overview}. This process commonly uses unsupervised clustering algorithms, such as k-means clustering.

\item \textit{Speaker Classification}: In this component, the speaker embeddings are classified into different speaker identities using a supervised classification algorithm, such as SVM or MLP \cite{anguera2012speaker,tranter2006overview}.

\item \textit{Re-segmentation}: This component is responsible for refining the initial segmentation by adjusting the segment boundaries based on the classification results. It helps in improving the accuracy of speaker diarization by reducing the errors made during the initial segmentation.

\end{itemize}
Various studies focus on traditional speaker diarization systems \cite{anguera2012speaker,tranter2006overview}. This paper will review the recent efforts toward deep learning-based speaker diarizations techniques. 
\subsubsection{Dataset}
\begin{itemize}
    \item  \textit{NIST SRE 2000 (Disk-8) or CALLHOME dataset}: The NIST SRE 2000 (Disk-8) corpus, also referred to as the CALLHOME dataset, is a frequently utilized resource for speaker diarization in contemporary research papers. Originally released in 2000, this dataset comprises conversational telephone speech (CTS) collected from diverse speakers representing a wide range of ages, genders, and dialects. It includes 500 sessions of multilingual telephonic speech, each containing two to seven speakers, with two primary speakers in each conversation. The dataset covers various topics, including personal and familial relationships, work, education, and leisure activities. The audio recordings were obtained using a single microphone and had a sampling rate of 8 kHz, with 16-bit linear quantization.
    \item \textit{Directions into Heterogeneous Audio Research (DIHARD) Challenge and dataset}: The DIHARD Challenge, organized by the National Institute of Standards and Technology (NIST), aims to enhance the accuracy of speech recognition and diarization in challenging acoustic environments, such as crowded spaces, distant microphones, and reverberant rooms. The challenge comprises tasks requiring advanced machine-learning techniques, including speaker diarization, recognition, and speech activity detection. The DIHARD dataset used in the challenge comprises over 50 hours of speech from more than 500 speakers, gathered from diverse sources like meetings, broadcast news, and telephone conversations. These recordings feature various acoustic challenges, such as overlapping speech, background noise, and distant or reverberant speech, captured through different microphone setups. To aid in the evaluation process, the dataset has been divided into separate development and evaluation sets. The assessment metrics used to gauge performance include diarization error rate (DER), as well as accuracy in speaker verification, identification, and speech activity detection.
    
    \item \textit{Augmented Multi-party Interaction (AMI) database}: The AMI database is a collection of audio and video recordings that capture real-world multi-party conversations in office environments. The database was developed as part of the AMI project, which aimed to develop technology for automatically analyzing multi-party meetings. The database contains over 100 hours of audio and video recordings of meetings involving four to seven participants, totaling 112 meetings. The meetings were held in multiple offices and were designed to reflect the kinds of discussions that take place in typical business meetings. The audio recordings were captured using close-talk microphones placed on each participant and additional microphones placed in the room to capture ambient sound. The video recordings were captured using multiple cameras placed around the room. In addition to the audio and video recordings, the database also includes annotations that provide additional information about the meetings, including speaker identities, speech transcriptions, and information about the meeting structure (e.g., turn-taking patterns). The AMI database has been used extensively in research on automatic speech recognition, speaker diarization, and other related speech and language processing topics.
    
    \item \textit{VoxSRC Challenge and VoxConverse corpus}: The VoxCeleb Speaker Recognition Challenge (VoxSRC) is an annual competition designed to assess the capabilities of speaker recognition systems in identifying speakers from speech recorded in real-world environments. The challenge provides participants with a dataset of audio and visual recordings of interviews, news shows, and talk shows featuring famous individuals. The VoxSRC encompasses several tracks, including speaker diarization, and comprises a development set (20.3 hours, 216 recordings) and a test set (53.5 hours, 310 recordings). Recordings in the dataset may feature between one and 21 speakers, with a diverse range of ambient noises, such as background music and laughter. To facilitate the speaker diarization track of the VoxSRC-21 and VoxSRC-22 competitions, VoxConverse, an audio-visual diarization dataset containing multi-speaker clips of human speech sourced from YouTube videos, is available, and additional details are provided on the project website \footnote{https://www.robots.ox.ac.uk/~vgg/data/voxconverse/}.
    
    \item \textit{LibriCSS}: The LibriCSS corpus is a valuable resource for researchers studying speech separation, recognition, and speaker diarization. The corpus comprises 10 hours of multichannel recordings captured using a 7-channel microphone array in a real meeting room. The audio was played from the LibriSpeech corpus, and each of the ten sessions was subdivided into six 10-minute mini-sessions. Each mini-session contained audio from eight speakers and was designed to have different overlap ratios ranging from 0\% to 40\%. To make research easier, the corpus includes baseline systems for speech separation and Automatic Speech Recognition (ASR) and a baseline system that integrates speech separation, speaker diarization, and ASR. These baseline systems have already been developed and made available to researchers.
    \item \textit{Rich Transcription Evaluation Series}: The Rich Transcription Evaluation Series dataset is a collection of speech data used for speaker diarization evaluation. The Rich Transcription Fall 2003 Evaluation (RT-03F) was the first evaluation in the series focused on "Who Said What" tasks. The dataset has been used in subsequent evaluations, including the Second DIHARD Diarization Challenge, which used the Jaccard index to compute the JER (Jaccard Error Rate) for each pair of segmentations. The dataset is essential for data-driven spoken language processing methods and calculates speaker diarization accuracy at the utterance level. The dataset includes rules, evaluation methods, and baseline systems to promote reproducible research in the field. The dataset has been used in various speaker diarization systems and their subtasks in the context of broadcast news and CTS data
    \item \textit{CHiME-5/6 challenge and dataset} The CHiME-5/6 challenge is a speech processing challenge focusing on distant multi-microphone conversational speech diarization and recognition in everyday home environments. The challenge provides a dataset of recordings from everyday home environments, including dinner recordings originally collected for and exposed during the CHiME-5 challenge. The dataset is designed to be representative of natural conversational speech. The challenge features two audio input conditions: single-channel and multichannel. Participants are provided with baseline systems for speech enhancement, speech activity detection (SAD), and diarization, as well as results obtained with these systems for all tracks. The challenge aims to improve the robustness of diarization systems to variations in recording equipment, noise conditions, and conversational domains.

    \item \textit{AMI dataset}: The AMI database is a comprehensive collection of 100 hours of recordings sourced from 171 meeting sessions held across various locations. It features two distinct audio sources – one recorded using lapel microphones for individual speakers and the other using omnidirectional microphone arrays placed on the table. It is an ideal dataset for evaluating speaker diarization systems integrated with the ASR module. AMI's value proposition is further enhanced by providing forced alignment data, which captures the timings at the word and phoneme levels and speaker labeling. Finally, it's worth noting that each meeting session involves a small group of three to five speakers.

\end{itemize}

\subsubsection{Models}
Speaker diarization has been a subject of research in the field of audio processing, with the goal of separating speakers in an audio recording. In recent years, deep learning has emerged as a powerful technique for speaker diarization, leading to significant advancements in this field. In this article, we will explore some of the recent developments in deep learning architecture for speaker diarization, focusing on different modules of speaker diarization as outlined in Figure \ref{fig:sd}. Through this discussion, we will highlight major advancements in each module.

\begin{itemize}

\item \textit{Segmentation and clustering}: Speaker diarization systems typically use a range of techniques for segmenting speech, such as identifying speaker change, uniform speaker segmentation, ASR-based word segmentation, and supervised speaker turn detection. However, each approach has its own benefits and drawbacks. Uniform speaker segmentation involves dividing speech into segments of equal length, which can be difficult to optimize to capture speaker turn boundaries and include enough speaker information. ASR-based word segmentation identifies word boundaries using automatic speech recognition, but the resulting segments may be too brief to provide adequate speaker information. Supervised speaker turn detection, on the other hand, involves a specialized model that can accurately identify speaker turn timestamps. While this method can achieve high accuracy, it requires labeled data for training. These techniques have been widely discussed in previous research, and choosing the appropriate one depends on the specific requirements of the application.
\begin{itemize}
    \item The authors in \cite{coria2021overlap} propose real-time speaker diarization system that combines incremental clustering and local diarization applied to a rolling window of speech data and is designed to handle overlapping speech segments. The proposed pipeline is designed to utilize end-to-end overlap-aware segmentation to detect and separate overlapping speakers.
    \item In another related work, authors in \cite{zhang2022towards} introduce a novel speaker diarization system with a generalized neural speaker clustering module as the backbone.
 
    \item In a recent study conducted by \citet{park2019auto}, a new framework for spectral clustering is proposed that allows for automatic parameter tuning of the clustering algorithm in the context of speaker diarization. The proposed technique utilizes normalized maximum eigengap (NME) values to determine the number of clusters and threshold parameters for each row in an affinity matrix during spectral clustering. The authors demonstrated that their method outperformed existing state-of-the-art methods on two different datasets for speaker diarization.

    \item Bayesian HMM clustering of x-vector sequences (VBx) diarization approach, which clusters x-vectors using a Bayesian hidden Markov model (BHMM) \cite{landini2022bayesian}, combined with a ResNet101  (\citet{he2016deep}) $x$-vector extractor achieves superior results on CALLHOME \cite{diez2020optimizing}, AMI \cite{carletta2006ami} and DIHARD II \cite{ryant2019second} datasets

\end{itemize}

\item \textit{Speaker Embedding Extraction and Classification}: 
\begin{itemize}
    \item Attentive Aggregation for Speaker Diarization \cite{kwon2021adapting}: This approach uses an attention mechanism to aggregate embeddings from multiple frames and generate speaker embeddings. The speaker embeddings are then used for clustering to identify speaker segments.
     \item  End-to-End Speaker Diarization with Self-Attention \cite{fujita2019end}: This method uses a self-attention mechanism to capture the correlations between the input frames and generates embeddings for each frame. The embeddings are then used for clustering to identify speaker segments.
     \item \citet{wang2022similarity} present an innovative method for measuring similarity between speaker embeddings in speaker diarization using neural networks. The approach incorporates past and future contexts and uses a segmental pooling strategy. Furthermore, the speaker embedding network and similarity measurement model are jointly trained. The paper extends this framework to target-speaker voice activity detection (TS-VAD) \cite{medennikov2020target}. The proposed method effectively learns the similarity between speaker embeddings by considering both past and future contexts.
     \item Time-Depth Separable Convolutions for Speaker Diarization \cite{koluguri2022titanet}: This approach uses time-depth separable convolutions to generate embeddings for each frame, which are then used for clustering to identify speaker segments. The method is computationally efficient and achieves state-of-the-art performance on several benchmark datasets.
\end{itemize}



\item \textit{Re-segmentation}: 
\begin{itemize}
    \item Numerous studies in this field centre around developing a re-segmentation strategy for diarization systems that can effectively handle both voice activity and overlapped speech detection. This approach can also be a post-processing step to identify and assign overlapped speech regions accurately. Notable examples of such works include those by \citet{bullock2020overlap} and \citet{bredin2021end}.
\end{itemize}

\item \emph{End-to-End Neural Diarization}: In addition to the above work, end-to-end speaker diarization systems have gained the attention of the research community due to their ability to handle speaker overlaps and their optimization to minimize diarization errors directly. In one such work, the authors propose end-to-end neural speaker diarization that does not rely on clustering and instead uses a self-attention-based neural network to directly output the joint speech activities of all speakers for each segment \cite{fujita2019end}. Following the trend, several other works propose enhanced architectures based on self-attention \cite{lin2020self,yu2022auxiliary}

\end{itemize}
    
    
    
\subsection{Speech-to-speech translation}
\subsubsection{Task Description}
Speech-to-text translation (ST) is the process of converting spoken language from one language to another in text form. Traditionally, this has been achieved using a cascaded structure that incorporates automatic speech recognition (ASR) and machine translation (MT) components. However, a more recent end-to-end (E2E)  method \cite{sung2019towards,salesky2019exploring,zhang2020adaptive,chen2020mam,han2021learning,zheng2021fused,ansari2020findings} has gained popularity due to its ability to eliminate issues with error propagation and high latency associated with cascaded methods \cite{sperber2020speech, chen2021specrec}. The E2E method uses an audio encoder to analyze audio signals and a text decoder to generate translated text.

One notable advantage of ST systems is that they allow for more natural and fluent communication than other language translation methods. By translating speech in real-time, ST systems can capture the subtleties of speech, including tone, intonation, and rhythm, which are essential for effective communication. Developing ST systems is a highly intricate process that involves integrating various technologies such as speech recognition, natural language processing, and machine translation. One significant obstacle in ST is the variation in accents and dialects across different languages, which can significantly impact the accuracy of the translation.

\subsubsection{Dataset}
There are numerous datasets available for the end-to-end speech translation task, with some of the most widely used ones being MuST-C \cite{cattoni2021must}, IWSLT \cite{scarton2019estimating}, and CoVoST 2 \cite{wang2020covost}. These datasets cover a variety of languages, including English, German, Spanish, French, Italian, Dutch, Portuguese, Romanian, Arabic, Chinese, Japanese, Korean, and Russian. For instance, TED-LIUM \cite{rousseau2012ted} is a suitable dataset for speech-to-text, text-to-speech, and speech-to-speech translation tasks, as it contains transcriptions and audio recordings of TED talks in English, French, German, Italian, and Spanish. Another open-source dataset is Common Voice, which covers several languages, including English, French, German, Italian, and Spanish. Additionally, VoxForge\footnote{http://www.voxforge.org/} is designed for acoustic model training and includes speech recordings and transcriptions in several languages, including English, French, German, Italian, and Spanish. LibriSpeech \cite{panayotov2015librispeech} is a dataset of spoken English specifically designed for speech recognition and speech-to-text translation tasks. Lastly, How2 \cite{duarte2021how2sign} is a multimodal machine translation dataset that includes speech recordings, text transcriptions, and video and image data, covering English, German, Italian, and Spanish. These datasets have been instrumental in training state-of-the-art speech-to-speech translation models and will continue to play a crucial role in further advancing the field.
\subsubsection{Models}
End-to-end speech translation models are a promising approach to direct the speech translation field. These models use a single sequence-to-sequence model for speech-to-text translation and then text-to-speech translation. In 2017, researchers demonstrated that end-to-end models outperform cascade models[3]. One study published in 2019 provides an overview of different end-to-end architectures and the usage of an additional connectionist temporal classification (CTC) loss for better convergence \cite{bahar2019comparative}. The study compares different end-to-end architectures for speech-to-text translation. In 2019, Google introduced Translatotron \cite{jia2022translatotron}, an end-to-end speech-to-speech translation system. Translatotron uses a single sequence-to-sequence model for speech-to-text translation and then text-to-speech translation. No transcripts or other intermediate text representations are used during inference. The system was validated by measuring the BLEU score, computed with text transcribed by a speech recognition system. Though the results lag behind a conventional cascade system, the feasibility of the end-to-end direct speech-to-speech translation was demonstrated \cite{jia2022translatotron}.

In a recent publication from 2020, researchers presented a study on an end-to-end speech translation system. This system incorporates pre-trained models such as Wav2Vec 2.0 and mBART, along with coupling modules between the encoder and decoder. The study also introduces an efficient fine-tuning technique, which selectively trains only $20\%$ of the total parameters \cite{ye2021end}. The system developed by the UPC Machine Translation group actively participated in the IWSLT 2021 offline speech translation task, which aimed to develop a system capable of translating English audio recordings from TED talks into German text.

E2E ST is often improved by pretraining the encoder and/or decoder with transcripts from speech recognition or text translation tasks \cite{di2019adapting,wang2020fairseq,zhang2020adaptive,xu2021stacked}. Consequently, it has become the standard approach used in various toolkits \cite{inaguma2020espnet,wang2020fairseq,zhao2020neurst,zheng2021fused}. However, transcripts are not always available, and the significance of pretraining for E2E ST is rarely studied. \citet{zhang2022revisiting} explored the effectiveness of E2E ST trained solely on speech-translation pairs and proposed an algorithm for training from scratch. The proposed system outperforms previous studies in four benchmarks covering 23 languages without pretraining. The paper also discusses neural acoustic feature modeling, which extracts acoustic features directly from raw speech signals to simplify inductive biases and enhance speech description. 

\subsection{Speech enhancement}
\subsubsection{Task Description}
In situations where there is ambient noise present, speech recognition systems can encounter difficulty in correctly interpreting spoken language signals, resulting in reduced performance \cite{du2014robust}. One possible solution to address this issue is the development of speech enhancement systems that can eliminate noise and other types of signal distortion from spoken language, thereby improving signal quality. These systems are frequently implemented as a preprocessing step to enhance the accuracy of speech recognition and can serve as an effective approach for enhancing the performance of ASR systems in noisy environments. This section will delve into the significance of speech enhancement technology in boosting the accuracy of speech recognition.
\subsubsection{Dataset}
One popular dataset for speech enhancement tasks is AISHELL-4, which comprises authentic Mandarin speech recordings captured during conferences using an 8-channel circular microphone array. In accordance with \cite{fu2021aishell}, AISHELL-4 is composed of 211 meeting sessions, each featuring 4 to 8 speakers, for a total of 120 hours of content. This dataset is of great value for research into multi-speaker processing owing to its realistic acoustics and various speech qualities, including speaker diarization and speech recognition

Another popular dataset used for speech enhancement is the dataset from Deep Noise Suppression (DNS) challenge \cite{reddy2020interspeech}, a large-scale dataset of noisy speech signals and their corresponding clean speech signals. The DNS dataset contains over $10,000$ hours of noisy speech signals and over $1,000$ hours of clean speech signals, making it useful for training deep learning models for speech enhancement.
The Voice Bank Corpus (VCTK)  is another dataset containing speech recordings from 109 speakers, each recording approximately $400$ sentences. The dataset contains clean and noisy speech recordings, making it useful for training speech enhancement models. These datasets provide realistic acoustics, rich natural speech characteristics, and large-scale noisy and clean speech signals, making them useful for training deep learning models.
\subsubsection{Models}
 Several Classical algorithms have been reported in the literature for speech enhancement, including spectral subtraction \cite{boll1979suppression}, Wiener and Kalman filtering \cite{lim1978all,scalart1996speech}, MMSE estimation \cite{ephraim1992bayesian}, comb filtering \cite{jin2009speech}, subspace methods \cite{hansen1997signal}. Phase spectrum compensation \cite{paliwal2011importance}. However, classical algorithms such as spectral subtraction and Wiener filtering approach the problem in the spectral domain and are restricted to stationary or quasi-stationary noise.
\begin{table*}
\centering
\caption{Performance of different speech enhancement algorithms on the Deep Noise Suppression (DNS) Challenge dataset. The table showcases improvements in PESQ-WB, PESQ-NB, SI-SDR-WB, and SI-SDR-NB metrics, and identifies the top-performing methods in each category.}
\resizebox{0.8\textwidth}{!}{%
\begin{tabular}{|l|cccc|l|}
\hline
\textbf{Model} & \textbf{PESQ-WB} & \textbf{PESQ-NB} & \textbf{SI-SDR-WB} & \textbf{SI-SDR-NB} & \textbf{Architecture} \\
\hline
FRCRN \cite{zhao2021monaural} & 3.23 & - & - & - & U-Net + CRN \\
Sudo rm -rf \cite{tzinis2022remixit} & 2.95 & - & 19.7 & - & UConvBlock + CNN \\
DCTCRN-P \cite{li2021real} & 2.82 & - & - & - & CNN \\
PoCoNet \cite{isik2020poconet} & 2.7885 & - & - & - & - \\
FullSubNet \cite{hao2021fullsubnet} & 2.777 & 3.305 & 17.29 & - & LSTM \\
RNN-Modulation \cite{vuong2021modulation} & 2.75 & - & - & - & GRU \\
Conv-TasNet-SNR \cite{koyama2020exploring} & 2.73 & - & - & - & CNN \\
Sudo rm-rf \cite{tzinis2022continual} & 2.69 & - & 18.6 & - & UConvBlock + CNN \\
RemixIT \cite{tzinis2022remixit} & 2.34 & - & 16.0 & - & UConvBlock \\
SN-Net \cite{zheng2021interactive} & - & 3.39 & - & 19.52 & CNN \\
DCCRN-E-Aug \cite{hu2020dccrn} & - & 3.214 & - & - & CNN + LSTM \\
DTLN \cite{westhausen2020dual} & - & 3.04 & 16.34 & - & LSTM \\
DCCRN-E \cite{hu2020dccrn} & - & 3.04 & - & - & CNN + LSTM \\
\hline
\end{tabular}}
\label{performance:se}
\end{table*}

Neural network-based approaches inspired from other areas such as computer vision \cite{hou2018audio,gabbay2017visual,afouras2018conversation} and generative adversarial networks \cite{wu2019speech,lin2019speech,routray2022phase,fu2019metricgan} or developed for general audio processing tasks \cite{wang2020complex,giri2019attention} have outperformed the classical approaches. Various neural network models based on different architectures, including fully connected neural networks \cite{xu2014regression}, deep denoising autoencoder \cite{lu2013speech}, CNN \cite{fu2016snr}, LSTM \cite{chen2015speech}, and Transformer \cite{koizumi2020speech} have effectively handled diverse noisy conditions.

Diffusion-based models have also shown promising results for speech enhancement \cite{lemercier2022storm,yen2022cold,lu2022conditional} and have led to the development of novel speech enhancement algorithms called Conditional Diffusion Probabilistic Model (CDiffuSE) that incorporates characteristics of the observed noisy speech signal into the diffusion and reverse processing \cite{lu2022conditional}. CDiffuSE is a generalized formulation of the diffusion probabilistic model that can adapt to non-Gaussian real noises in the estimated speech signal. Another diffusion-based model for speech enhancement is StoRM \cite{lemercier2022storm}, which stands for Stochastic Regeneration Model. It uses a predictive model to remove vocalizing and breathing artifacts while producing high-quality samples using a diffusion process, even in adverse conditions. StoRM has shown great ability at bridging the performance gap between predictive and generative approaches for speech enhancement. Furthermore, authors in \cite{yen2022cold} propose cold diffusion process is an advanced iterative version of the diffusion process to recover clean speech from noisy speech. According to the authors, it can be utilized to restore high-quality samples from arbitrary degradations. Table \ref{performance:se} summarizing the performance of different speech enhancement algorithms on the Deep
Noise Suppression (DNS) Challenge dataset using different metrics.

\subsection{Audio Super Resolution}
\subsubsection{Task Description}
Audio super-resolution is a technique that involves predicting the missing high-resolution components of low-resolution audio signals. Achieving this task can be difficult due to the continuous nature of audio signals. Current methods typically approach super-resolution by treating audio as discrete data and focusing on fixed scale factors. In order to accomplish audio super-resolution, deep neural networks are trained using pairs of low and high-quality audio examples. During testing, the model predicts missing samples within a low-resolution signal. Some recent deep network approaches have shown promise by framing the problem as a regression issue either in the time or frequency domain \cite{8462049}. These methods have been able to achieve impressive results.
\subsubsection{Datasets}
This section provides an overview of the diverse datasets utilized in Audio Super Resolution literature. One of the most frequently used datasets is the MUSDB18, specifically designed for music source separation and enhancement. This dataset encompasses more than 150 songs with distinct tracks for individual instruments. Another prominent dataset is UrbanSound8K, which comprises over, $8000$ environmental sound files collected from 10 different categories, making it ideal for evaluating Audio Super Resolution algorithms in noisy environments. Furthermore, the VoiceBank dataset is another essential resource for evaluating Audio Super Resolution systems, comprising over 10,000 speech recordings from five distinct speakers. This dataset offers a rich source of information for assessing speech processing systems, including Audio Super Resolution. Another dataset, LibriSpeech, features more than 1000 hours of spoken words from several books and speakers, making it valuable for evaluating Audio Super Resolution algorithms to enhance the quality of spoken words. Finally, the TED-LIUM dataset, which includes over 140 hours of speech recordings from various speakers giving TED talks, provides a real-world setting for evaluating Audio Super Resolution algorithms for speech enhancement. By using these datasets, researchers can evaluate Audio Super Resolution systems for a wide range of audio signals and improve the generalizability of these algorithms for real-world scenarios.
\subsubsection{Models}
Audio super-resolution has been extensively explored using deep learning architectures \cite{rakotonirina2021self,yoneyama2022nonparallel,8462049,lee2021nu,han2022nu,birnbaum2019temporal,abdulatif2022cmgan,nguyen2022tunet,kim2022learning,liu2022neural}. One notable paper by \citet{rakotonirina2021self} proposes a novel network architecture that integrates convolution and self-attention mechanisms for audio super-resolution. Specifically, they use Attention-based Feature-Wise Linear Modulation (AFiLM) \cite{rakotonirina2021self} to modulate the activations of the convolutional model. In another recent work by \citet{yoneyama2022nonparallel}, the super-resolution task is decomposed into domain adaptation and resampling processes to handle acoustic mismatch in unpaired low- and high-resolution signals. To address this, they jointly optimize the two processes within the CycleGAN framework.

Moreover, the Time-Frequency Network (TFNet) \cite{8462049} proposed a deep network that achieves promising results by modeling the task as a regression problem in either time or frequency domain. To further enhance audio super-resolution, the paper proposes a time-frequency network that combines time and frequency domain information. Finally, recent advancements in diffusion models have introduced new approaches to neural audio upsampling. Specifically, \citet{lee2021nu}, and \citet{han2022nu} propose NU-Wave 1 and 2 diffusion probabilistic models, respectively, which can produce high-quality waveforms with a sampling rate of 48kHz from coarse 16kHz or 24kHz inputs. These models are a promising direction for improving audio super-resolution.

\subsection{Voice Activity Detection (VAD)}
\subsubsection{Task Description}
Due to the increasing sophistication of mobile devices like smartphones, speech-controlled applications have become incredibly popular. These apps offer a hands-free method for controlling home devices, facilitating telephony, and allowing drivers to safely use their vehicle's infotainment systems while on the go. However, accurately distinguishing between noise and human speech is critical for these applications to work without interruption. To overcome this issue, Voice Activity Detection (VAD) systems have been created to recognize speech presence or absence, thus ensuring consistent and effective operation.
\subsubsection{Datasets}
Voice activity detection models can be trained and evaluated using various datasets, each with unique features. The TIMIT dataset is popular, providing, $6300$ phonetically transcribed utterances from 630 speakers. On the other hand, CHiME-$5$ is designed for speech separation and recognition in real-world environments and includes multichannel recordings of $20$ speakers in locations such as cafés, buses, and pedestrian areas. Despite its primary purpose, CHiME-5 is widely used for voice activity detection. AURORA-$4$ is specifically designed to evaluate the robustness of ASR systems and contains over $10,000$ in noisy speech utterances recorded in environments like car noise, babble noise, and street noise. It is also extended to VAD for evaluating challenging scenarios. DEMAND is a suitable dataset for evaluating VAD algorithms as it includes over 1200 artificially created noise signals with various noise types like white noise, pink noise, and café noise. Finally, VoxCeleb contains over 100,000 utterances from more than 6,000 speakers, primarily designed for speaker recognition systems evaluation, but it can also be used for voice activity detection.
\subsubsection{Models}
Recent advances in deep learning have greatly improved the performance of voice activity detection (VAD), particularly in noisy environments \cite{rho2022vad,mihalache2022using}. To further improve VAD accuracy, researchers have explored various deep learning architectures, including NAS-VAD \cite{rho2022vad} and self-attentive VAD \cite{jo2021self}. NAS-VAD employs neural architecture search to reduce the need for human effort in network design and has demonstrated superior performance in terms of AUC and F1-score compared to other models. Similarly, self-attentive VAD uses a self-attention mechanism to capture long-term dependencies in input signals and has also outperformed other models on the TIMIT dataset. Additionally, a deep neural network (DNN) system has been proposed for automatic speech detection in audio signals \cite{mihalache2022using}. This system uses MLPs, RNNs, and CNNs, with CNNs delivering the best performance. Furthermore, a hybrid acoustic-lexical deep learning approach has been proposed for deception detection, combining both acoustic and lexical features.
\subsection{Speech Quality Assessment}
\subsubsection{Task Description}
Speech quality assessment is a crucial process that involves the objective evaluation of speech signals using various metrics and measures. The primary aim of this assessment is to determine the level of intelligibility and comprehensibility of speech to a human listener. Although human evaluation is considered the gold standard for assessing speech quality, it can be time-consuming, expensive, and not scalable. Mean opinion score (MOS) is the most commonly used and reliable method of obtaining human judgments for speech quality estimation. Accurate speech quality assessment is essential in the development and design of real-world applications such as ASR, Speech Enhancement, and VoIP.
\subsubsection{Datasets}
The speech quality assessment algorithms are evaluated using several datasets, each with unique characteristics. The TIMIT Acoustic-Phonetic Continuous Speech Corpus \cite{garofolo1993timit} has clean speech recordings and artificially generated degraded versions for speech synthesis and quality assessment research. The NOIZEUS dataset \cite{hu2007evaluation} is designed for evaluating noise reduction and speech quality assessment algorithms, with clean speech and artificially degraded versions containing various types of noise and distortion. The ETSI Aurora databases \cite{macho2002evaluation} are used for evaluating speech enhancement techniques and quality assessment algorithms, containing speech recordings with different types of distortions like acoustic echo and background noise. Furthermore, for training and validation, the clean speech recordings from the DNS Challenge \cite{reddy2020interspeech} can be used along with the noise dataset such as FSDK50 \cite{fonseca2021fsd50k} for additive noise degradation.
\subsubsection{Models}
Current objective methods such as Perceptual Evaluation of Speech Quality (PESQ) \cite{rix2001perceptual} and  Perceptual Objective Listening Quality Assessment (POLQA) \cite{beerends2013perceptual} for evaluating the quality of speech mostly rely on the availability of the corresponding clean reference. These methods fail in real-world scenarios where the ground truth clean reference is unavailable. In recent years, several attempts to automatically estimate the MOS using neural networks for performing quality assessment and predicting ratings or scores have attracted much attention \cite{soni2016novel,ooster2019improving,catellier2020wawenets,dong2020attention,dong2020pyramid,cauchi2019non}. These approaches outperform traditional approaches without the need for a clean reference. However, they lack robustness and generalization capabilities, limiting their use in real-world applications. 
The authors in \cite{ooster2019improving} explore Deep machine listening for Estimating Speech Quality (DESQ) for predicting the perceived speech quality based on phoneme posterior probabilities obtained using a deep neural network. 

In recent years, there have been several quality assessment frameworks developed to estimate speech quality, such as NORESQA \cite{manocha2021noresqa} based on non-matching reference (NMR). NORESQA takes inspiration from the human ability to assess speech quality even when the content is non-matching. Additionally, NORESQA introduces two new metrics - NORESQA-score, which is based on SI-SDR for speech, and NORESQA-MOS, which evaluates the Mean Opinion Score (MOS) of a speech recording using non-matching references. A recent extension to NORESQA, known as NORESQA-MOS, has been proposed in \cite{manocha2022speech}. The primary difference between these frameworks is that while NORESQA estimates speech quality using non-matching references through NORESQA-score and NORESQA-MOS, NORESQA-MOS is specifically designed to assess the MOS of a given speech recording using NMRs.

\subsection{Speech Separation}
\subsubsection{Task Description}
Speech separation refers to separating a mixed audio signal into its sources, including speech, music, and background noise. The problem is often referred to as the cocktail party problem \cite{haykin2005cocktail}, as it mimics the difficulty of listening to a conversation in a noisy room with multiple speakers. This problem is particularly relevant in real-world scenarios such as phone conversations, meetings, and live events, where various extraneous sounds may contaminate speech. Traditionally, speech separation has been studied as a signal-processing problem, where researchers have focused on developing algorithms to separate sources based on their spectral characteristics \cite{zeremdini2015comparison,vincent2018audio}. However, recent advances in machine learning have led to a new approach that formulates speech separation as a supervised learning problem \cite{hershey2016deep,wang2018alternative,luo2018real}. This approach has seen a significant improvement in performance with the advent of deep neural networks, which can learn complex relationships between input features and output sources.

\subsubsection{Datasets}
The WSJ0-2mix dataset comprises mixtures of two Wall Street Journal corpus (WSJ) speakers. It consists of a training set of 30,000 mixtures and a test set of 5000 mixtures, and it has been widely used to evaluate speech separation algorithms. CHiME-4 is a dataset that contains recordings of multiple speakers in real-world environments, such as a living room, a kitchen, and a café and is designed to test algorithms in challenging acoustic environments. TIMIT-2mix is a dataset based on the TIMIT corpus, consisting of mixtures of two speakers, and includes a training set of 462 mixtures and a test set of 400 mixtures. The dataset provides a more controlled environment than CHiME-4 to test speech separation algorithms. LibriMix is derived from the LibriSpeech corpus and includes mixtures of up to four speakers, with a training set of 100,000 mixtures and a test set of 1,000 mixtures, providing a more realistic and challenging environment than WSJ0-2mix. Lastly, the MUSDB18 dataset contains mixtures of music tracks separated into individual stems, including vocals, drums, bass, and other instruments. It consists of a training set of 100 songs and a test set of 50 songs. Despite not being specifically designed for that purpose, it has been used as a benchmark for evaluating speech separation algorithms.

\subsubsection{Models}
Deep Clustering++ \cite{hershey2016deep}, first proposed in 2015, employs deep neural networks to extract features from the input signal and cluster similar feature vectors in a latent space to separate different speakers. The model's performance is improved using spectral masking and a permutation invariant training method. The advantage of this model is its ability to handle multiple speakers, but it also has a high computational cost. Chimera++ \cite{wang2018alternative} is another effective model that combines deep clustering with mask-inference networks in a multi-objective training scheme. The model is trained using a multitask learning approach, optimizing speech enhancement and speaker identification. Chimera++ can perform speech enhancement and speaker identification but has a relatively long training time.

TasNet v2 \cite{luo2018real} employs a convolutional neural network (CNN) to process the input signal and generate a time-frequency mask for each source. The model is trained using an invariant permutation training (PIT) method \cite{kolbaek2017multitalker}, which enables it to separate multiple sources accurately. TasNet v2 achieves state-of-the-art performance in various speech separation tasks with high separation accuracy, but its disadvantage is its relatively high computational cost. The variant of TasNet based on CNNs is proposed in \cite{luo2019conv}. The model is called Conv-TasNet and can generate a time-frequency mask for each source to obtain the separated source's signal. Compared to previous models, Conv-TasNet has faster processing time but lower accuracy.

\begin{table*}[ht]
\centering
\caption{Table comparing the performance of different speech separation methods using SI-SDRi metrics on various speech separation benchmarks.}
\resizebox{\textwidth}{!}{%
\begin{tabular}{llcccccccc}
\hline
Model & Architecture & WSJ0-2mix & WSJ0-3mix & WSJ0-5mix & Libri2Mix & Libri5Mix & Libri10Mix & Libri20Mix & WHAM \\ \hline
Separate And Diffuse \cite{lutati2023separate} & Diffusion & 23.9 & 20.9 & - & 21.5 & 14.2 & 9 & 5.2 & - \\
MossFormer (L) \cite{zhao2023mossformer} & Transformer & 22.8 & 21.2 & - & - & - & - & - & - \\
MossFormer (M) \cite{zhao2023mossformer} & Transformer & 22.5 & 20.8 & - & - & - & - & - & 17.3 \\
SepFormer \cite{subakan2021attention} & Transformer & 22.3 & 19.5 & - & - & - & - & - & - \\
Sandglasset \cite{lam2021sandglasset} & Transformer + LSTM & 21.0 & 19.5 & - & - & - & - & - & - \\
Hungarian PIT \cite{dovrat2021many} & RNN & - & - & 13.22 & - & 12.72 & 7.78 & 4.26 & - \\
TDANet (L) \cite{li2022efficient} & Transformer + CNN & - & - & - & 17.4 & - & - & - & 15.2 \\
TDANet \cite{li2022efficient} & Transformer + CNN & - & - & - & 16.9 & - & - & - & 14.8 \\
Sepit \cite{lutati2022sepit} & CNN & 22.4 & 20.1 & - & - & 13.7 & 8.2 & - & - \\
Gated DualPathRNN \cite{nachmani2020voice} & CNN + LSTM & 20.12 & 16.85 & 10.56 & - & - & - & - & - \\
Dual-path RNN \cite{luo2020dual} & LSTM & 18.8 & - & - & - & - & - & - & - \\
Conv-Tasnet \cite{luo2019conv} & CNN & 15.3 & - & - & - & - & - & - & - \\ \hline
\end{tabular}}
\label{performance:SS}
\end{table*}

In recent research, encoder-decoder architectures have been explored for effectively separating source signals. One promising approach is the Hybrid Tasnet architecture \cite{yang2019improved}, which utilizes an encoder to extract features from the input signal and a decoder to generate the independent sources. This hybrid architecture captures both short-term and long-term dependencies in the input signal, leading to improved separation performance. However, it should be noted that this model's higher computational cost should be considered when selecting an appropriate separation method.

Dual-path RNN \cite{luo2020dual} uses RNN architecture to perform speech separation. The model uses a dual-path structure \cite{luo2020dual} to capture low-frequency and high-frequency information in the input signal. Dual-path RNN achieves impressive performance in various speech separation tasks. The advantage of this model is its ability to capture low-frequency and high-frequency information, but its disadvantage is its high computational cost. Gated DualPathRNN \cite{nachmani2020voice} is a variant of Dual-path RNN that employs gated recurrent units (GRUs) to improve the model's performance. The model uses a gating mechanism to control the flow of information in the recurrent network, allowing it to capture long-term dependencies in the input signal. Gated DualPathRNN achieves state-of-the-art performance in various speech separation tasks. The advantage of this model is its ability to capture long-term dependencies, but its disadvantage is its higher computational cost than other models.

Wavesplit \cite{zeghidour2021wavesplit} employs a Wave-U-Net \cite{stoller2018wave} architecture to perform speech separation. The model uses a fully convolutional neural network to extract features from the input signal and generate a time-frequency mask for each source. Wavesplit achieves impressive performance in various speech separation tasks. The advantage of this model is its high separation accuracy and relatively fast processing time, but its disadvantage is its relatively high memory usage.

Numerous studies have investigated the application of Transformer architecture in the context of speech separation. One such study is SepFormer \cite{subakan2021attention}, which has yielded encouraging outcomes on the WSJ0-2mix and WSJ0-3mix datasets, as evidenced by the data presented in \Cref{performance:SS}. Additionally, MossFormer \cite{zhao2023mossformer} is another cutting-edge architecture that has successfully pushed the boundaries of monaural speech separation across multiple speech separation benchmarks. It is worth noting that although both models employ attention mechanisms, MossFormer integrates a blend of convolutional modules to further amplify its performance.

Diffusion models have been proven to be highly effective in various machine learning tasks related to computer vision, as well as speech-processing tasks. The recent development of DiffSep \cite{scheibler2022diffusion} for speech separation, which is based on score-matching of a stochastic differential equation, has shown competitive performance on the VoiceBank-DEMAND dataset. Additionally, Separate And Diffuse \cite{lutati2023separate}, another diffusion-based model that utilizes a pretrained diffusion model, currently represents the state-of-the-art performance in various speech separation benchmarks (refer to \Cref{performance:SS}). These advancements demonstrate the significant potential of diffusion models in advancing the field of machine learning and speech processing.

\subsection{Spoken Language Understanding}
\subsubsection{Task Description}
Spoken Language Understanding (SLU) is a rapidly developing field that brings together speech processing and natural language processing to help machines comprehend human speech and respond appropriately. The ultimate goal of SLU is to bridge the gap between human and machine understanding. Typically, SLU tasks involve identifying the domain or topic of a spoken utterance, determining the speaker's intent or goal in making the utterance, and filling in any relevant slots or variables associated with that intent. For example, consider the spoken utterance, "\textit{What is the weather like in San Francisco today}?" An SLU system would need to identify the domain (weather), the intent (obtaining current weather information), and the specific slot to be filled (location-San Francisco) to generate an appropriate response. By improving SLU capabilities, we can enable more effective communication between humans and machines, making interactions more natural and efficient.

Data-driven methods are frequently utilized to achieve these tasks, employing large datasets to train models capable of accurately recognizing and interpreting spoken language. Among these methods, machine learning techniques, such as deep neural networks, are widely employed, given their exceptional ability to handle complex and ambiguous speech data. The SLU task may be subdivided into the following categories for greater clarity.
\begin{itemize}
    \item\textit{ Keyword Spotting}: Keyword Spotting (KS) is a technique used in speech processing to identify specific words or phrases within spoken language. It involves analysing audio recordings and detecting instances of pre-defined keywords or phrases. This technique is commonly used in applications such as voice assistants, where the system needs to recognize specific commands or questions from the user.
    \item \textit{Intent Classification}: Intent Classification (IC) is a spoken language understanding task that involves identifying the intent behind a spoken sentence. It is usually implemented as a pipeline process, with a speech recognition module followed by text processing that classifies the intents. However, end-to-end intent classification using speech has numerous advantages compared to the conventional pipeline approach using AST followed by NLP modules.
    \item \textit{Slot Filling}: Slot Filling (SF) is a widely used technique in Speech Language Understanding (SLU) that enables the extraction of important information, such as names, dates, and locations, from a user's speech. The process involves identifying the specific pieces of information that are relevant to the user's request and placing them into pre-defined slots. For instance, if a user asks for the weather in a particular city, the system will identify the city name and fill it into the appropriate slot, thereby providing an accurate and relevant response.
\end{itemize}
\subsubsection{Dataset}
\begin{itemize}
\item Keyword Spotting Datasets:
\begin{itemize}
    \item \textit{\citet{coucke2019efficient}}: This dataset is a speech command recognition dataset that consists of 105,000 spoken commands in English, with each command being one of 35 keywords. The dataset is designed to be highly varied and challenging, with a diverse set of speakers and background noise conditions.
    \item \textit{\citet{leroy2019federated}}: This dataset is a federated learning-based keyword spotting dataset, it is composed of data from multiple sources that are trained together without sharing the raw data. The dataset consists of audio recordings from multiple devices and environments, with the goal of improving the robustness of KS across different devices and settings
    \item \textit{Auto-KWS} \cite{wang2021auto}: This dataset is automatically generated using TTS approach. The dataset consists of 1000 keywords spoken by 100 different synthetic voices, with variations in accent, gender, and age.
    \item  \textit{Speech Commands} \cite{warden2018speech}: This data is a large-scale dataset for KS task that consists of over $100,000$ spoken commands in English, with each command belonging to 35 different keywords. The dataset is specifically designed to be highly varied and challenging, with a diverse set of speakers and background noises. It is commonly used as a benchmark dataset for KS research.
\end{itemize}
\item  Intent Classification and Slot Filling
\begin{itemize}
    \item \textit{ATIS} \cite{hemphill1990atis}: The Airline Travel Information System (ATIS) dataset is a collection of spoken queries and responses related to airline travel, such as flight reservations, flight status, and airport information. The dataset is annotated with both intent labels (e.g. “flight booking”, “flight status inquiry") and slot labels (e.g. depart city, arrival city, date). The ATIS dataset has been used extensively as a benchmark for natural language understanding models.
    \item \textit{SNIPS} \cite{coucke2018snips}: SNIPS is a dataset of voice commands designed for building a natural language understanding system. It consists of thousands of examples of spoken requests, each annotated with the intent of the request (e.g. “play music”, “set an alarm”, etc.). The dataset is widely used for training IC and SF models. 
    \item \textit{Fluent Speech Commands} \cite{lugosch2019speech}: It is a dataset of voice commands for controlling smart home devices, such as lights, thermostats, and locks. The dataset consists of over 1,5000 spoken commands, each labeled with the intended devices and action (e.g. “turn on the living room lights”, “set the thermostat to 72 degrees”). The dataset is designed to have variations in speaker accent, background noise, and device placement.
    \item \textit{MIT-Restaurant and MIT-Movie} \cite{liu2013asgard}: These are two datasets created by researchers at MIT for training natural language understanding models from restaurant and movie information requests. The dataset contains spoken and text-based queries, each labeled with the intent of the request (e.g. “find a nearby Italian restaurant”,” get information about the movie Inception”) and relevant slot information (e.g. restaurant type, movie name, etc). The datasets are widely used for benchmarking natural language understanding models. 
\end{itemize}
\end{itemize}

\subsubsection{Models}

\begin{itemize}
    \item\textit{ Keyword Spotting}: The state-of-the-art techniques for keyword spotting in speech involve deep learning models, such as CNNs \cite{rostami2022keyword} and transformers \cite{berg2021keyword}. Wav2Keyword is one of the popular model based on Wav2Vec2.0 architecture \cite{seo2021wav2kws} and have achieved SOTA results on Speech Commands data V1 and V21. Another model that achieves SOTA classification accuracy on the Google Speech commands dataset is Keyword Transformer (KWT) \cite{seo2021wav2kws}. KWT uses a transformer model and achieves $98.6\%$ and $97.7\%$ accuracy on the 12 and 35-word tasks, respectively. KWT also has low latency and can be used on mobile devices. 
    \item The DIET architecture, as introduced in \cite{bunk2020diet}, is a transformer-based multitask model that addresses intent classification and entity recognition simultaneously. DIET allows for the seamless integration of various pre-trained embeddings such as BERT, GloVe, and ConveRT. Results from experiments show that DIET outperforms fine-tuned BERT and has the added benefit of being six times faster to train.
    \item \citet{chang2022exploration} investigated the effectiveness of prompt tuning on the GSLM architecture and showcased its competitiveness on various SLU tasks, such as KS, IC, and SF. Impressively, this approach achieves comparable results with fewer trainable parameters than full fine-tuning. Despite being a popular and effective technique in numerous NLP tasks, prompt tuning has not received much attention in the speech community. Additionally, other researchers have pursued a different path by utilizing pre-trained wav2vec2.0 and different adapters \cite{li2023evaluating} to attain state-of-the-art outcomes.
\end{itemize}

\begin{table*}[ht]
\centering
\caption{Comprehensive performance analysis of various models for Keyword Spotting (KS) and Slot Filling (SF) tasks, evaluated on two benchmark datasets: Google Speech Commands for KS and ATIS for SF.}
\resizebox{\textwidth}{!}{%
\begin{tabular}{cccccccc}
\toprule
\multicolumn{5}{c}{\textbf{Keyword Spotting on Google Speech Commands (Accuracy \% $\uparrow$)}}                                      & \multicolumn{3}{c}{\textbf{Slot Filling on ATIS (F1 $\uparrow$)}} \\ \midrule
Model             & Reference & Google Speech Commands V1 12 & Google Speech Commands V2 12 & Google Speech Commands V2 35 & Model                       & Reference    & ATIS      \\ \midrule
TripletLoss-res15 & \cite{vygon2021learning}      & 98.56                        & 98.37                        & 97.0                         & CTRAN                       & \cite{rafiepour2023ctran}         & 0.9846    \\
Wav2KWS           & \cite{seo2021wav2kws}      & 97.9                         & 98.5                         & 97.8                         & Bi-model with a decoder     & \cite{wang2018bi}         & 0.9689    \\
KWT-3             & \cite{berg2021keyword}      & 97.49 ±0.15                  & 98.56 ±0.07                  & 97.69 ±0.09                  & Joint BERT                  & \cite{chen2019bert}         & 0.961     \\
KWT-1             & \cite{berg2021keyword}      & 97.27 ±0.08                  & 98.43±0.08                   & 97.74 ±0.03                  & Joint BERT + CRF            & \cite{chen2019bert}         & 0.96      \\
KWT-2             & \cite{berg2021keyword}      & 97.26±0.18                   & 98.08±0.10                   & 96.95±0.14                   & SF-ID                       & \cite{niu2019novel}         & 0.958     \\
Attention RNN     & \cite{rybakov2020streaming}      & 95.6                         & 96.9                         & 93.9                         & Capsule-NLU                 & \cite{zhang2018joint}         & 0.952     \\ \bottomrule
\end{tabular}}
\label{performance:ks}
\end{table*}

Despite the remarkable progress made in the field of SLU, accurately comprehending human speech in real-life situations continues to pose significant challenges. These challenges are amplified by the presence of diverse accents, dialects, and linguistic variations. In a notable study, Vanzo et al. \cite{vanzo2016robust} emphasize the significance of SLU in facilitating effective human-robot interaction, particularly within the context of house service robots. The authors delve into the specific obstacles encountered in this domain, which encompass handling noisy and unstructured speech, accommodating various accents and speech variations, and deciphering complex commands involving multiple actions. To overcome these obstacles, ongoing research endeavors are dedicated to developing innovative solutions that enhance the precision and efficacy of SLU systems. By addressing these challenges, the aim is to enable more robust and accurate speech comprehension in diverse real-life scenarios.

Recent studies, including the comprehensive analysis of the performance of different models and techniques for Keyword Spotting (KS) and Slot Filling (SF) tasks on Google Speech Commands and ATIS benchmark datasets (\cref{performance:ks}), have furnished valuable insights into the strengths and limitations of such approaches in SLU. Capitalizing on these findings and leveraging the latest advances in deep learning and speech recognition could help us continue to expand the frontiers of spoken language understanding and drive further innovation in this domain.
\subsection{Audio/visual multimodal speech processing}
The process of speech perception in humans is intricate and involves multiple sensory modalities, including auditory and visual cues. The generation of speech sounds involves articulators such as the tongue, lips, and teeth, whose movements are critical for producing different speech sounds and visible to others. The importance of visual cues becomes more pronounced for individuals with hearing impairments who depend on lip-reading to comprehend spoken language, while individuals with normal hearing can also benefit from visual cues in noisy environments.

When investigating language comprehension and communication, it is essential to consider both auditory and visual information, as studies have demonstrated that visual information can assist in distinguishing between acoustically similar sounds that differ in articulatory characteristics. A comprehensive understanding of the interaction between these sensory modalities can lead to the development of assistive technologies for individuals with hearing impairments and enhance communication strategies in challenging listening environments.
\subsubsection{Task Description}
The tasks under audiovisual multimodal processing can be subdivided into the following categories.
\begin{itemize}
    \item \textit{Lip-reading}: Lip-reading is a remarkable ability that allows us to comprehend spoken language from silent videos. However, it is a challenging task even for humans. Recent advancements in deep learning technology have enabled the development of neural network-based lip-reading models to accomplish this task with high accuracy. These models take silent facial videos as input and produce the corresponding speech audio or characters as output. The potential applications of automatic lip-reading models are vast and diverse, including enabling videoconferencing in noisy environments, using surveillance videos as long-range listening devices, and facilitating conversations in noisy social settings. Developing these models could significantly improve our daily lives.
    \item \textit{Audiovisual speech separation}: Recent years have witnessed a growing interest in audiovisual speech separation, driven by the remarkable human capacity to selectively focus on a specific sound source amidst background noise, commonly known as the "cocktail party effect." This phenomenon poses a significant challenge in computer speech recognition, prompting the development of automatic speech separation techniques aimed at isolating individual speech sources from complex audio signals. In a noteworthy study by Ephrat et al. (2018) \citet{ephrat2018looking}, the authors proposed that audiovisual speech separation surpasses audio-only approaches by leveraging visual cues from a speaker's face to resolve ambiguity in speech signals. By integrating visual information, the model's ability to disentangle overlapping speech signals is enhanced. The implications of automatic speech separation extend across diverse applications, including assistive technologies for individuals with hearing impairments and head-mounted devices designed to facilitate effective communication in noisy meeting scenarios.
    \item \textit{Talking face generation}: Generating a realistic talking face of a target character, synchronized with a given speech and ensuring smooth transitions between facial images, is the objective of talking face generation. This task has garnered substantial interest and poses a significant challenge due to the dynamic nature of facial movements, which depend on both visual information (input face image) and acoustic information (input speech audio) to achieve accurate lip-speech synchronization. Despite its challenges, talking face generation holds immense potential for various applications, including teleconferencing, creating virtual characters with specific facial expressions, and enhancing speech comprehension. In recent years, significant advancements have been made in the field of talking face generation, as evidenced by notable studies \cite{song2018talking, zhou2019talking, chen2019hierarchical, eskimez2020end, eskimez2021speech}.
\end{itemize}

\subsubsection{Datasets}
Several datasets are widely used for audiovisual multimodal research, including VoxCeleb, TCD-TIMID \cite{harte2015tcd} , etc. We briefly discuss some of them in the following section.
\begin{itemize}
    \item \emph{TCD-TIMID \cite{harte2015tcd}:} This is an extensive and diverse audiovisual dataset that encompasses both audio and video recordings of 600 distinct sentences spoken by 60 participants. The dataset features a wide range of speakers with different genders, accents, and backgrounds, making it highly suitable for talker-independent speech recognition research. The audio recordings are of exceptional quality, captured using high-fidelity microphones with a sampling rate of 48kHz. Meanwhile, the video footage is of 720p resolution and includes depth information for every frame
    \item \emph{LipReading in the Wild (LRW) \cite{chung2017lip}:} The LRW is a comprehensive audiovisual dataset that encompasses 500 distinct words spoken by more than 1000 speakers. This dataset has been segmented into distinct training, evaluation, and test sets to facilitate efficient research. Additionally, the LRW-1000 dataset \cite{8756582} represents a subset of LRW, featuring a 1000-word vocabulary. Researchers can benefit from pre-trained weights included with this dataset, simplifying the evaluation process. Overall, these datasets are highly regarded in the scientific community for their size and versatility in supporting research related to speech recognition and natural language processing
    \item \emph{LRS2 and LRS3 \footnote{https://www.robots.ox.ac.uk/\~vgg/data/lip\_reading/lrs2.html}}: The LRS2 and LRS3 datasets are additional examples of audiovisual speech recognition datasets that have been gathered from videos captured in real-world settings. Each of these datasets has its own distinct train/test split and includes cropped face tracks as well as corresponding audio clips sourced from British television. Both datasets are considered to be of significant value to researchers in the field of speech recognition, particularly those focused on audiovisual analysis.
    \item \emph{GRID \cite{cooke2006audio}:} This dataset comprises high-fidelity audio and video recordings of more than 1000 sentences spoken by 34 distinct speakers, including 18 males and 16 females. The sentences were gathered using the prompt "put red at G9 now" and are widely employed in research related to audio-visual speech separation and talking face synthesis. The dataset is considered to be of exceptional quality and is highly sought after in the scientific community.
\end{itemize}

\subsubsection{Models}
In recent years, there has been a remarkable surge in the development of algorithms tailored for multimodal tasks. Specifically, significant attention has been devoted to the advancement of neural networks for Text-to-Speech (TTS) applications \cite{ren2019fastspeech, ren2020fastspeech, ren2021portaspeech, kim2021conditional}. The integration of visual and auditory modalities through multimodal processing has played a pivotal role in enhancing various tasks relevant to our daily lives. Lip-reading, for instance, has witnessed notable progress in recent years, whether accompanied by audio or not. Son et al. have made a significant contribution to this field with their hybrid model \cite{son2017lip}. Combining convolutional neural networks (CNN), long short-term memory (LSTM) networks, and an attention mechanism, their model captures correlations between lip videos and audio, enabling accurate character generation. Additionally, the authors introduce a new dataset called LRS, which facilitates the development of lip-reading models.

Another noteworthy model, LiRA \cite{ma2021lira}, focuses on self-supervised learning for lip-reading. It leverages lip image sequences and audio waveforms to derive high-level representations during the pre-training stage, achieving word-level and sentence-level lip-reading capabilities. In the realm of capturing human emotions expressed through acoustic signals, Ephrat et al. \cite{ephrat2017improved} propose an innovative model that frames the task as an acoustic regression problem instead of a visual-to-text modeling approach. Their work emphasizes the advantages of this perspective. Furthermore, Vid2Speech \cite{ephrat2017vid2speech}, a CNN-based model, takes facial image sequences as input and generates corresponding speech audio waveforms. It employs a two-tower CNN model that processes facial grayscale images while calculating optical flow between frames. Additionally, other models such as those based on mutual information maximization \cite{zhao2020mutual} and spatiotemporal fusion \cite{zhang2019spatio} have been proposed for the lip-reading task, further expanding the methodologies explored in this domain.

In an early attempt to develop algorithms for audiovisual speech separation, the authors of \cite{ephrat2018looking} proposed a CNN-based architecture that encodes facial images and speech spectrograms to compute a complex mask for speech separation. Additionally, they introduced the AVspeech dataset in this work. AV-CVAE \cite{nguyen2021deep} utilizes a conditional VAE to detect the lip movements of the speaker and predict separated speech. In a deviation from speech signals, \cite{montesinos2021cappella} focuses on audiovisual singing separation and employs a two-stream CNN architecture, Y-Net \cite{mehta2018net}, to process audio and video separately. This work introduces a large dataset of solo singing videos for audiovisual singing separation. The VisualSpeech \cite{gao2021visualvoice} architecture takes a face image sequence and mixed audio of lip movement as input and predicts a complex mask. It also proposes a cross-modal embedding space to facilitate the correlation of audio and visual modalities. Finally, FaceFilter \cite{chung2020facefilter} uses still images as visual information, and other methods for the audiovisual speech separation task are proposed in \cite{afouras2018conversation,michelsanti2021overview,gabbay2017visual}.

The rise of Deepfake videos on the internet has led to a surge in demand for creating realistic talking faces for various applications, such as video production, marketing, and entertainment. Previously, the conventional approach involved manipulating 3D meshes to create specific faces, which was time-consuming and limited to certain identities. However, recent advancements in deep generative models have made significant progress. For example, DAVS \cite{zhou2019talking} introduced an end-to-end trainable deep neural network capable of learning a joint audiovisual representation, which uses adversarial training to disentangle the latent space. Another architecture proposed by ATVGnet \cite{chen2019hierarchical} consists of an audio transformation network (AT-net) and a visual generation network (VG-net) for processing acoustic and visual information, respectively. This method introduced a regression-based discriminator, a dynamically adjustable pixel-wise loss, and an attention mechanism. In \cite{zhu2018arbitrary}, a novel framework for talking face generation was presented, which discovers audiovisual coherence through an asymmetrical mutual information estimator. Furthermore, the authors in \cite{eskimez2020end} proposed an end-to-end approach based on generative adversarial networks that use noisy speech for talking face generation. In addition, alternative methods based on conditional recurrent adversarial networks and speech-driven talking face generation were introduced in \cite{song2018talking,eskimez2021speech}. 

\section{Advanced Transfer Learning Techniques for Speech Processing}
\subsection{Domain Adaptation}
\subsubsection{Task Description}
Domain adaptation is a field that deals with adapting a model trained on a labeled dataset from a source domain to a target domain, where the source domain differs from the target domain. The goal of domain adaptation is to reduce the performance gap between the source and target domains by minimizing the difference between their distributions. In speech processing, domain adaptation has various applications such as speech recognition \cite{bousquet2019robustness,nidadavolu2019cycle,lee2019coral+,chowdhury2022domain,hu2022domain}, speaker verification \cite{xia2019cross,chen2021self,wang2019vae,zhang2023meta,himawan2019deep}, and speech synthesis \cite{xin2020cross,yue2022exploring}. This section explores the use of domain adaptation in these tasks by reviewing recent literature on the subject. Specifically, we discuss the techniques used in domain adaptation, their effectiveness, and the challenges that arise when applying them to speech processing.

\subsubsection{Models}
Various techniques have been proposed to adapt a deep learning model for speech processing tasks. An example of a technique is reconstruction-based domain adaptation, which leverages an additional reconstruction task to generate a communal representation for all the domains. The Deep Reconstruction Classification Network (DRCN) \cite{ghifary2016deep} is an illustration of such an approach, as it endeavors to address both tasks concurrently: (i) classification of the source data and (ii) reconstruction of the input data. Another technique used in domain adaptation is the domain-adversarial neural network architecture, which aims to learn domain-invariant features using a gradient reversal layer \cite{9688269,9892596,8461423}.

 Different domain adaptation techniques are successfully applied to different speech processing tasks, such as speaker recognition \cite{li2022coral++,hu2022domain,bousquet2019robustness,nidadavolu2019cycle} and verification \cite{chen2021self,chen2020adversarial,zhang2023meta,li2022editnet,zhu2022multi}, where the goal is to verify the identity of a speaker using their voice. One approach for domain adaptation in speaker verification is to use adversarial domain training to learn speaker-independent features insensitive to variations in the recording environment \cite{chen2020adversarial}. 
 
 Domain adaptation has also been applied to speech recognition \cite{mani2020asr,hwang2022large,yue2022exploring,sukhadia2023domain} to improve speech recognition accuracy in a target domain. One recent approach for domain adaptation in ASR is prompt-tuning \cite{dingliwal2021prompt}, which involves fine-tuning the ASR system on a small amount of data from the new domain. Another approach is to use adapter modules for transducer-based speech recognition systems \cite{majumdar2023damage,sathyendra2022contextual}, which can balance the recognition accuracy of general speech and improve recognition on adaptation domains. The Machine Speech Chain integrates both end-to-end (E2E) ASR and neural text-to-speech (TTS) into one circle \cite{yue2022exploring}. This integration can be used for domain adaptation by fine-tuning the E2E ASR on a small amount of data from the new domain and then using the TTS to generate synthetic speech in the new domain for further training.

In addition to domain adaptation techniques used in speech recognition, there has been growing interest in adapting text-to-speech (TTS) models to specific speakers or domains. This research direction is critical, especially in low-resource settings where collecting sufficient training data can be challenging. Several recent works have proposed different approaches for speaker and domain adaptation in TTS, such as AdaSpeech \cite{chen2021adaspeech, yan2021adaspeech, wu2022adaspeech}.
 
\subsection{Meta Learning}
 \subsubsection{Task Description}
Meta-learning is a branch of machine learning that focuses on improving the learning algorithms used for tasks such as parameter initialization, optimization strategies, network architecture, and distance metrics. This approach has been demonstrated to facilitate faster fine-tuning, better performance convergence, and the ability to train models from scratch, which is especially advantageous for speech-processing tasks. Meta-learning techniques have been employed in various speech-processing tasks, such as low-resource ASR \cite{hsu2020meta,indurthi2020end}, SV \cite{zhang2021meta}, TTS \cite{huang2022meta} and domain generalization for speaker recognition \cite{kang2020domain}.

Meta-learning has the potential to improve speech processing tasks by learning better learning algorithms that can adapt to new tasks and data more efficiently. Meta-learning can also reduce the cost of model training and fine-tuning, which is particularly useful for low-resource speech processing tasks. Further investigation is required to delve into the full potential of meta-learning in speech processing and to develop more effective meta-learning algorithms for different speech-processing tasks.

\subsubsection{Models}
In low-resource ASR, meta-learning is used to quickly adapt unseen target languages by formulating ASR for different languages as different tasks and meta-learning the initialization parameters from many pretraining languages \cite{hsu2020meta,singh2022improved}. The proposed approach, MetaASR \cite{hsu2020meta}, significantly outperforms the state-of-the-art multitask pretraining approach on all target languages with different combinations of pretraining languages. In speaker verification, meta-learning is used to improve the meta-learning training for SV by introducing two methods to improve the backbone embedding network \cite{chen2021improved}. The proposed methods can obtain consistent improvements over the existing meta-learning training framework \cite{kye2020meta}.

Meta-learning has proven to be a promising approach in various speech-related tasks, including low-resource ASR and speaker verification. In addition to these tasks, meta-learning has also been applied to few-shot speaker adaptive TTS and language-agnostic TTS, demonstrating its potential to improve performance across different speech technologies. Meta-TTS \cite{huang2022meta} is an example of a meta-learning model used for a few-shot speaker adaptive TTS. It can synthesize high-speaker-similarity speech from a few enrolment samples with fewer adaptation steps. Similarly, a language-agnostic meta-learning approach is proposed in \cite{lux2022language} for low-resource TTS.

\subsection{Parameter-Efficient Transfer Learning}
\label{sec:pelt}
Transfer learning has played a significant role in the recent progress of speech processing. Fine-tuning pre-trained large models, such as those trained on  LibriSpeech ~\cite{panayotov2015librispeech} or Common Voice ~\cite{ardila2019common}, has been widely used for transfer learning in speech processing. However, fine-tuning all parameters for each downstream task can be computationally expensive. To overcome this challenge, researchers have been exploring parameter-efficient transfer learning techniques that optimize only a fraction of the model parameters, aiming to improve training efficiency. This article investigates these parameter-efficient transfer learning techniques in speech processing, evaluates their effectiveness in improving training efficiency without sacrificing performance, and discusses the challenges and opportunities associated with these techniques, highlighting their potential to advance the field of speech processing.

\subsubsection{Adapters} 
\begin{figure*}[t]
    \centering
    \includegraphics[width=\columnwidth]{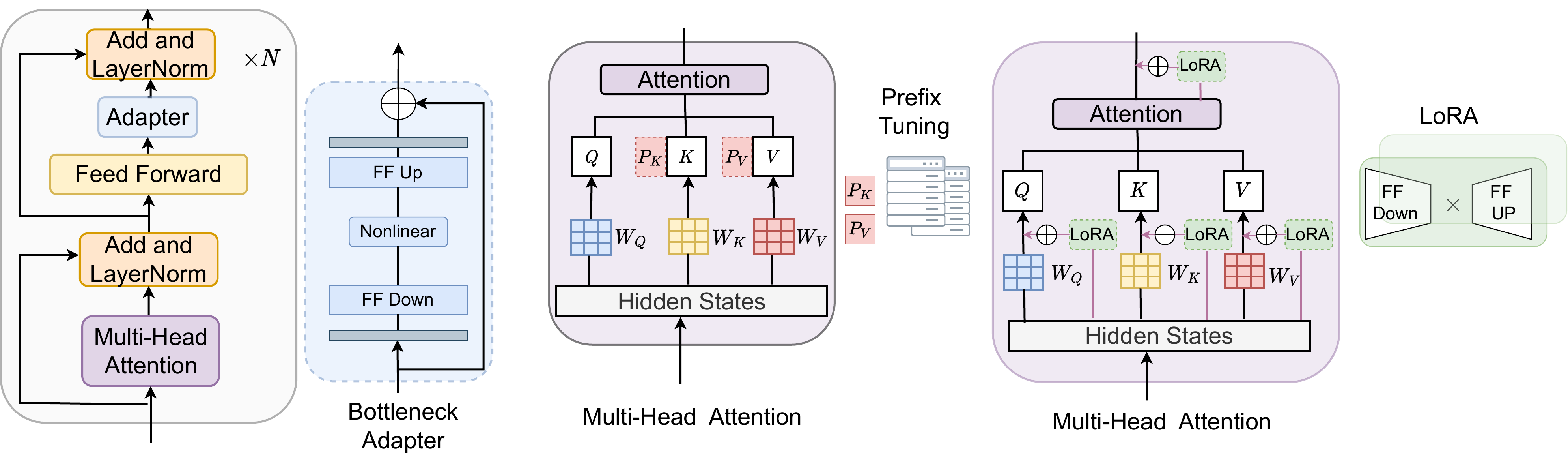}
    \caption{Transformer architecture and Adapter, Prefix Tuning, and LoRA.}
    \label{fig:adaptarchi}
\end{figure*}

\begin{figure}[h]
    \centering
    \includegraphics[width=0.6\columnwidth]{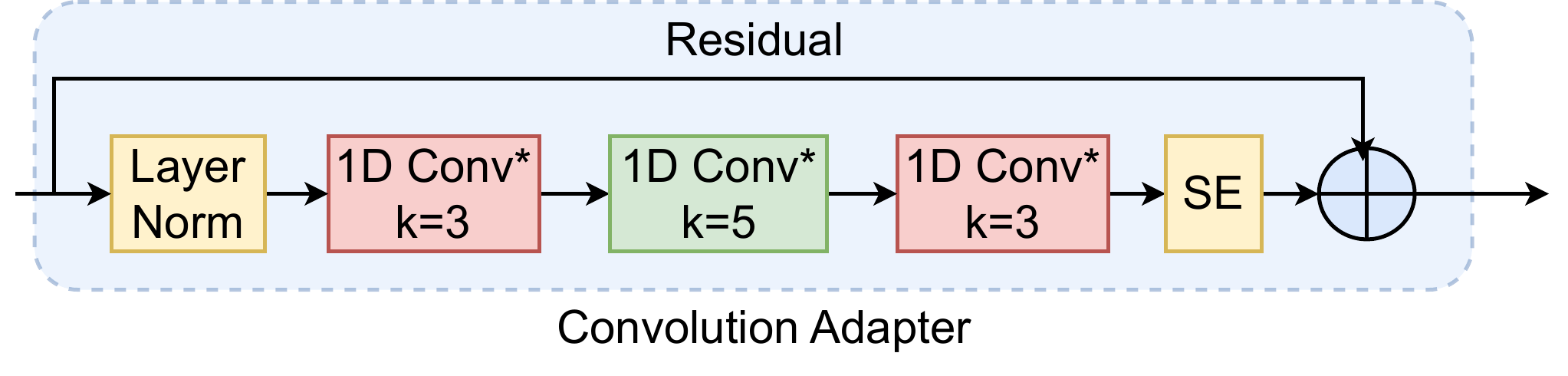}
    \caption{The architecture of 1D convolution layer-based lightweight adapter. $k$ is the kernel size of 1D convolution. $*$ denotes depth-wise convolution.}
    \label{fig:convadapt}
\end{figure}

In recent years, retrofitting adapter modules with a few parameters to pre-trained models has emerged as an effective approach in speech processing. This involves optimizing the adapter modules while keeping the pre-trained parameters frozen for downstream tasks. Recent studies (Li et al., 2023; Liu et al., 2021) \cite{li2023evaluating,DBLP:journals/corr/abs-2105-01051} have shown that adapters often outperform fine-tuning while using only a fraction of the total parameters. Different adapter architectures are available, such as bottleneck adapters (Houlsby et al., 2019)\cite{pmlr-v97-houlsby19a}, tiny attention adapters (Zhao et al., 2022)\cite{zhao2022tiny}, prefix-tuning adapters (Li and Liang, 2021)\cite{li-liang-2021-prefix}, and LoRA adapters (Hu et al., 2022)\cite{hu2022lora}, among others
Next, we will review the different approaches for parameter-efficient transfer learning. The different approaches are illustrated in  \Cref{fig:adaptarchi} and \Cref{fig:convadapt}

\vspace{1em}
\noindent \textbf{Adapter Tuning.} Adapters are a type of neural module that can be retrofitted onto a pre-trained language model, with significantly fewer parameters than the original model. One such type is the bottleneck or standard adapter (Houlsby et al., 2019; Pfeiffer et al., 2020) \cite{houlsby2019parameter, pfeiffer2020adapterfusion}. The adapter takes an input vector $h \in \mathbf{R}^{d}$ and down-projects it to a lower-dimensional space with dimensionality $m$ (where $m<d$), applies a non-linear function $g(\cdot)$, and then up-projects the result back to the original $d$-dimensional space. Finally, the output is obtained by adding a residual connection.
\begin{equation}
    \bm{h} \leftarrow \bm{h} + g(\bm{h} \bm{W}_{\text{down}}) \bm{W}_{\text{up}}
\end{equation}
where matrices $\bm{W}_{\text{down}}$ and $\bm{W}_{\text{up}}$ are used as down and up projection matrices, respectively, with $\bm{W}{\text{down}}$ having dimensions $\mathbb{R}^{d \times m}$ and $\bm{W}_{\text{up}}$ having dimensions $\mathbb{R}^{m \times d}$. Previous studies have empirically shown that a two-layer feedforward neural network with a bottleneck is effective. In this work, we follow the experimental settings outlined in \cite{pfeiffer2020adapterfusion} for the adapter, which is inserted after the feedforward layer of every transformer module, as depicted in \Cref{fig:adaptarchi}.

\vspace{1em}
\noindent \textbf{Prefix tuning.} Recent studies have suggested modifying the attention module of the Transformer model to improve its performance in natural language processing tasks. This approach involves adding learnable vectors to the pre-trained multi-head attention keys and values at every layer, as depicted in Figure \ref{fig:adaptarchi}. Specifically, two sets of learnable prefix vectors, $\bm{P_K}$ and $\bm{P_V}$, are concatenated with the original key and value matrices $\bm{K}$ and $\bm{V}$, while the query matrix $\bm{Q}$ remains unchanged. The resulting matrices are then used for multi-head attention, where each head of the attention mechanism is computed as follows:
\begin{equation}
\text{head}_{i} = \text{Attn}(\bm{Q}\bm{W}_{Q}^{(i)},[\bm{P}_{K}^{(i)},\bm{K}\bm{W}_{Q}^{(i)}],[\bm{P}_{V}^{(i)},\bm{V}\bm{W}_{Q}^{(i)}]) 
\end{equation}
where Attn($\cdot$) is scaled dot-product attention given by:
\begin{equation}
\label{attn}
    \text{Attn}(\bm{Q},\bm{K},\bm{V}) = \text{softmax} (\frac{\bm{Q}\bm{K}^{T}}{\sqrt{d_{k}}})\bm{V}
\end{equation}
The attention heads in each layer are modified by prefix tuning, with only the prefix vectors $\bm{P}{K}$ and $\bm{P}{V}$ being updated during training. This approach provides greater control over the transmission of acoustic information between layers and effectively activates the pre-trained model's knowledge.

\vspace{1em}
\noindent \textbf{LoRA.} LoRA is a novel approach proposed by Hu et al. (2021) \cite{hu2021lora}, which aims to approximate weight updates in the Transformer by injecting trainable low-rank matrices into its layers. In this method, a pre-trained weight matrix $W \in \mathbb{R}^{d \times k}$ is updated by a low-rank decomposition $\bm{W} + \Delta \bm{W} = \bm{W} + \bm{W}{\text{down}} \bm{W}{\text{up}}$, where $\bm{W}{\text{down}} \in \mathbb{R}^{d \times r}$, $\bm{W}{\text{up}} \in \mathbb{R}^{r \times k}$ are tunable parameters and $r$ represents the rank of the decomposition matrices, with $r<d$. Specifically, for a given input $\bm{x}$ to the linear projection in the multi-headed attention layer, LoRA modifies the projection output $\bm{h}$ as follows:
\begin{equation}
\label{lora}
    \bm{h} \leftarrow \bm{h} + s\cdot \bm{x} \bm{W}_{\text{down}}\bm{W}_{\text{up}}
\end{equation}
In this work,  LoRA is integrated into four locations of the multi-head attention layer, as illustrated in \Cref{fig:adaptarchi}. Thanks to its lightweight nature, the pre-trained model can accommodate many small modules for different tasks, allowing for efficient task switching by replacing the modules. Additionally, LoRA incurs no inference latency and achieves a convergence rate that is comparable to that of training the original model, unlike fully fine-tuned models \cite{hu2021lora}.

\noindent \textbf{Convolutional Adapter.} CNNs have become increasingly popular in the field of speech processing due to their ability to learn task-specific information and combine channel-wise information within local receptive fields. To further improve the efficiency of CNNs for speech processing tasks, Li et al. (2023) \cite{li2023evaluating} proposed a lightweight adapter, called the ConvAdapter, which uses three 1D convolutional layers, layer normalization, and a squeeze-and-excite module (Zhang et al., 2017) ~\cite{https://doi.org/10.48550/arxiv.1709.01507}, as shown in \cref{fig:convadapt}. By utilizing depth-wise convolution, which requires fewer parameters and is more computationally efficient, the authors were able to achieve better performance while using fewer resources. In this approach, the ConvAdapter is added to the same location as the Bottleneck Adapter (\Cref{fig:adaptarchi}).

\begin{table*}[h]
    \centering
       \caption{The study evaluated various parameter-efficient training methods on pre-trained Word2Vec 2.0, including full fine-tuning, on the SURE benchmark. The fraction of trainable parameters were represented by percentages, with the number of KS task's trainable parameters given. Results are reported using weighted-f1 as the metric (w-f1) on MELD, with the best performance in bold and the second best underlined. To avoid data imbalance, the researchers opted for using weighted-f1 as the metric. The study cites Li et al. (2023) \cite{li2023evaluating} as a reference.}
    \resizebox{0.9\textwidth}{!}{%
    \begin{tabular}{@{}llcccccccc@{}} 
    \toprule
    \multirow{2}{*}{Method}& \multirow{2}{*}{\#Parameters} &\multicolumn{2}{c}{SER (acc \% / w-f1) $\uparrow{}$} & \multicolumn{2}{c}{SR (acc \%) $\uparrow{}$} &\multicolumn{3}{c}{ASR (wer) $\downarrow$} &\multicolumn{1}{c} {KS (acc \%) $\uparrow$}\\
    \cmidrule(lr){3-4}\cmidrule(lr){5-6}\cmidrule(lr){7-9} \cmidrule(lr){10-10}
    {}& {}&ESD  & MELD & ESD  & VCTK  & ESD& FLEURS& LS & Speech Command \\
    \midrule
    {Fine Tuning} & 315,703,947 & \textbf{96.53} & 42.93 & 99.00 & 92.36 & 0.2295 & \textbf{0.135} &\textbf{0.0903 }&99.08\\
    {Adapter}& 25,467,915 (8.08\%) & \underline{94.07}  & 41.58 & 98.87  & 96.32 & \underline{0.2290} & 0.214 & 0.2425&99.19 \\
    {Prefix Tuning} & 1,739,787 (\textbf{0.55\%})  & 90.00  & \underline{44.21} & \textbf{99.73}  & \textbf{98.49} & \textbf{0.2255} & 0.166 & 0.1022&\underline{98.86}\\
    {LoRA}& 3,804,171 (1.20\%) & 90.00  & \textbf{47.05} & 99.00  & 97.61 & 0.2428 & \underline{0.149} & \underline{0.1014}&98.28 \\
    {ConvAdapter}& 2,952,539 (\underline{0.94\%}) & 91.87  & 46.30 & \underline{99.60}  & \underline{97.61} & 0.2456 & 0.2062 & 0.2958 &\textbf{98.99} \\\bottomrule
\end{tabular}
}
    \label{sure:Task1}
\end{table*}

\begin{table*}[h]
\centering
\caption{Results on SURE  benchmark for full fine-tuning and other parameter-efficient training methods on pre-trained Wav2Vec 2.0 for IC and PR tasks on  \textbf{FS}: Fluent Speech \cite{lugosch2019speech} and \textbf{LS}: LibriSpeech \cite{panayotov2015librispeech} datasets, respectively.}
\resizebox{0.9\textwidth}{!}{
\begin{tabular}{llllllll}
\hline
\multicolumn{1}{c}{\multirow{3}{*}{Method}} & \multicolumn{2}{c}{IC}                                                     & \multicolumn{2}{c}{PR}                                                     & \multicolumn{3}{c}{SF}                                                                               \\ \cline{2-8} 
\multicolumn{1}{c}{}                        & \multicolumn{1}{c}{\multirow{2}{*}{\#Parameters}} & \multicolumn{1}{c}{FS} & \multicolumn{1}{c}{\multirow{2}{*}{\#Parameters}} & \multicolumn{1}{c}{LS} & \multicolumn{1}{c}{\multirow{2}{*}{\#Parameters}} & \multicolumn{2}{c}{SNIPS}                        \\ \cline{3-3} \cline{5-5} \cline{7-8} 
\multicolumn{1}{c}{}                        & \multicolumn{1}{c}{}                              & ACC\% $\uparrow$       & \multicolumn{1}{c}{}                              & PER $\downarrow$     & \multicolumn{1}{c}{}                              & \multicolumn{1}{c}{F1 \% $\uparrow$} & \multicolumn{1}{c}{CER $\downarrow$} \\ \hline
Fine-Tuning                                 & 315707288                                         & 99.60                  & 311304394                                         & 0.0577                 & 311375119                                         & 93.89                  & 0.1411                  \\
Adapter                                     & 25471256  (8.06\%)                                & 99.39                  & 25278538 (8.01\%)                                 & 0.1571                 & 25349263 (8.14\%)                                 & 92.60                   & 0.1666                  \\
Prefix Tuning                               & 1743128  (0.55\%)                                 & 93.43                  & 1550410 (0.49\%)                                  & 0.1598                 & 1621135 (0.50\%)                                  & 62.32                  & 0.6041                  \\
LoRA                                        & 3807512 (1.20\%)                                  & 99.68                  & 3614794 (1.16\%)                                  & 0.1053                 & 3685519 (1.18\%)                                  & 90.61                      & 0.2016                       \\
ConvAdapter                                 & 3672344 (1.16\%)                                  & 95.60                  & 3479626 (1.11\%)                                  & 0.1532                 & 3550351 (1.14\%)                                  & 59.27                      & 0.6405                    \\ \hline
\end{tabular}}
\label{sure:Task2}
\end{table*}

\begin{table}[ht]
\centering
\caption{Results on the SURE benchmark for the TTS task. MCD and WER are the metrics used to compare fine-tuning and other parameter-efficient approaches.}
\resizebox{0.6\textwidth}{!}{
\begin{tabular}{lccccc}
\hline
\multicolumn{1}{c}{\multirow{2}{*}{Method}} & \multirow{2}{*}{Parameters (\%)} & \multicolumn{2}{c}{LTS} & \multicolumn{2}{c}{L2ARCTIC} \\ \cline{3-6} 
\multicolumn{1}{c}{}                        &                                  & MCD $\downarrow$       & WER   $\downarrow$      & MCD $\downarrow$           & WER $\downarrow$          \\ \hline
Fine-tuning                                  & 35802977                         & 6.2038     & 0.2655     & 6.71469       & 0.2141       \\
Adapter                                  & 659200                           & 6.1634     & 0.3143     & 6.544         & 0.2504       \\
Prefix                                      & 153600                           & 6.2523     & 0.3334     & 7.4264        & 0.3244       \\
LoRA                                        & 81920                            & 6.8319     & 0.3786     & 7.0698        & 0.3291       \\
Convadapter                                 & 108800                           & 6.9202     & 0.3365     & 6.9712        & 0.3227       \\ \hline
\end{tabular}}
\label{sure:TTS}
\end{table}

\Cref{sure:Task1}, \Cref{sure:Task2}, and \Cref{sure:TTS} present the results of various speech processing tasks in the SURE benchmark. The findings demonstrate that the adapter-based methods perform comparably well in fine-tuning. However, there is no significant advantage of any particular adapter type over others for these benchmark tasks and datasets.

\subsubsection{Knowledge Distillation (KD)} Knowledge distillation involves training a smaller model to mimic the behavior of a larger and more complex model. This can be done by training the smaller model to predict the outputs of the larger model or, by using, the larger model's hidden representations as input to the smaller model. Knowledge distillation is effective in reducing the computational cost of training and inference.

\citet{Cho20a} conducted knowledge distillation (KD) by directly applying it to the downstream task. One way to improve this approach is to use KD as pre-training for various downstream tasks, thus allowing for knowledge reuse. A noteworthy result achieved by \citet{Denisov20a} was using KD in pretraining. However, they achieved this by initializing an utterance encoder with a trained ASR model's backbone, followed by a trained NLU backbone. Knowledge distillation can be applied directly into a wav2vec 2.0 encoder without ASR training and a trained NLU module to enhance this method. \citet{Kim21a} implemented a more complex architecture, utilizing KD in both the pretraining and fine-tuning stages.

\subsubsection{Model Compression} Researchers have also explored various architectural modifications to existing models to make them more parameter-efficient. One such approach is \emph{pruning}~\cite{Frantar2023SparseGPTML,DBLP:journals/corr/abs-1910-04732}, where motivated by lottery-ticket hypothesis (LTH)~\cite{DBLP:journals/corr/abs-1803-03635}, the task-irrelevant parameters are masked based on some threshold defined by importance score, such as some parameter norm. Another form of compression could be \emph{low-rank factorization}~\cite{Hsu2022LanguageMC}, where the parameter matrices are factorized into lower-rank matrices with much fewer parameters. Finally, \emph{quantization} is a popular approach to reduce the model size and improve energy efficiency with a minimal performance penalty. It involves transforming 32-bit floating point model weights into integers with fewer bit-counts~\cite{DBLP:journals/corr/abs-2011-10680}---8-bit, 4-bit, 2-bit, and even 1-bit---through scaling and shifting. At the same time, the quantization of the activation is also handled based on the input.

\citet{DBLP:journals/corr/abs-2106-05933} iteratively prune and subsequently fine-tune wav2vec2.0 on downstream tasks to obtained improved results over fine-tuned wav2vec2.0. \citet{9053878} employ low-rank transformers to excise the model size by half and increase the inference speed by 1.35 times. \citet{peng-etal-2021-shrinking} employ KD and quantization to make wav2vec2.0 twice as fast, twice as energy efficient, and 4.8 times smaller at the cost of a 7\% increase in WER. Without the KD step, the model is 3.6 times smaller with mere 0.1\% WER degradation.

\section{Conclusion and Future Research Directions}
The rapid advancements in deep learning techniques have revolutionized speech processing tasks, enabling significant progress in speech recognition, speaker recognition, and speech synthesis. This paper provides a comprehensive review of the latest developments in deep learning techniques for speech-processing tasks. We begin by examining the early developments in speech processing, including representation learning and HMM-based modeling, before presenting a concise summary of fundamental deep learning techniques and their applications in speech processing. Furthermore, we discuss key speech-processing tasks, highlight the datasets used in these tasks, and present the latest and most relevant research works utilizing deep learning techniques.


We envisage several lines of development in speech processing:
\begin{enumerate}
    \item \emph{Large Speech Models:} In addition to the advancements made with wav2vec2.0, further progress in the field of ASR and TTS models involves the development of larger and more comprehensive models, along with the utilization of larger datasets. By leveraging these resources, it becomes possible to create TTS models that exhibit enhanced naturalness and human-like prosody. One promising approach to achieve this is through the application of adversarial training, where a discriminator is employed to distinguish between machine-generated speech and reference speech. This adversarial framework facilitates the generation of TTS models that closely resemble human speech, providing a significant step forward in achieving more realistic and high-quality synthesized speech. By exploring these avenues, researchers aim to push the boundaries of speech synthesis technology, ultimately enhancing the overall performance and realism of TTS systems.
    \item \emph{Multilingual Models:} Self-supervised learning has emerged as a transformative approach in the field of speech recognition, particularly for low-resource languages characterized by scarce or unavailable labeled datasets. The recent development of the XLS-R model, a state-of-the-art self-supervised speech recognition model, represents a significant milestone in this domain. With a remarkable scale of over 2 billion parameters, the XLS-R model has been trained on a diverse dataset spanning 128 languages, surpassing its predecessor in terms of language coverage. The notable advantage of scaling up larger multilingual models like XLS-R lies in the substantial performance improvements they offer. As a result, these models are poised to outperform single-language models and hold immense promise for the future of speech recognition. By harnessing the power of self-supervised learning and leveraging multilingual datasets, the XLS-R model showcases the potential for addressing the challenges posed by low-resource languages and advancing the field of speech recognition to new heights.
    \item \emph{Multimodal Speech Models:} Traditional speech and text models have typically operated within a single modality, focusing solely on either speech or text inputs and outputs. However, as the scale of generative models continues to grow exponentially, the integration of multiple modalities becomes a natural progression. This trend is evident in the latest developments, such as the unveiling of groundbreaking language models like GPT-4~\cite{OpenAI2023GPT4TR} and Kosmos-I~\cite{Huang2023LanguageIN}, which demonstrate the ability to process both images and text jointly. These pioneering multimodal models pave the way for the emergence of large-scale architectures that can seamlessly handle speech and other modalities in a unified manner. The convergence of multiple modalities within a single model opens up new avenues for comprehensive understanding and generation of multimodal content, and it is highly anticipated that we will witness the rapid development of large multimodal models tailored for speech and beyond in the near future.
    \item \emph{In-Context Learning:} Utilizing mixed-modality models opens up possibilities for the development of in-context learning approaches for a wide range of speech-related tasks. This paradigm allows the tasks to be explicitly defined within the input, along with accompanying examples. Remarkable progress has already been demonstrated in large language models (LLMs), including notable works such as InstructGPT~\cite{Ouyang2022TrainingLM}, FLAN-T5~\cite{Chung2022ScalingIL}, and LLaMA~\cite{Touvron2023LLaMAOA}. These models showcase the efficacy of in-context learning, where the integration of context-driven information empowers the models to excel in various speech tasks. By leveraging mixed-modality models and incorporating contextual cues, researchers are advancing the boundaries of speech processing capabilities, paving the way for more versatile and context-aware speech systems.
    \item \emph{Controllable Speech Generation:}An intriguing application stemming from the aforementioned concept is controllable text-to-speech (TTS), which allows for fine-grained control over various attributes of the synthesized speech. Attributes such as tone, accent, age, gender, and more can be precisely controlled through in-context text guidance. This controllability in TTS opens up exciting possibilities for personalization and customization, enabling users to tailor the synthesized speech to their specific requirements. By leveraging advanced models and techniques, researchers are making significant strides in developing controllable TTS systems that provide users with a powerful and flexible speech synthesis experience.
    \item \emph{Parameter-efficient Learning:} With the increasing scale of LLMs and speech models, it becomes imperative to adapt these models with minimal parameter updates. This necessitates the development of specialized adapters that can efficiently update these emerging mixed-modality large models. Additionally, model compression techniques have proven to be practical solutions in addressing the challenges posed by these large models. Recent research~\cite{DBLP:journals/corr/abs-2106-05933, 9053878, peng-etal-2021-shrinking} has demonstrated the effectiveness of model compression, highlighting the sparsity that exists within these models, particularly for specific tasks. By employing model compression techniques, researchers can reduce the computational requirements and memory footprint of these models while preserving their performance, making them more practical and accessible for real-world applications.
    \item \emph{Explainability:} Explainability remains elusive to these large networks as they grow. Researchers are steadfast in explaining these networks' functioning and learning dynamics. Recently, much work has been done to learn the fine-tuning and in-context learning dynamics of these large models for text under the neural-tangent-kernel (NTK) asymptotic framework~\cite{Malladi2022AKV}. Such exploration is yet to be done in the speech domain. More yet, explainability could be built-in as inductive bias in architecture. To this end, brain-inspired architectures~\cite{millet2022toward} are being developed, which may shed more light on this aspect of large models.
    \item \emph{Neuroscience-inspired Architectures:}In recent years, there has been significant research exploring the parallels between speech-processing architectures and the intricate workings of the human brain~\cite{millet2022toward}. These studies have unveiled compelling evidence of a strong correlation between the layers of speech models and the functional hierarchy observed in the human brain. This intriguing finding has served as a catalyst for the development of neuroscience-inspired speech models that demonstrate comparable performance to state-of-the-art (SOTA) models~\cite{millet2022toward}. By drawing inspiration from the underlying principles of neural processing in the human brain, these innovative speech models aim to enhance our understanding of speech perception and production while pushing the boundaries of performance in the field of speech processing.
    \item \emph{Text-to-Audio Models for Text-to-Speech:} Lately, transformer and diffusion-based text-to-audio (TTA) model development is turning into an exciting area of research. Until recently, most of these models~\cite {Liu2023AudioLDMTG,Kreuk2022AudioGenTG,yang2022diffsound,ghosal2023texttoaudio,wang2023audit} overlooked speech in favour of general audio. In the future, however, the models will likely strive to be equally performant in both audio and speech. To that end, current TTS methods will likely be an integral part of those models. Recently, \citet{bark} have aimed at striking a good balance between general audio and speech, although their implementation is not public, nor have they provided any detailed paper.
\end{enumerate}

\section*{Acknowledgement}
This research is supported by the Ministry of Education, Singapore, under its AcRF Tier-2 grant (Project no. T2MOE2008, and Grantor reference no. MOE-T2EP20220-0017), and A*STAR under its RIE 2020 AME programmatic grant (project reference no. RGAST2003. Any opinions, findings and conclusions or recommendations expressed in this material are those of the author(s) and do not reflect the views of the Ministry of Education, Singapore.
\bibliographystyle{ACM-Reference-Format}
\bibliography{ref.bib}
\end{document}